\documentclass[a4paper,11pt]{article}
\pdfoutput=1 
\usepackage{jcappub} 
\usepackage[T1]{fontenc} 
\usepackage{graphicx}
\usepackage{dcolumn}
\usepackage{bm}
\usepackage{hyperref}
\usepackage{aas_macros}

\title{Understanding WIMP-Baryon Interactions with Direct Detection: A Roadmap}
\author[1]{Vera Gluscevic,\note{Corresponding author.}}
\author[2, 3]{Annika H. G. Peter}

\affiliation[1]{School of Natural Sciences, Institute for Advanced Study,\\Einstein Drive, Princeton NJ 08540, USA}
\affiliation[2]{CCAPP and Department of Physics, The Ohio State University,\\191 W. Woodruff Ave., Columbus, OH 43210, USA}
\affiliation[3]{Department of Astronomy, The Ohio State University, \\140 W. 18th Ave., Columbus, OH 43210, USA}

\emailAdd{verag@ias.edu}
\emailAdd{apeter@physics.osu.edu}

\abstract{We study prospects of dark-matter direct-detection searches for probing non-relativistic effective theory for WIMP-baryon scattering. We simulate a large set of noisy recoil-energy spectra for different scattering scenarios (beyond the standard momentum-independent contact interaction), for Generation 2 and futuristic experiments. We analyze these simulations and quantify the probability of successfully identifying the operator governing the scattering, if a WIMP signal is observed. We find that the success rate depends on a combination of factors: the WIMP mass, the mediator mass, the type of interaction, and the experimental energy window. For example, for a 20 GeV WIMP, Generation 2 is only likely to identify the right operator if the interaction is Coulomb-like, and is unlikely to do so in any other case. For a WIMP with a mass of 200 GeV or higher, success is almost guaranteed. We also find that, regardless of the scattering model and the WIMP parameters, a single Generation 2 experiment is unlikely to successfully discern the momentum dependence of the underlying operator on its own, but prospects improve drastically when experiments with different target materials and energy windows are analyzed jointly.  Furthermore, we examine the quality of parameter estimation and degeneracies in the multi-dimensional parameter space of the effective theory. We find in particular that the resulting WIMP mass estimates can be severely biased if data are analyzed assuming the standard (momentum-independent) operator while the actual operator has momentum-dependence. Finally, we evaluate the ultimate reach of direct detection, finding that the prospects for successful operator selection prior to reaching the irreducible backgrounds are excellent, if the signal is just below the current limits, but slim if Generation 2 does not report WIMP detection.}
\begin{document}
\newcommand{\mwimp}{$m_\chi$}
\maketitle
\flushbottom
\section{Introduction}
\label{sec:intro}

More than five-sixths of the gravitational force in the universe is unaccounted for by baryonic matter and is sourced by an unknown particle (or particles) comprising ``dark matter'' (DM). The weakly interacting massive particle (WIMP) is the most widely sought class of candidates for DM \cite{Jungman:1995df}. The effort to characterize its particle properties includes direct, indirect, and collider searches. 

In direct-detection searches---the focus of this work---the experiments typically rely on underground low-background detectors to identify WIMP-nucleon scattering events in different target materials. Even though there is not yet definitive proof of WIMP detection, these experiments have already reached high sensitivity to a range of WIMP masses (roughly between 1 GeV and 1 TeV), and planning and/or construction of ``Generation 2'' (G2) experiments is currently underway. Proposals for G2 experiments that would be coming online in the next several years cover a wide range of detector technologies. The experiments most relevant to this work have both background discrimination on an event-by-event basis and spectral sensitivity to nuclear recoils.  Most notable among them are ton-scale noble-liquid detectors: LZ \cite{2011arXiv1110.0103M} (two-phase xenon time-projection chamber (TPC) proposed by the LUX collaboration \cite{Faham:2014hza}); Xenon1T \cite{2012arXiv1206.6288A} (two-phase xenon TPC; under construction as the next stage the Xenon experiment \cite{2012PhRvL.109r1301A}); DarkSide \cite{2013APh....49...44A} (two-phase argon TPC, with an existing prototype DarkSide50); ArDM \cite{2011JPhCS.308a2006M} (two-phase argon TPC, currently under construction); the DEAP/CLEAN family of experiments \cite{deap/clean} (single-phase argon and neon detectors, with existing prototype and the first stage DEAP-3600 under construction \cite{2012JPhCS.375a2027B}); PandaX \cite{2014arXiv1405.2882C} (xenon-based detector, with a current stage I pathfinder); and XMASS \cite{2013NIMPA.716...78A} (xenon-based detector). Additionally, there are proposals for G2 cryogenic solid-state detectors, such as, for example, SuperCDMS \cite{2013arXiv1310.8327C}\footnote{\label{loer}B. Loer, \url{http://www.pa.ucla.edu/sites/default/files/webform/loer_cdms_UCLA2014v2.pdf}} (germanium-based interleaved Z-sensitive ionization phonon (iZIP) detector, with the first stage already operational \cite{2014arXiv1402.7137A,2014PhRvL.112d1302A}), which specialize in low-mass WIMP searches by achieving very low recoil-energy thresholds.   There are design studies for the future generation of ten-ton-scale experiments (including, for example, the final stage of LZ, and the DARWIN experiment \cite{2012JPhCS.375a2028B}), which aim to reach the neutrino-background floor \cite{2009NJPh...11j5011S, 2010APh....34...90G, 2012PDU.....1...94B}. Past that point, further advances in direct searches will be fundamentally limited by the presence of irreducible backgrounds, signifying the threshold of a new regime in direct detection that is beyond the scope of this work.
Additionally, there are a number of new (and planned) experiments with either no background rejection capability, or with no (or limited) energy resolution (see, for example, Refs.~\cite{cdex2013,2013PhRvD..88a2002A,2013NIMPA.712...27B,
2013arXiv1311.3310M,Cushman:2013zza,2012PhLB..711..153A,
2012PhRvD..86e2001B}).  Since the focus of this work is using the information from spectral analysis, we do not consider these experiments here. 

In light of the large experimental effort, the goal of this work is quantitative understanding of the information extractable from direct-detection data, in particular about the microphysics of WIMP-baryon scatterings. Traditionally, direct-detection searches have focused on elastic momentum-independent (spin-dependent and spin-independent) contact interactions.  These interactions arise generically in the minimal supersymmetric standard model and other minimal extensions to the Standard Model of particle physics (for reviews, see \cite{Jungman:1995df, hooper2007}).  However, they constitute only a small subset of all possibilities.  Recently, a number of authors have explored a full set of allowed non-relativistic effective operators that describe WIMP-nucleon scattering \cite{2010JCAP...01..006C, 2010JCAP...11..042F,2010JCAP...01..020F,2012PhRvD..85l3507M, 2013JCAP...02..004F,2013arXiv1308.6288A}.  These operators have a diversity of momentum dependence and produce nuclear-recoil spectra different from those appearing in the case of supersymmetry-inspired contact interactions.  

In this work, we evaluate the potential of G2 experiments to identify the true underlying interaction operator, and distinguish it from the canonical case of momentum-independent heavy-mediator-governed contact interaction. We also investigate the ``ultimate'' reach of direct detection, as defined by the neutrino-background floor.  We specifically focus on the kinds of experiments that have good spectral resolution, because the key observational feature in this quest is the shape of the nuclear-recoil energy spectrum.
We structure this work to address the following questions: 
\begin{itemize}
\item If the signal is just below the reach of current experiments, how likely is the G2 data to discern the dominant operator mediating the WIMP-nucleon scattering interaction? What experimental capabilities are optimal for this task? 
\item What is the quality we can expect from parameter estimation under different scattering scenarios?
\item How robust will WIMP-mass estimates be to assumptions about the mediator mass and the momentum dependence of the scattering operator?
\item What is the ultimate reach of  direct searches, prior to becoming limited by the irreducible background of astrophysical neutrinos?
\end{itemize}

This paper is organized as follows. In \S\ref{sec:dd}, we briefly review the basic calculations behind direct detection analysis. In \S\ref{sec:eft}, we summarize the effective-theory (EFT) approach in categorizing the non-relativistic operators to which direct detection is sensitive, and we select a subset of operators to focus on. In \S\ref{sec:method}, we describe our simulations and details of our analysis method. In \S\ref{sec:results}, we present the results that aim to answer the questions posed above. Summary, discussion, and concluding remarks are in \S\ref{sec:conclusions}.
\section{Direct Detection Analysis: General Setup}
\label{sec:dd}
If WIMPs make up the Milky Way's DM halo and the stellar disk rotates with respect to the halo, the Earth is always in the way of a WIMP ``wind'' due to solar motion around the Galactic center with velocity $v_\mathrm{lag}$=220 km/sec \cite{kerr1986}.  This velocity scale implies that direct-detection experiments need to be sensitive to nuclear-recoil kinetic energies $E_R$ roughly of order 1-100 keV.  For most direct-detection experiments (with the notable exception of bubble chambers), the main data-product goal is the measurement of the nuclear-recoil energy spectrum within the energy window of the experiment. 
The shape of this spectrum (the differential event rate), given by $dR/dE_R$, is controlled by several factors: characteristics of the experiment (in particular, the mass of the target nucleus), particle-physics parameters (WIMP mass, type and strength of the scattering interaction, and the nuclear response), and astrophysical parameters (the velocity of the WIMP wind, the WIMP velocity distribution, and the Galactic escape velocity at the Earth's position). In this work, we assume that DM consists of only one species---a WIMP---and focus on spin-independent elastic scattering.

To obtain the event rate $dR/dE_R$ as a function of $E_R$, per unit time and unit target mass, we can start from the standard expression for a scattering rate, $n_\chi\sigma v$. In our case, the number density of scatterers $n_\chi = {\rho_\chi}/{m_\chi}$ is given by the local DM mass-density $\rho_\chi=0.3\pm 0.1$ GeV/cm$^3$ \cite{2012ApJ...756...89B} and the WIMP mass $m_\chi$. The total number of recoil events in the detector with a target of fiducial mass $m$ is then proportional to the total number of scattering centers in the target, $N_T \equiv m/m_N$, where $m_N$ is the mass of one target nucleus. For simplicity, we assume that the factor of efficiency of detection of an event at a given energy is included in the calculation of $m$ and we do not consider it here separately.\footnote{In the following, the fact that we do not consider energy dependence of the efficiency factor drives us to make slightly more conservative assumptions about what the actual energy windows for G2 experiments will be; however, the effect of this simplification on the main results and conclusions of this work is negligible.} We now need to integrate over the WIMP velocity distribution in the lab frame $f(\mathrm{\mbox{\bf{v}}})$, and also take into account that the cross-section $\sigma$ is generally a function of $E_R$ and use differential cross-section $d\sigma/dE_R$. Assuming elastic scattering and perfect energy resolution, the full expression for the observed recoil rate is
\begin{equation}
\frac{dR}{dE_R}(E_R) = \frac{\rho_\chi}{m_\chi}N_T\int\limits_{v_{\mathrm{min}}}^{v_{\mathrm{esc, lab}}}vf(\mathrm{\mbox{\bf{v}}}) d^3{v} \frac{d\sigma}{dE_R} (E_R),
\label{eq:drdq_general}
\end{equation}
where 
\begin{equation}
\mu_N=\frac{m_Nm_\chi}{m_N+m_\chi}
\end{equation}
is the WIMP-nucleus reduced mass, and the lower limit of the integral,
\begin{equation}
v_\mathrm{min} = \sqrt{(m_NE_R)/(2\mu_N^2)},
\end{equation}
represents a minimum WIMP speed necessary to produce a recoil of a given energy. The upper limit $v_\mathrm{esc, lab}$ is a hard cutoff on the speed distribution, or the Galactic escape velocity at Earth's position boosted to the lab frame by $v_\mathrm{lag}$. Observations of local stellar dynamics provide a value for the escape velocity in the Galactic rest frame, $v_\mathrm{esc}=533^{+54}_{-41}$~km/sec \cite{2014A&A...562A..91P}, which we adopt here. Given the rate above, the ``expected number of events'' $\left<N\right>$ is calculated as
\begin{equation}
\left<N\right>\equiv RT_\mathrm{obs},
\label{eq:nexp}
\end{equation} 
where $T_\mathrm{obs}$ is the exposure (in kg--years). Similarly, the total rate of events in the experimental energy window $E_R\in \left[E_{R\mathrm{, min}}, E_{R\mathrm{, max}}\right]$ is
\begin{equation}
R\equiv\int_{E_{R\mathrm{, min}}}^{E_{R\mathrm{, max}}} dE_R\frac{dR}{dE_R}(E_R).
\end{equation}

As for the rest of the astrophysical ingredients, we fix the WIMP velocity distribution in the Galactic frame $f_g$ to the isotropic Maxwell-Boltzmann distribution of the Standard Halo Model \cite{2013arXiv1310.7039P}, 
\begin{equation}
 f_g(\mathrm{\mbox{\bf{v}}}_g) = \left\{
     \begin{array}{lr}
       \frac{\rho_\chi/m_\chi}{(2\pi v_\mathrm{rms}^2)^{3/2}}e^{-v_g^2/2v_\mathrm{rms}^2}&\mathrm{for}~v_g<v_\mathrm{esc}\\
       0&\mathrm{for}~v_g\geq v_\mathrm{esc}
     \end{array}
   \right.
\end{equation}
where $\mathrm{\mbox{\bf{v}}}_g$ is measured in the Galactic frame, and $v_\mathrm{rms}$ is a one-dimensional velocity dispersion of WIMPs, set here to the standard value of $155$ km/sec \cite{peter2011}.\footnote{Note that, taken to the lab frame, $f_g$ becomes $f$.}

As for the particle-physics ingredients, so long as the momentum exchange is much smaller than the masses of the particles involved, an elastic 2--to--2 scattering can be described by a differential cross-section of the following form \cite{Jungman:1995df}
\begin{equation}
\frac{d\sigma}{dE_R}=\frac{A^2m_NF^2(E_R)}{2v^2\mu_p^2}\sigma_{SI},
\label{eq:dsdq}
\end{equation}
where $A$ is the atomic number of the target nucleus, $\mu_p$ is WIMP-proton reduced mass, $v$ is the incoming WIMP's speed in the lab frame, $F(E_R)$ is the nuclear form factor, and $\sigma_{SI}$ is the usually quoted (spin-independent) cross-section for scattering, constrained by the recent LUX results \cite{Faham:2014hza} to $\sigma_{SI}\lesssim 10^{-45}\mathrm{cm}^2$, for a WIMP of mass $33$ GeV$/c^2$.

In the following, we turn to explore a broader range of possible interactions that could govern WIMP-nucleon scattering. In particular, we adopt the EFT approach of Ref.~\cite{2010JCAP...11..042F} to parametrize generic extensions of Eq.~(\ref{eq:dsdq}) that account for possible energy (or momentum) dependence of the scattering cross-section, in the case where the mediator for the interaction is either heavy or light (compared to the momentum exchange).
\section{Non-relativistic effective operators for WIMP-nuclear scattering}
\label{sec:eft}
\begin{figure}[tbp]
\centering
\includegraphics[width=.6\textwidth,keepaspectratio=true]{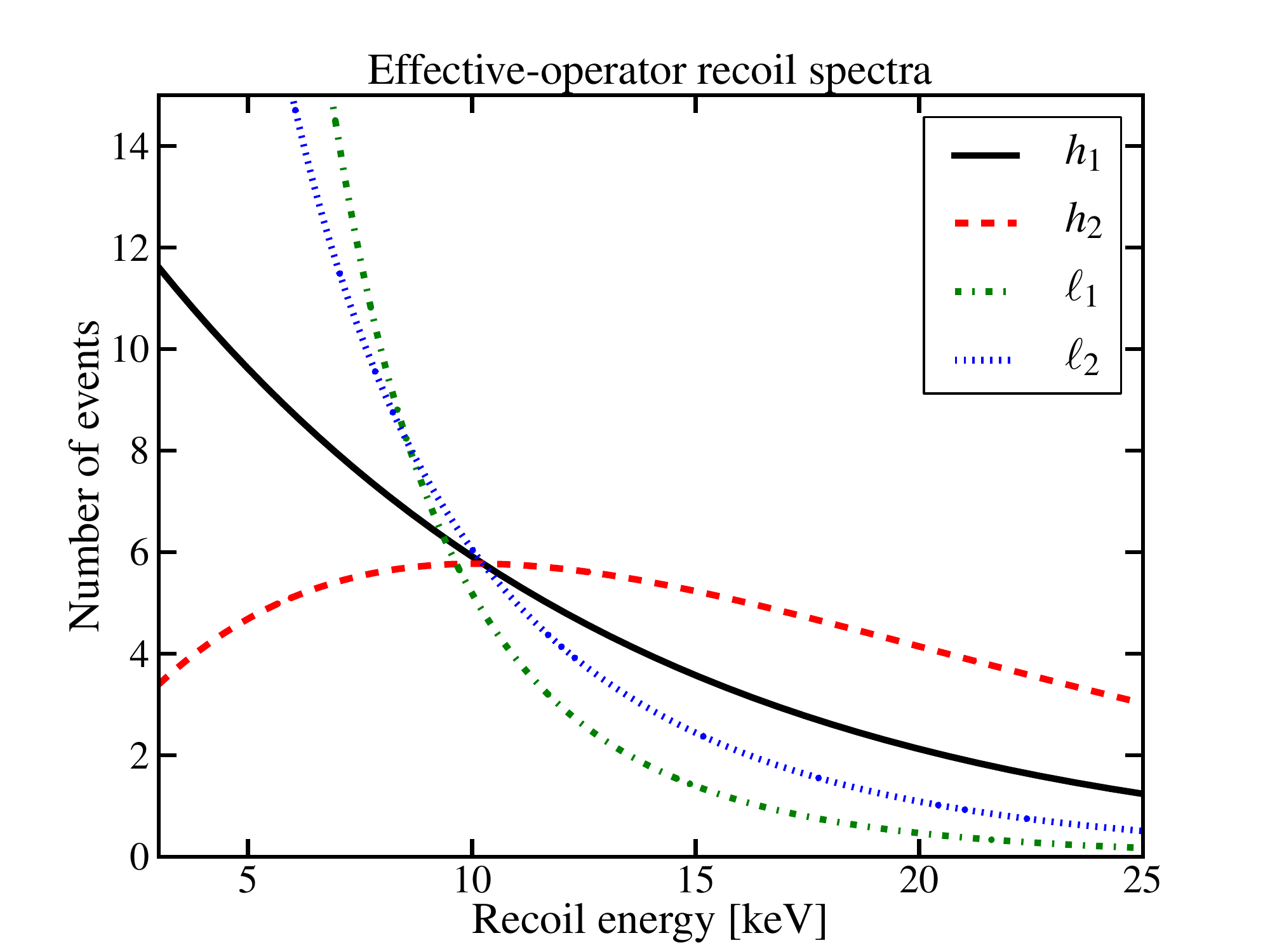}
\caption{Recoil-energy spectra for different effective operators. Different operators of Eq.~(\ref{eq:Veff}) give rise to the different recoil-energy spectra illustrated here and denoted by the corresponding Wilson coefficients: $h$'s correspond to the heavy-mediator cases (with $h_1$ being the canonical momentum-independent and spin-independent interaction), and $\ell$'s correspond to the light-mediator cases, discussed in detail in the text. These spectra are calculated for a 50 GeV WIMP, for G2 ``Xe'' experiment described in \S\ref{sec:sims}, and for coefficients set to their current upper limits, obtained from the recent LUX results \cite{Faham:2014hza}. \label{fig:operators}}
\end{figure}
Direct-detection searches and the corresponding theoretical literature predominantly focus on a simple picture in which WIMPs interact with nuclei in a momentum-independent way and the interaction is mediated by a heavy particle.  In this picture, the entire momentum (or energy) dependence of the differential cross-section (shown in Eq.~(\ref{eq:dsdq})) is contained within the form factor which captures the composite nature of the nucleus\footnote{The interaction depends either on the distribution of mass within the nucleus (spin-independent case), or on the distribution of spins (spin-dependent case), and also on the momentum transfer relative to the de Broglie wavelength of the nucleus. }. As pointed out in Refs.~\cite{2010JCAP...11..042F,2013JCAP...02..004F}, the space of allowed operators in a non-relativistic effective description of the scattering process is much broader than this.  It is therefore critical to place this broader picture in the context of direct-detection searches, and understand the implications it might have on the interpretation of any future detection signal. In addition, since the parameters of the EFT tie back to the underlying high-energy description, the experiment-driven EFT approach enables us to categorize and probe a landscape of possible theories in a systematic way. 

Motivated by these points, we explore a variety of WIMP-nucleus scattering scenarios associated with different energy-dependent effective operators. To capture the diversity of phenomenologies that might arise, we choose a subset of effective operators discussed in Ref. \cite{2010JCAP...11..042F} and focus on spin-independent case only. A full set of allowed effective operators is presented in Ref. \cite{2013JCAP...02..004F}, but we leave a comprehensive exploration of the entire set for future work. Ref. \cite{2014arXiv1406.0524C} is a first look at future prospects for detecting WIMPs in some of these other effective-operator scenarios. 

In this study, we adopt the approach of Ref.~\cite{2010JCAP...11..042F}, which parametrized recoil signals with direct observables: Wilson coefficients (or the effective couplings) of an EFT, each associated with a single momentum-dependent term in the scattering cross-section. We consider two distinct scenarios: scattering through a mediator particle with a mass much larger than the momentum transfer (Wilson coefficients denoted by $h$), and through a mediator with a much smaller mass (denoted by $\ell$), using the non-relativistic limit (since the WIMP speed is expected to be only of the order of hundreds of km/sec in Earth's rest frame). Taking only the operators suppressed by at most one power of the recoil energy and focusing on a static case, the effective potential reads
\begin{equation}
V_\mathrm{eff} = h_1\delta^3(\vec{r}) - h_2\vec{s}_\chi  \cdot \vec\nabla\delta^3(\vec{r}) + \ell_1\frac{1}{4\pi r} + \ell_2 \frac{\vec s_\chi \cdot \vec r}{4\pi r^3},
\label{eq:Veff}
\end{equation}
where $s_\chi$ is the spin of the WIMP, and $\ell$'s and $h$'s are the two light- and heavy-mediator cases, respectively. The scattering cross-section is then
\begin{multline}
\frac{d\sigma}{dE_R} = \frac{A^2F^2(E_R)m_N}{2\pi v^2}\\\times \left(\left|h_1+\frac{\ell_1}{2m_NE_R}\right|^2 + \frac{1}{4}\left|h_2\sqrt{2m_NE_R} + \frac{\ell_2}{\sqrt{2m_NE_R}}\right|^2\right).
\label{eq:dsdq_eft}
\end{multline}

The first term of Eq.~(\ref{eq:Veff}) is the canonical case of a spin-independent contact interaction, related to the standard cross-section of Eq.~(\ref{eq:dsdq}) as
\begin{equation}
h_1^2 \equiv \frac{\sigma_{SI}\pi}{\mu_p^2}.
\end{equation} 
The $h_2$ term is a T-violating operator \cite{2013JCAP...02..004F} that can arise from several scenarios, such as coupling of the dark electric dipole moment to a new gauge boson, or heavy scalar exchange. The $\ell_1$ term (the Coulomb potential) can arise through exchange of a new light boson with a mass smaller than $E_R$. The $\ell_2$ term can be due to the WIMP dipole coupling to the nucleus monopole. Recoil-energy spectra produced by each of these four terms are illustrated in Figure \ref{fig:operators}. 

We note that, in general, for each new operator an appropriate form factor should be calculated, as shown in Refs.~ \cite{2013JCAP...02..004F,2013arXiv1308.6288A,
2014arXiv1405.6690A,2014arXiv1401.3739G}. For our choice of operators, the canonical spin-independent form factor is appropriate, so we adopt the Woods-Saxon form factor \cite{Jungman:1995df} for this study.  
\section{Method}
\label{sec:method}
\subsection{Simulations}
\label{sec:sims}
\begin{table*}[tbp]
  \setlength{\extrarowheight}{2pt}
  \setlength{\tabcolsep}{5pt}
  \begin{center}
	\begin{tabular}{c|m{3cm}m{3cm}m{2cm}m{3cm}}
	$m_\chi$ & $h_1$ [$10^{-10}$ GeV$^{-2}$] & $h_2$ [$10^{-9}$ GeV$^{-3}$] & $\ell_1$ [$10^{-13}$] & $\ell_2$ [$10^{-11}$ GeV$^{-1}$]\\
	\hline\hline
	20 GeV & 22 & 100& 38& 19\\
	50 GeV & 14 & 56& 32& 14\\
	200 GeV & 23& 85& 54& 23\\
	\end{tabular}
  \end{center}
\caption{Current upper limits for the four coupling coefficients described in \S\ref{sec:eft}, calculated from no WIMP detection with LUX data \cite{Faham:2014hza}, for three benchmark WIMP masses considered in this work. }
\label{tab:limits}
\end{table*}
\begin{table*}[tbp]
  \setlength{\extrarowheight}{3pt}
  \setlength{\tabcolsep}{10pt}
  \begin{center}
	\begin{tabular}{c|m{1cm}m{2cm}m{2.cm}m{1.5cm}m{1.7cm}}
	Experiment label & Target & Nuclear mass [AMU] & Energy window [keV] & Exposure [kg-yr] \\
	\hline\hline
	Xe & Xe & 131  & 5-40 & 2000 \\
	Ar & Ar & 40 & 30-100 & 1000 \\
	Ge & Ge & 73 & 8-100 & 100  \\
	SiLT & Si & 28 & 0.1-100 & 0.2  \\
	GeLT & Ge & 73 & 0.3-100 & 4  \\
	Xe ultimate & Xe & 131 & 3-40 & 10 000 \\
	\end{tabular}
  \end{center}
\caption{Experimental parameters used for our simulations. The first five represent G2 experiments, while the last one represents the ``ultimate'' experiment that can reach the neutrino-background floor. Target-mass fiducialization and efficiency factors are folded into the exposure for each experiment, chosen to agree with the projected exclusion curves of Ref.~\cite{Cushman:2013zza} for the no-background and perfect-energy-resolution case.}
\label{tab:experiments}
\end{table*}
\begin{figure*}
\centering
\includegraphics[width=.32\textwidth,keepaspectratio=true]{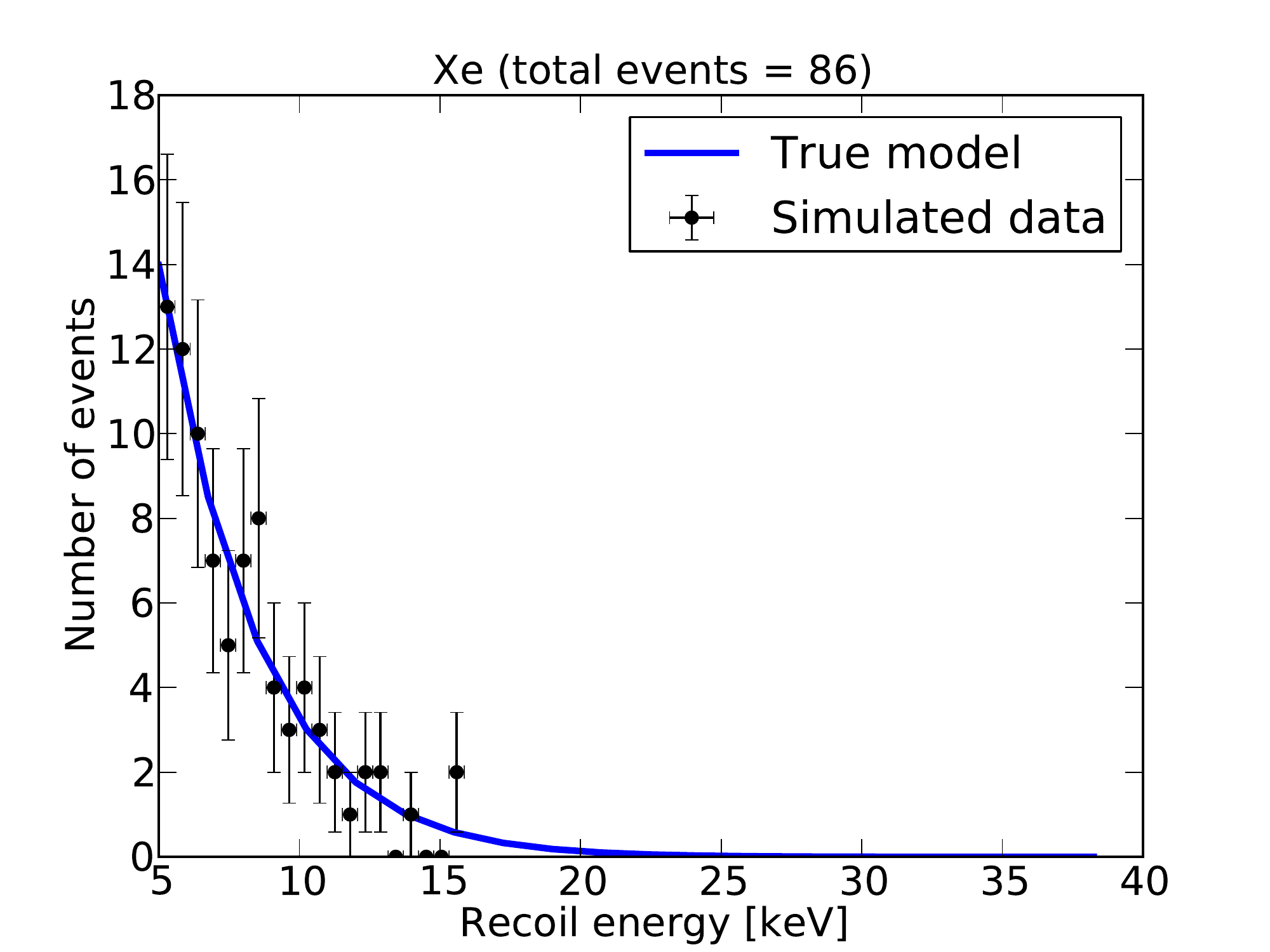}
\includegraphics[width=.32\textwidth,keepaspectratio=true]{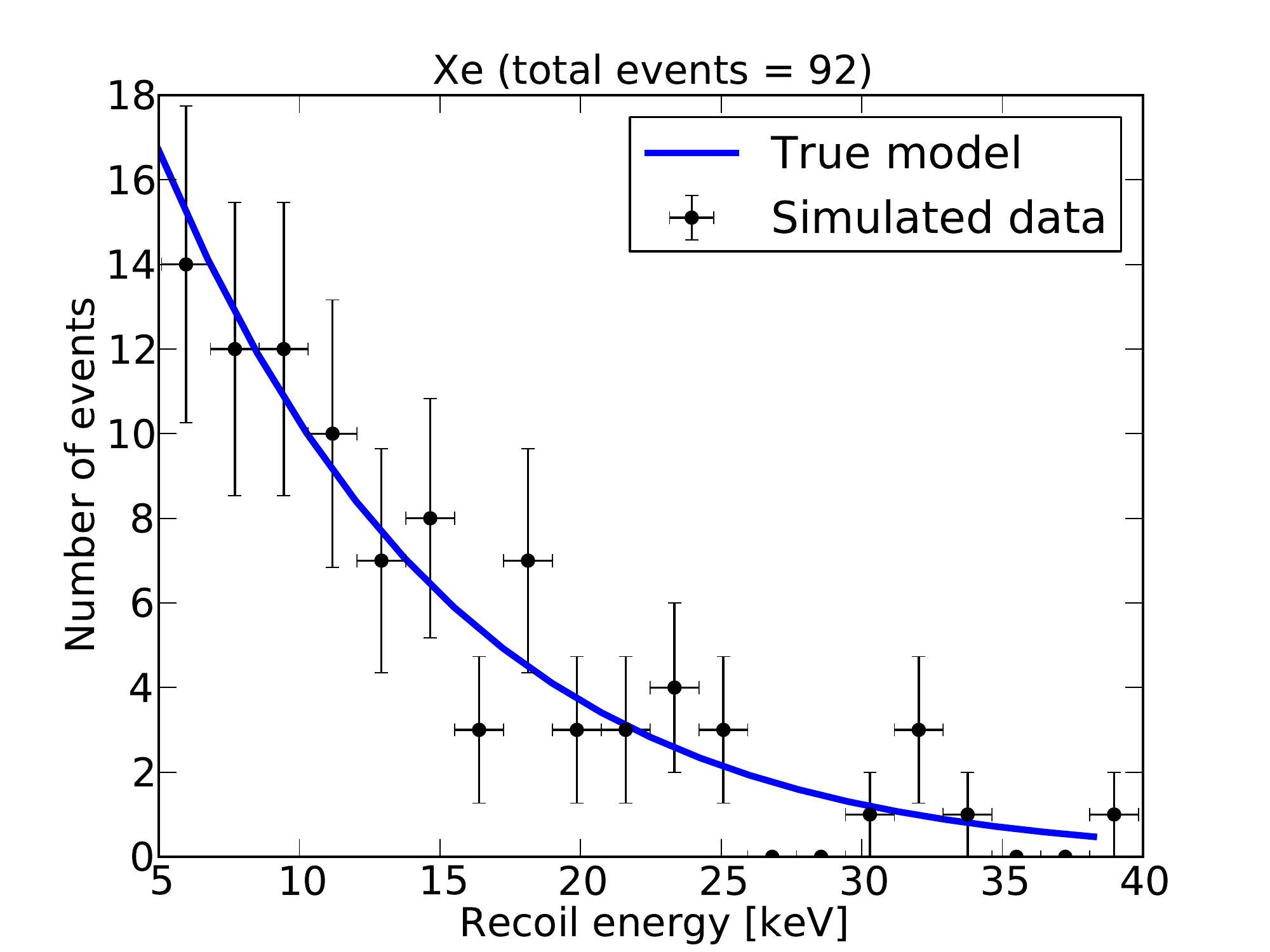}
\includegraphics[width=.32\textwidth,keepaspectratio=true]{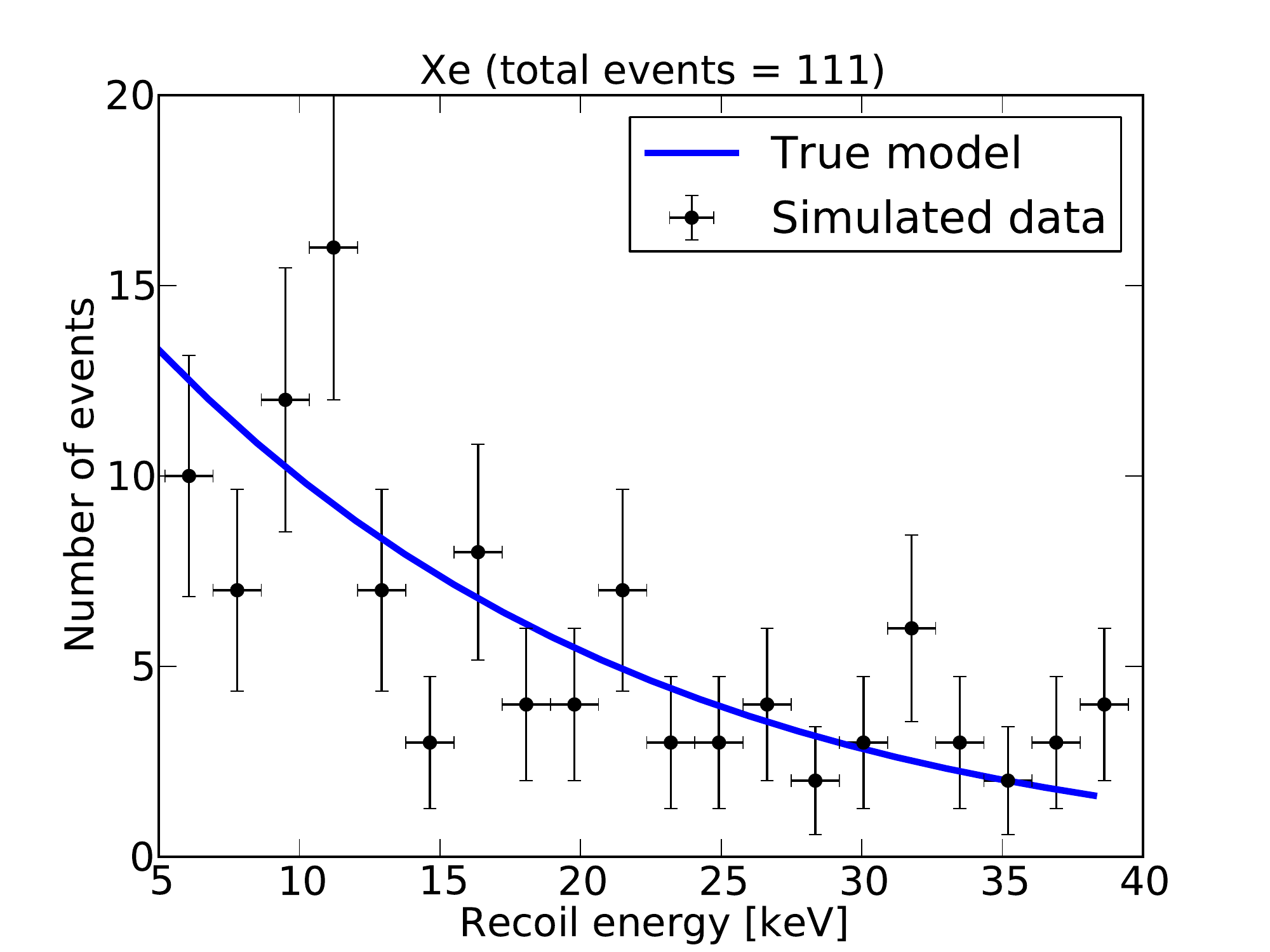}
\caption{Examples of the simulated recoil-energy spectra with Poisson-noise error bars are shown for three benchmark WIMP masses (from left to right): 20 GeV, 50 GeV, and 200 GeV. In these plots, $h_1$ is the underlying operator model. The expected number of events for each of these cases is 88, 93, and 113, respectively. We bin the events for the purposes of visualization only.\label{fig:simulated_spectra_h1}}
\end{figure*}
\begin{figure*}
\centering
\includegraphics[width=.32\textwidth,keepaspectratio=true]{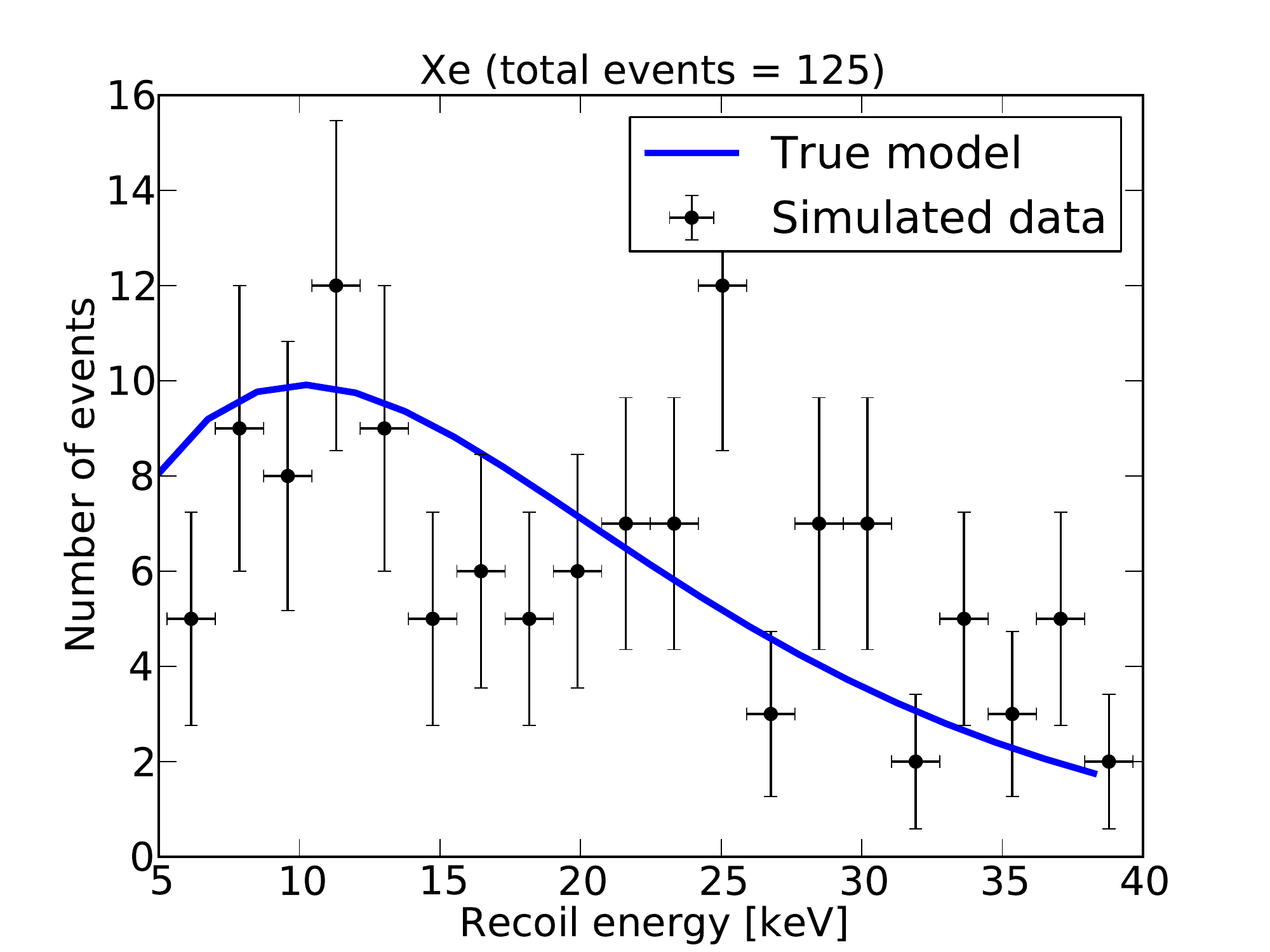}
\includegraphics[width=.32\textwidth,keepaspectratio=true]{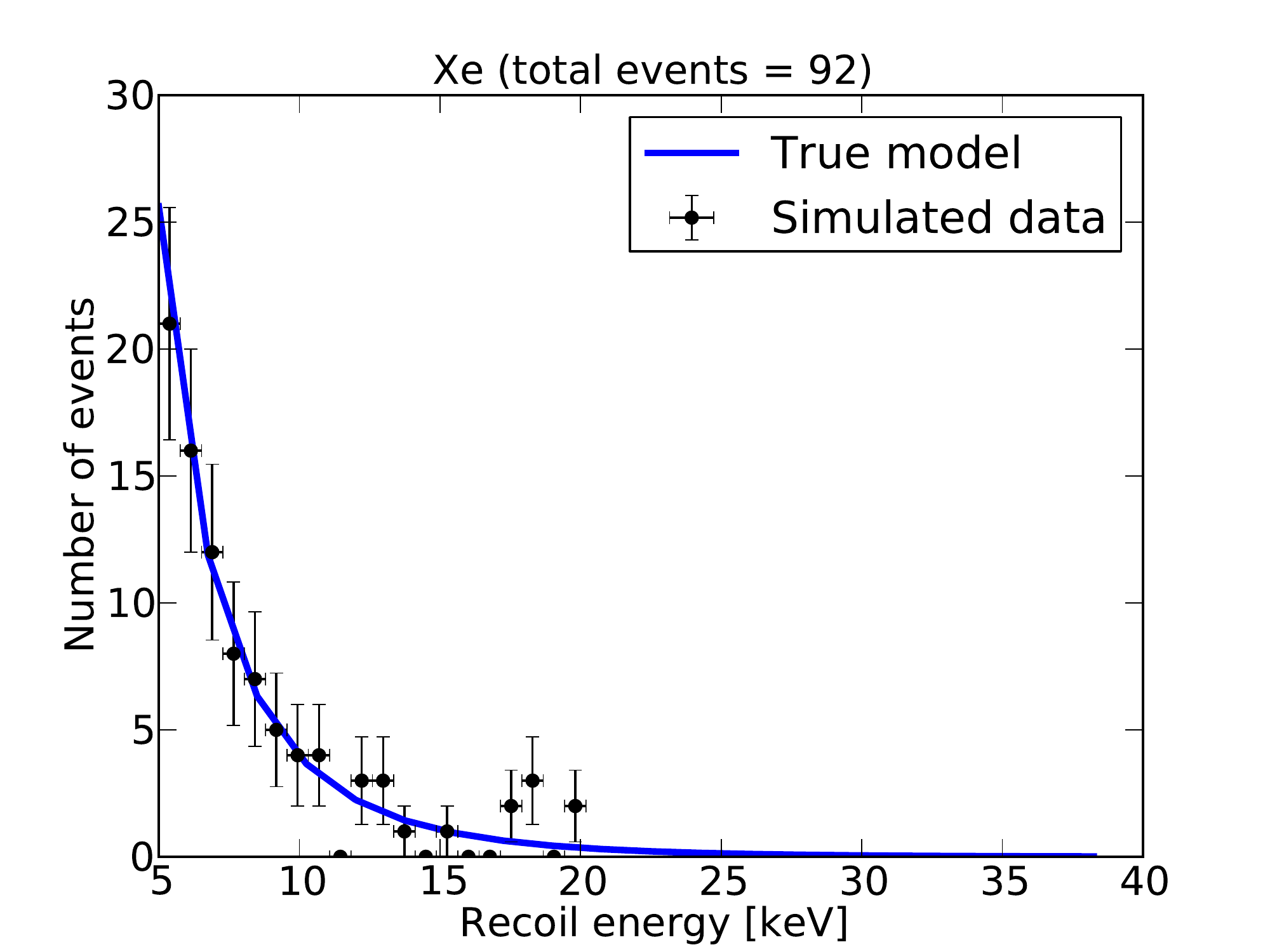}
\includegraphics[width=.32\textwidth,keepaspectratio=true]{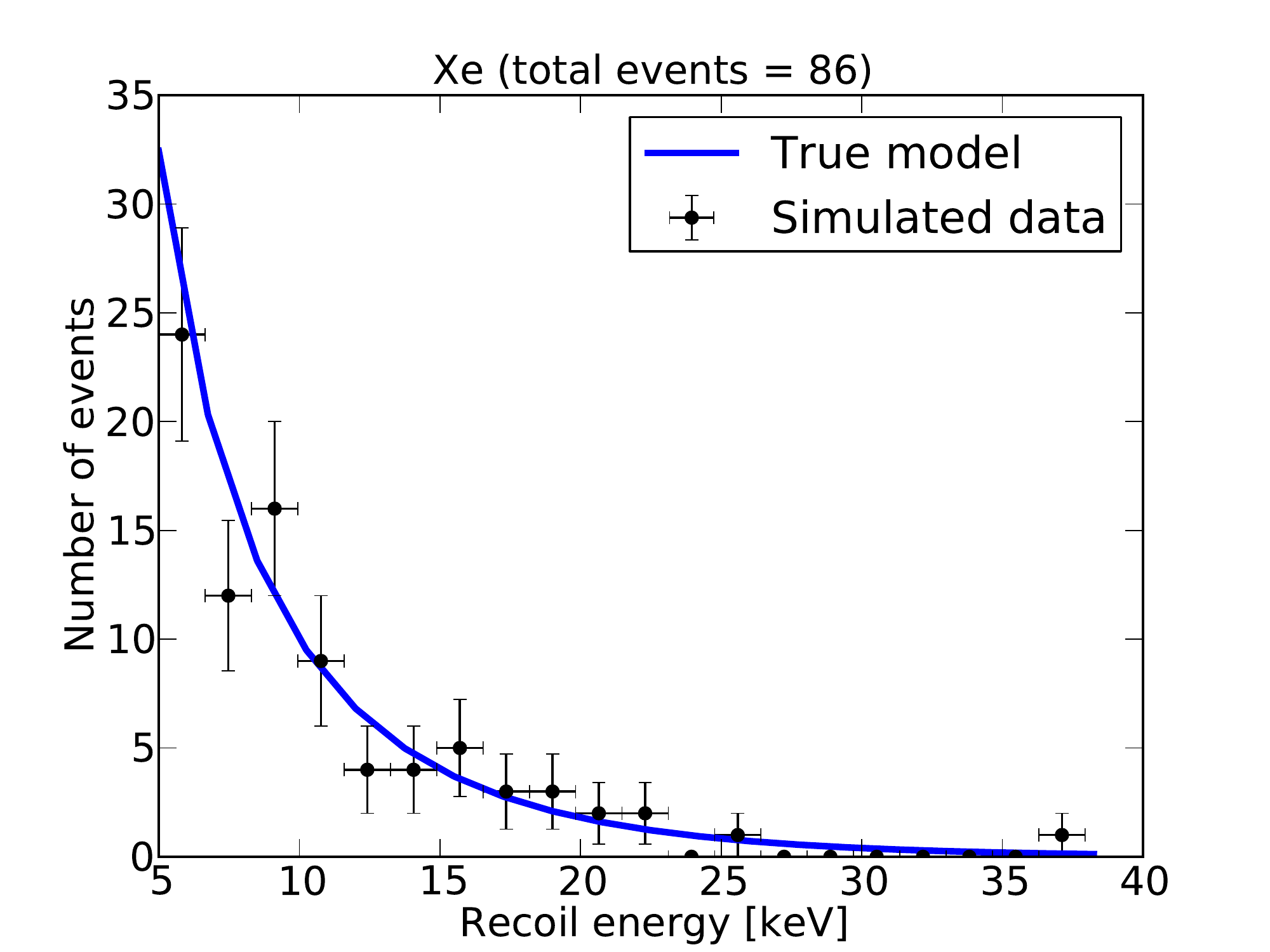}
\caption{Examples of the simulated recoil-energy spectra with Poisson-noise error bars, for WIMP mass of 50 GeV, for the following operator models (left to right): $h_2$, $\ell_1$, and $\ell_2$. The expected number of events in these three cases is 123, 91, and 91, respectively. We bin the events for the purposes of visualization only.\label{fig:simulated_spectra_all}}
\end{figure*}
Our goal is to evaluate the information direct searches can deliver about DM mass and its interaction with nuclei, should a WIMP signal be detected, and in situations where different EFT scenarios describe the scattering. Since we are doing a prediction exercise, it is important to account for the fact that Poisson noise can significantly affect the outcome. This is especially true for G2 experiments where only on the order of 100 events might be recorded (per experiment, integrated over the entire energy window), in the best case where the signal is just beyond the reach of present-day experiments.

To understand the allowed parameter space for the effective operators, we first evaluate the the current upper limits for $h_1$, $h_2$, $\ell_1$, and $\ell_2$, imposed by the recent LUX results \cite{Faham:2014hza}. We do so by assuming that the observed recoil rate of Eq.~(\ref{eq:drdq_general}) is produced by a single operator at a time, and by taking into account that less than 1 WIMP event was observed with LUX. From these conditions, we can calculate the exclusion curve corresponding to each of the four coefficients. The values in Table \ref{tab:limits} list a set of benchmark values obtained from the exclusion curves calculated with this procedure.

Next, we choose the experimental parameters (target material, fiducial target mass, exposure, and lower and upper limit of the energy window) in such a way as to capture the general scope of G2 and ``ultimate'' direct searches (rather than to represent any single experiment in detail); see the list in Table \ref{tab:experiments}. For example, the choice of parameters for the experiment we call ``Xe'' is made to be roughly representative of the plans for the leading ton-scale xenon-based experiments, such as LZ \cite{2011arXiv1110.0103M} and Xenon1T \cite{2012arXiv1206.6288A}; ``Ar'' is chosen to represent argon-based analogues of these, such as DarkSide, DEAP, and ArDM \cite{2013APh....49...44A,2013JInst...8C9005B,2014arXiv1406.0462G}; ``Ge'', ``SiLT'', and ``GeLT'' are designed to capture the projected sensitivity for G2 SuperCDMS \cite{ben_loer_ucla2014} 
with, respectively, germanium target with the nominal energy-threshold mode of operation, silicon target in the high-voltage (HV, or low-threshold, denoted here as LT) operating mode, and germanium target in HV mode. The exposures (in kg-years) include the factor of target-mass fiducialization and efficiency, and we chose them in agreement with the projected exclusion plots of Ref.~\cite{Cushman:2013zza}, given the energy windows listed in Table \ref{tab:experiments} (which are also chosen following the references listed above). Finally, we consider a 10-ton liquid-xenon experiment with a very low energy threshold, motivated by the projected performance of DARWIN and the final stage of LZ. Such an experiment would hit the astrophysical neutrino floor for WIMPs of all masses above a few GeV, and in this sense represents the ``ultimate'' reach of direct detection, before entering a regime of background-dominated measurements.  

In this work we consider perfect energy resolution and assume zero backgrounds. Previous studies have found that relaxation of the energy-resolution criterion does not significantly degrade parameter estimation \cite{2013arXiv1310.7039P}.  To account for the fact that for some of the experiments the no-background assumption does not hold (especially for the low-threshold operation mode of germanium- and silicon-based detectors), we chose their exposures such to be equivalent to the background-free case. We use this simplification to probe the limit of the performance of these experiments, regardless of what future measurements might show in terms of the backgrounds. We discuss these assumptions in more detail in \S\ref{sec:conclusions}. 

We create many noisy recoil-energy spectra realizations for each chosen set of particle parameters $\Theta\equiv\{m_\chi, h_1,h_2,\ell_1,\ell_2\}$, using the experimental parameters of Table \ref{tab:experiments}. Throughout this work, for the sake of simplicity, we assume the standard astrophysical scenario (with parameter values from \S\ref{sec:dd}), and discuss this assumption in more detail in \S\ref{sec:conclusions}. Each realization of the data for a given experiment and for an underlying $\Theta$ consists of a list of $N$ observed recoil events recorded at energies $\{E_R\}$.  To create such data set, we first randomly draw a number $N$ from a Poisson probability distribution,
\begin{equation}
P(N) = \frac{\left<N\right>^{N}e^{-\left<N\right>}}{N!},
\label{eq:P}
\end{equation}
and then assign energies to these events following the probability distribution for a single recoil to be observed at a set $E_R$,
\begin{equation}
P_1(\Theta, E_R) = \frac{dR/dE_R(E_R)}{R}.
\label{eq:P1}
\end{equation}
For a fixed choice of the dominant operator, for each value of its coupling, and for each of the three benchmark WIMP masses: 20 GeV, 50 GeV, and 200 GeV, we create one hundred data realizations for every experiment listed in Table \ref{tab:experiments}. Each of the energy spectra is analyzed as described in the following \S\ref{sec:bayes}. Examples of the simulated spectra are shown in Figures \ref{fig:simulated_spectra_h1} and \ref{fig:simulated_spectra_all}. 

In our simulations, we choose several benchmark values for Wilson coefficients. In most cases, we wish to evaluate the best-case scenario where the signal is as large as currently allowed and dominated by a single operator. Therefore, we ``turn on'' one operator at a time and set the corresponding coupling coefficient to its upper bound, while all other coefficients of Eq.~(\ref{eq:dsdq_eft}) are set to zero. This is the default setup of the simulations discussed in \S\ref{sec:results}, unless explicitly stated otherwise in the text.
\subsection{Bayesian Inferrence}
\label{sec:bayes}
\begin{table}[t]
  \setlength{\extrarowheight}{3pt}
  \setlength{\tabcolsep}{12pt}
  \begin{center}
	\begin{tabular}{m{4cm}|m{4cm}}
	Parameter name & Prior range\\
		\hline\hline
	 $m_\chi$ [GeV] & 1-1000\\
	 $h_1$ [$10^{-10}$ GeV$^{-2}$] & 0.1-10 000 \\
	 $h_2$ [$10^{-9}$ GeV$^{-3}$] & 0.1-10 000 \\
	 $\ell_1$ [$10^{-13}$] & 0.1-10 000 \\
	 $\ell_2$ [$10^{-11}$ GeV$^{-1}$] & 0.1-10 000\\
	\end{tabular}
  \end{center}
\caption{Prior ranges used for each of our \textsc{MultiNest} runs. Since the coupling coefficients are multipliers of the likelihood, and since we wish to explore several orders of magnitude in their values, all prior types are set to flat in log space.
\label{tab:priors}}
\end{table}
A natural choice of framework we adopt here to analyze our simulated data sets is Bayesian inference.  Bayesian inference was successfully used to constrain DM parameter space by a number of authors (see, for example, Refs.~\cite{2011JCAP...04..012A, 2011JCAP...09..022A}), and is also well suited for model selection, which particularly concerns this study. This framework relies on evaluating the \textit{posterior probability distribution} in a given parameter space (some subset of $\Theta$, in this case). The posterior $\mathcal{P}$ is given by Bayes' theorem,
\begin{equation}
\mathcal{P}(\Theta | \{E_R\},M) = \frac{\mathcal{L}(\{E_R\} | \Theta, M)p(\Theta, M)}{\mathcal{E}(\Theta, M)},
\label{eq:bayes_theorem}
\end{equation}
where $\mathcal{L}$ is the \textit{likelihood}, i.e.~the probability of data given the model (i.e., the underlying effective operator, or ``the hypothesis'') and particular values of the corresponding parameters; $p$ is the \textit{prior probability} that represents the status of our knowledge before considering data. Table \ref{tab:priors} lists our choice of priors \footnote{We verified that the choice of priors does not influence our results, so long as the prior range is large enough. This led to the choices listed in Table  \ref{tab:priors}, adopted for all posterior calculations presented in this study.}. The normalization  $\mathcal{E}$ is the integral of the likelihood over the entire prior space, or the \textit{evidence},
\begin{equation}
\mathcal{E}(\Theta, M)=\int d\Theta \mathcal{L}(\{E_R\} | \Theta, M)p(\Theta, M).
\label{eq:evidence}
\end{equation}
Note that $\Theta$ in Eqs.~(\ref{eq:bayes_theorem}) and (\ref{eq:evidence}) represents a subset of the full parameter set, relevant for a chosen model $M$ (for example, for the $h_1$ model, $\Theta=\{m_\chi, h_1\}$). Likelihood for a single data set is calculated as \cite{2013arXiv1310.7039P}
\begin{equation}
\mathcal{L}=P(N)\prod_{i=1}^{N} P_1(\Theta, E_R^i),
\end{equation}
where $P$ and $P_1$ are given by Eqs.~(\ref{eq:P}) and (\ref{eq:P1}), respectively. Note that this is not a binned likelihood, and the information from every recorded event in incorporated in its calculation.
The combined likelihood for multiple experiments is the product of the likelihoods for individual experiments. 

To perform model selection, we evaluate the probability of a model $M_1$ as the following ratio of evidences
\begin{equation}
p(M_1) = \frac{E(\Theta, M_1)}{\sum\limits_{i}^\mathrm{}E(\Theta,M_i)},
\label{eq:E_ratio}
\end{equation}
where the denominator is the sum of evidences for all the models considered in this analysis. For instance, in this work, the sum is taken over the four single-operator models for WIMP-nucleon scattering.
This expression assumes flat priors on models; i.e.,~it assumes that we have no \textit{a priori} reason to consider one operator a more likely hypothesis than any other. Calculated this way, Bayesian evidence is a relative measure of ``goodness-of-fit'', in the sense that Eq.~(\ref{eq:E_ratio}) quantifies how well the hypothesis of one specific operator dominating the signal compares to the hypotheses that any of the other operators (out of the four operators we consider here) dominate the signal.  If we inquire instead how likely it is that several operators are involved in the scattering, we need to consider the evidence for that specific multi-operator model, and compare it to what we think is the full set of model possibilities. 

Using an implementation of the multimodal nested-sampling algorithm \textsc{MultiNest} \cite{2009MNRAS.398.1601F, pymultinest}, we analyze each simulated data realization separately, and also jointly with data from other experiments. The analysis includes the following: 
\begin{itemize}
\item \textit{Model selection}: evaluation of the evidence ratio of Eq.~(\ref{eq:E_ratio}) for each of the four operator models, in each data realization.
\item \textit{Parameter estimation for a single-operator model}: calculation of the marginalized posterior probabilities where data was fit with two free parameters ($m_\chi$ and the coupling coefficient corresponding to the assumed operator model).
\item \textit{Agnostic parameter estimation}: calculation of the marginalized posteriors for the full $\Theta$, where the total of five free parameters was fit to each data realization.
\end{itemize}

For most of the presented analysis, we use the following \textsc{MultiNest} parameters: \texttt{efr}$=0.3$, \texttt{tol}$=0.1$, and $2000$ live points. For the purposes of this study, we created on the order of 8000 data realizations, about 42000 \textsc{MultiNest} runs, yielding a total of $3\times 10^8$ posterior evaluations, overall requiring $10^5$ CPU hours.
\section{Results}
\label{sec:results}
\subsection{Distinguishing underlying interactions: single-operator scenarios}
\label{sec:distops_single}
\begin{table*}[t]
\setlength{\extrarowheight}{3pt}
\setlength{\tabcolsep}{12pt}
\begin{center}
\begin{tabular}{c|m{3cm}m{3cm}m{3cm}}
WIMP mass & 20 GeV& 50 GeV& 200 GeV\\
\hline\hline
Xe& (88, 91, 87, 90)& (93, 123, 91, 91)& (113, 155, 91, 94)\\
Ar& (0, 1, 0, 0)& (3, 4, 2, 2)& (7, 10, 3, 5)\\
Ge& (2, 2, 3, 2)& (2, 3, 3, 3)& (4, 6, 3, 3)\\
SiLT& (0, 0, 5, 0)& (0, 0, 2, 0)& (0, 0, 1, 0)\\
GeLT& (0, 0, 27, 2)& (0, 0, 8, 1)& (0, 0, 6, 0)\\
\end{tabular}
\end{center}
\caption{Number of expected events for the single-operator simulations described in \S\ref{sec:distops_single} (where one operator is used at a time, and the corresponding coupling coefficient set to its current upper limit). Each entry of the table is a list of $\left<N\right>$ (calculated from Eq.~(\ref{eq:nexp})) for the four operator models: $(h_1, h_2, \ell_1, \ell_2)$. The experiments in consideration are listed in Table \ref{tab:experiments}), and the numbers are shown for the three benchmark WIMP masses. 
\label{tab:nexp}}
\end{table*}
\begin{figure*}
\centering
\includegraphics[width=.48\textwidth,keepaspectratio=true]{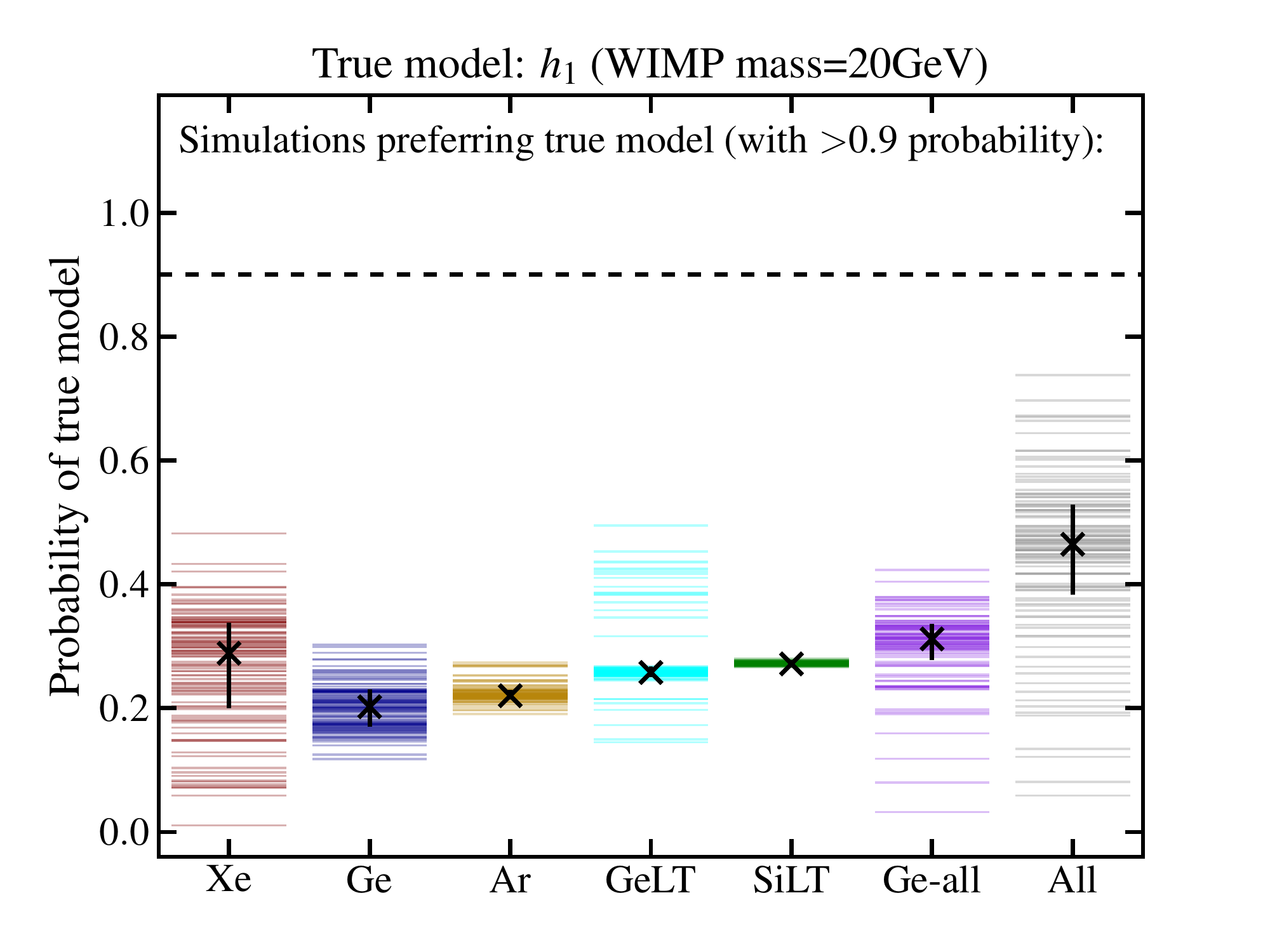}
\includegraphics[width=.48\textwidth,keepaspectratio=true]{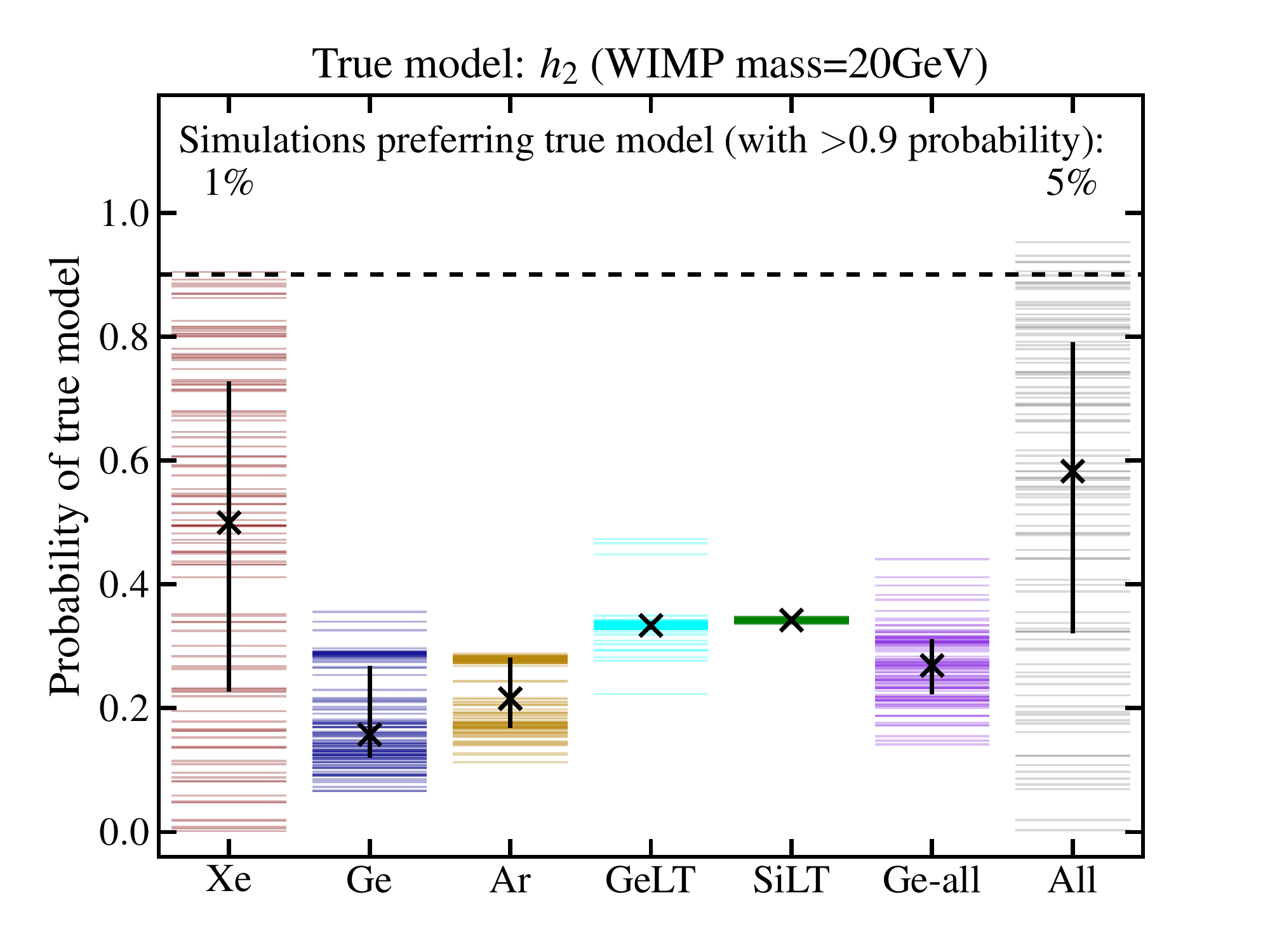}
\includegraphics[width=.48\textwidth,keepaspectratio=true]{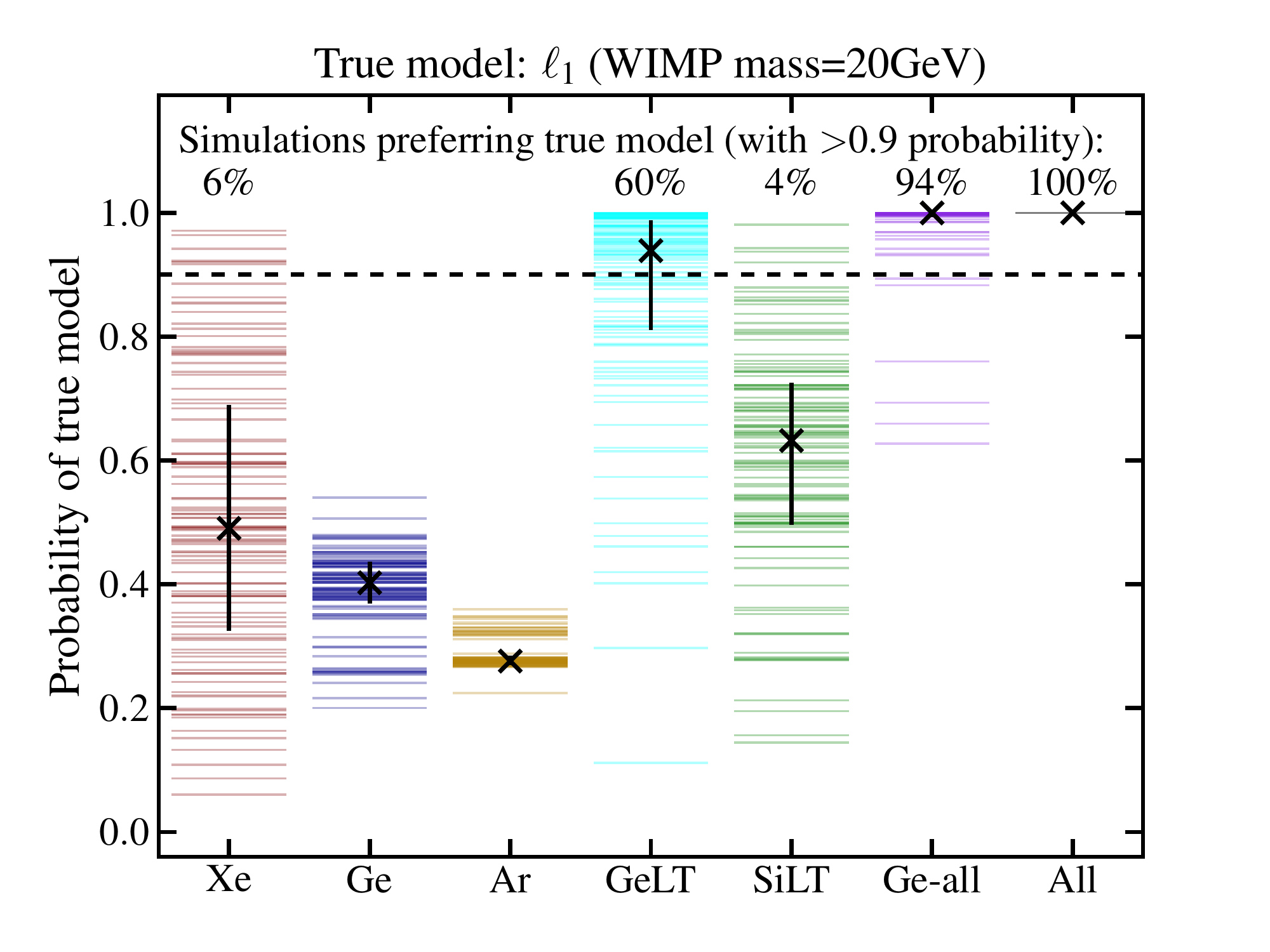}
\includegraphics[width=.48\textwidth,keepaspectratio=true]{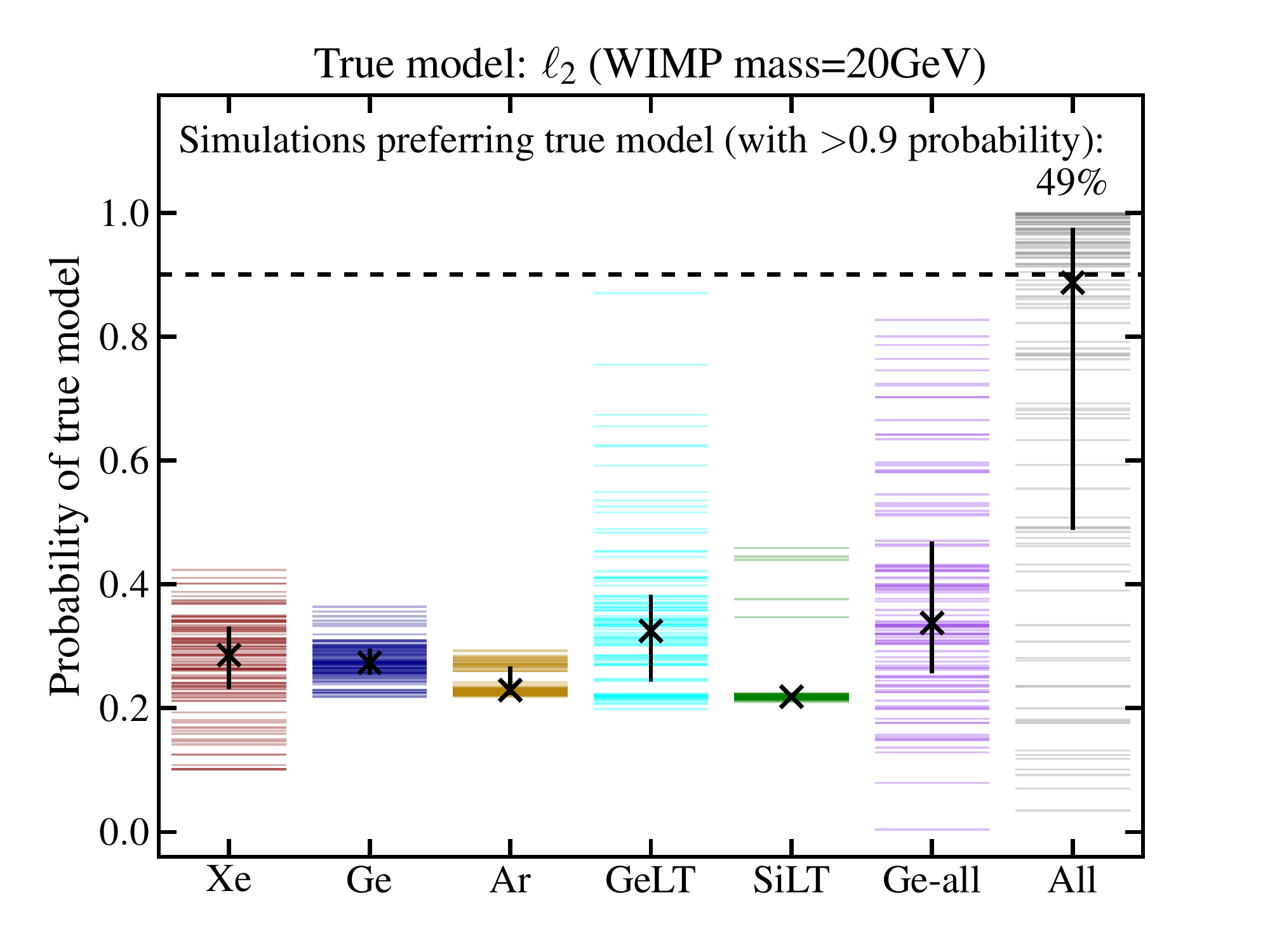}
\caption{Probability of successful model selection with G2 experiments for WIMP mass of 20 GeV. "Ge-all" refers to a combined analysis of ``Ge''+``GeLT''+``SiLT'' experiments, and ``All'' denotes a combination of all five G2 experiments.  Each panel represents model-selection results from the analysis of 100 noisy simulations (per experiment) generated under a single-operator model (where the respective Wilson coefficient is set to its current upper limit). The four operators of \S\ref{sec:eft} are treated as four separate models in this analysis and compared against each other to calculate the probability of the true underlying model for each data realization; this probability is on the vertical axis.  The percentage on top of each column shows in what fraction of the simulations was the right model chosen with more than $90\%$ probability (this is labeled as``successful model selection''). The ``$\times$'' marks the median probability, and the vertical line denotes the span of $50\%$ of the simulations around it, 
as a visual guide for a typical experimental outcome. See text for more details. \label{fig:lines_m20}}
\end{figure*}
\begin{figure*}
\centering
\includegraphics[width=.48\textwidth,keepaspectratio=true]{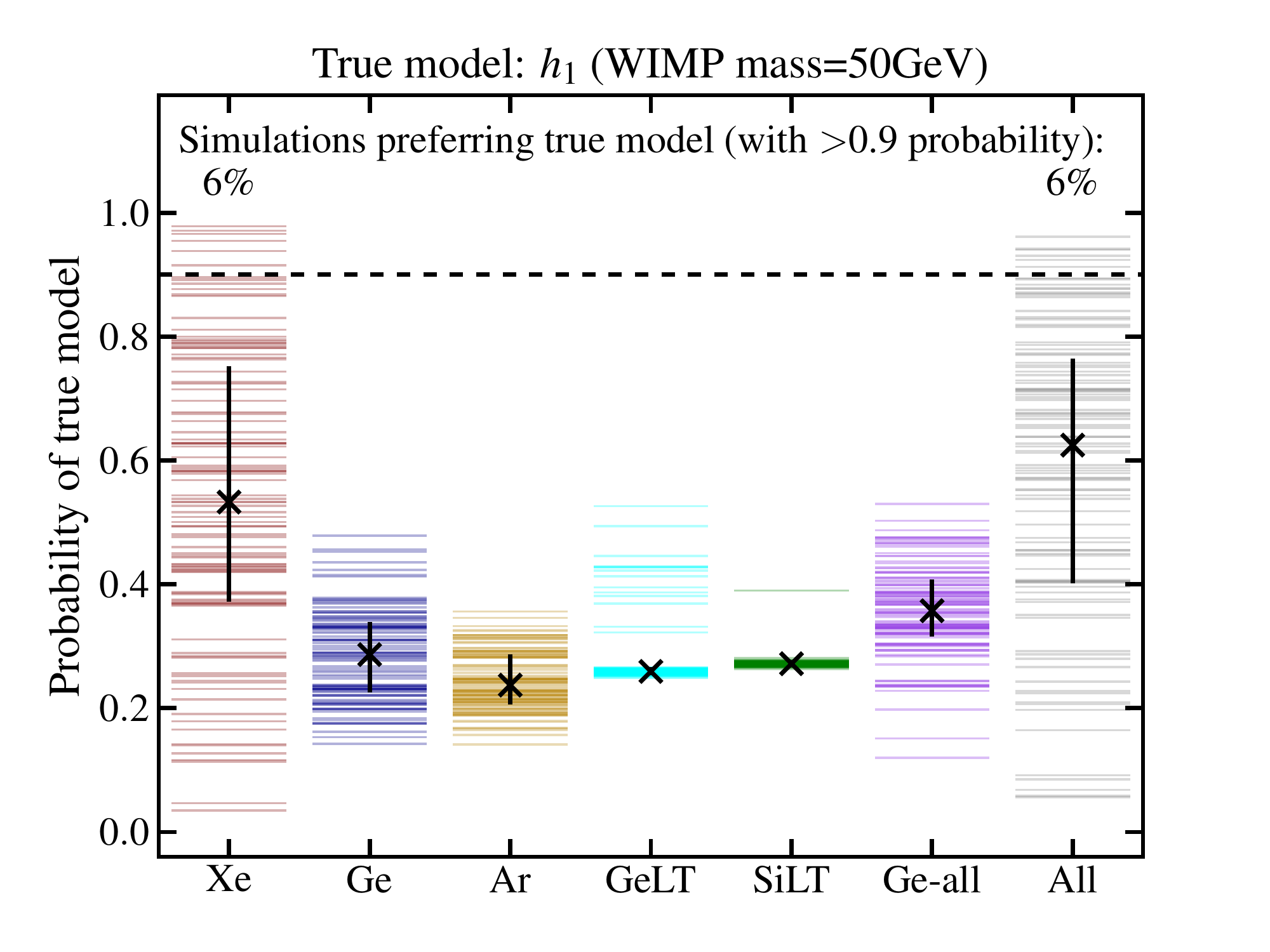}
\includegraphics[width=.48\textwidth,keepaspectratio=true]{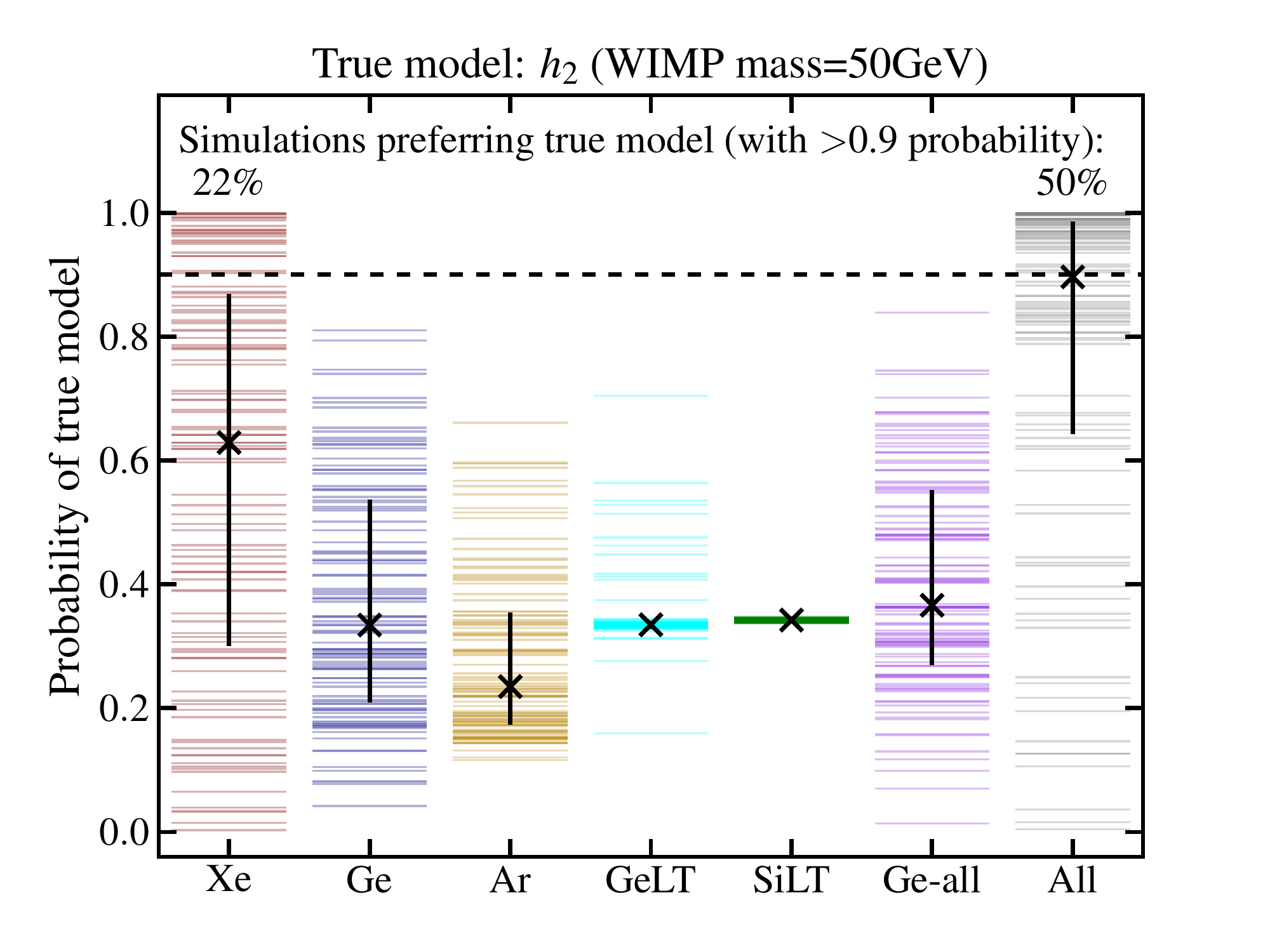}
\includegraphics[width=.48\textwidth,keepaspectratio=true]{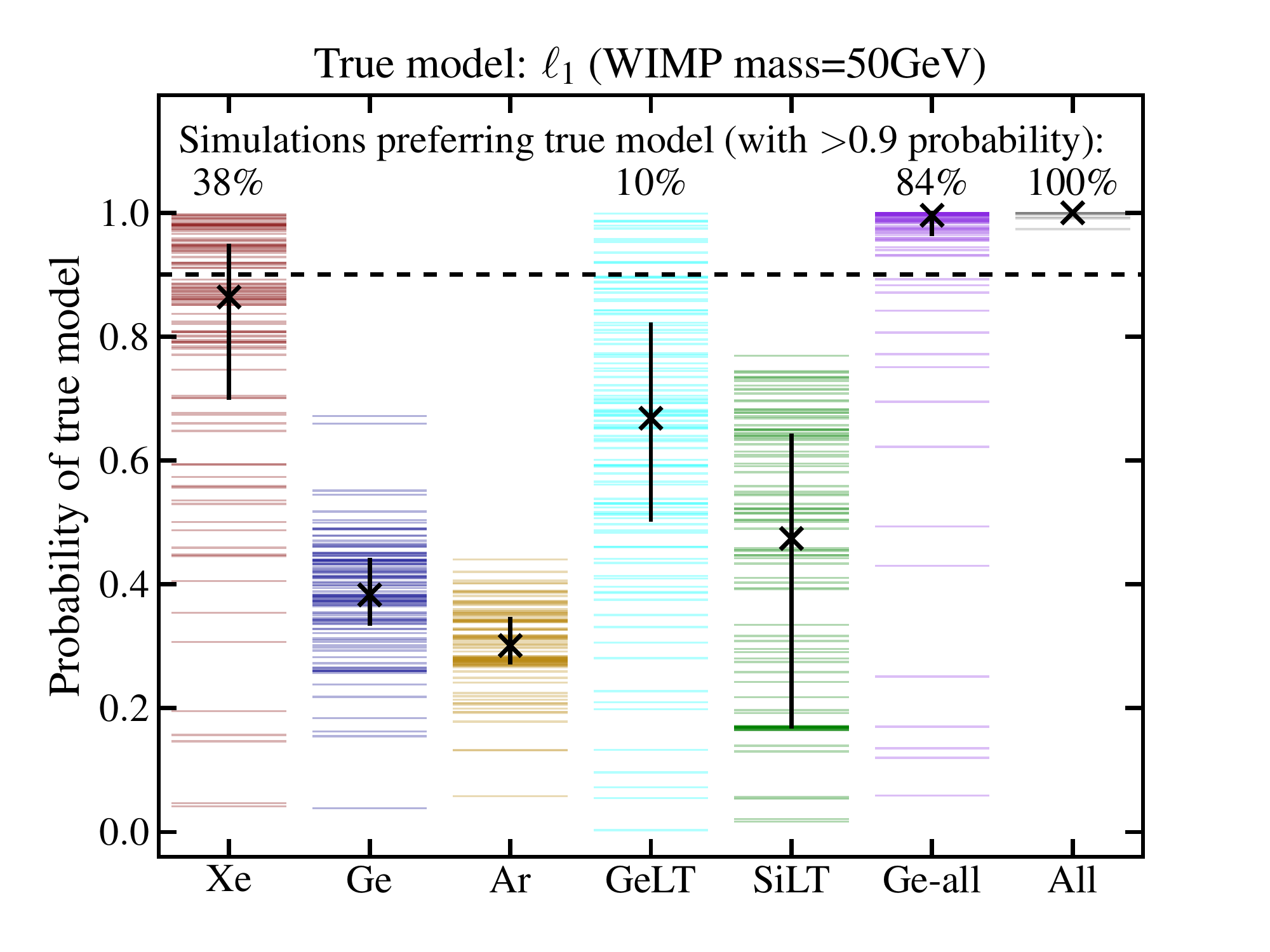}
\includegraphics[width=.48\textwidth,keepaspectratio=true]{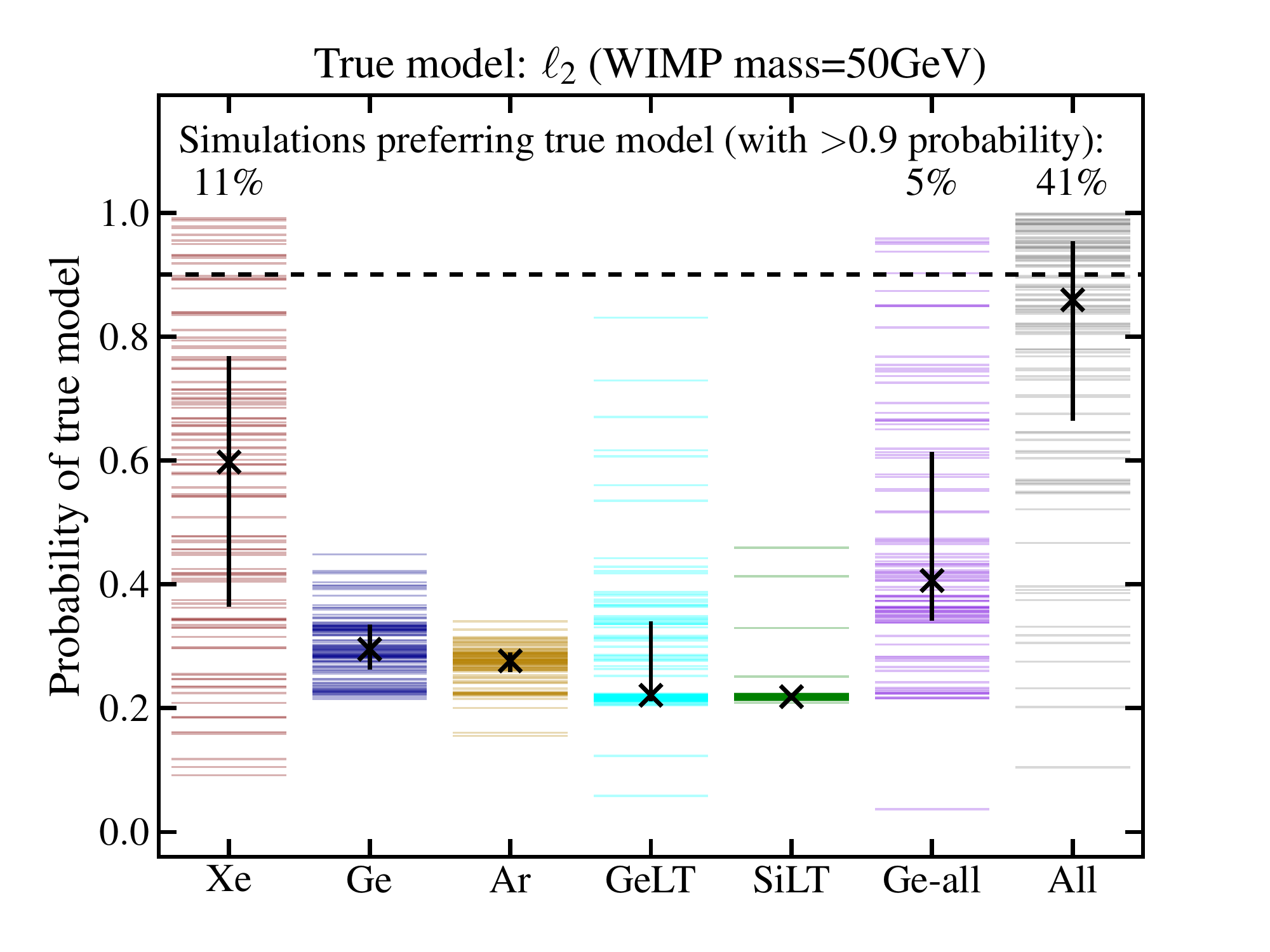}
\caption{Probability of successful model selection with G2 experiments for WIMP mass of 50 GeV (the same description as Figure \ref{fig:lines_m20}; see text for more details.).\label{fig:lines_m50}}
\end{figure*}
\begin{figure*}
\centering
\includegraphics[width=.48\textwidth,keepaspectratio=true]{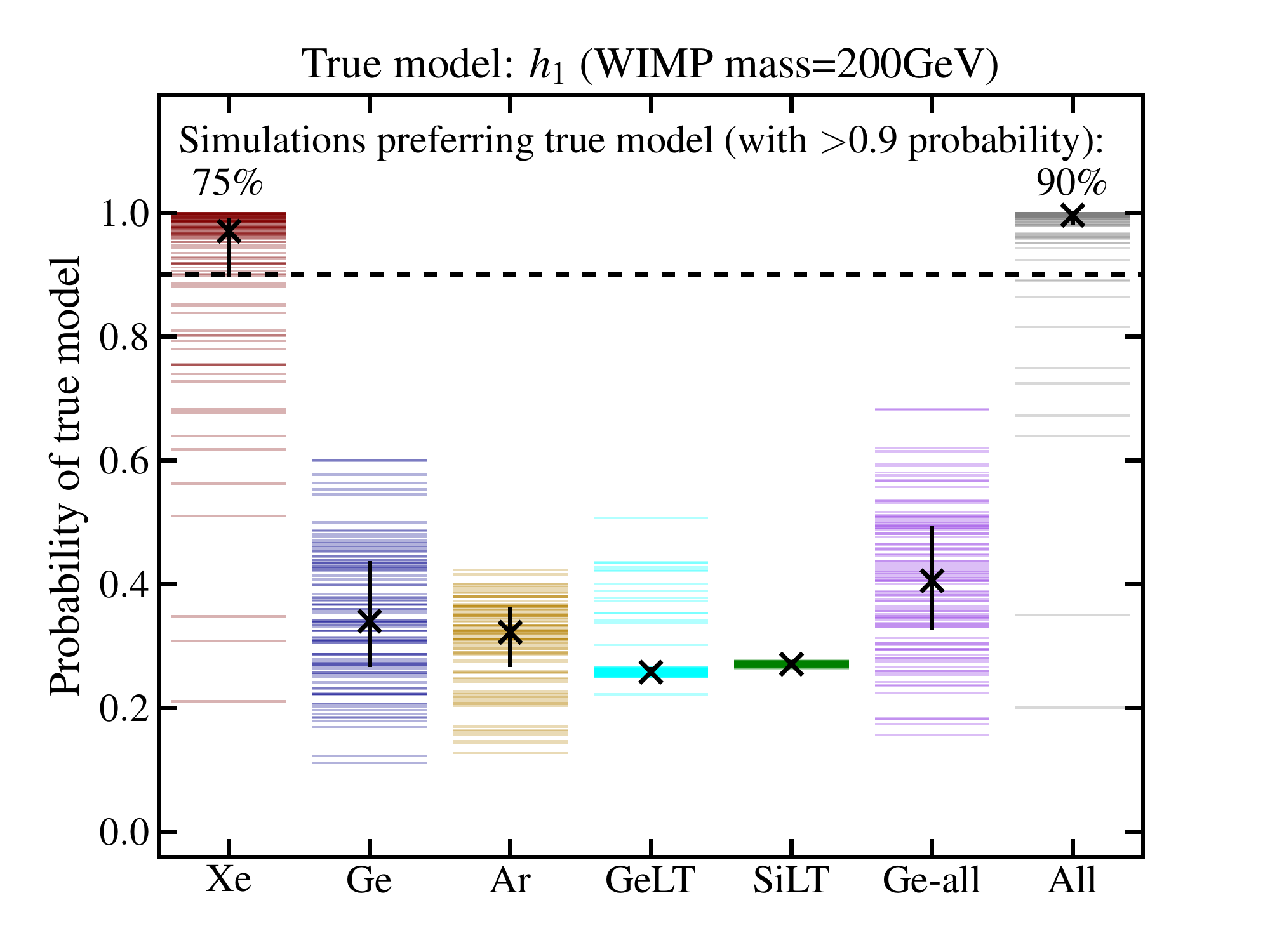}
\includegraphics[width=.48\textwidth,keepaspectratio=true]{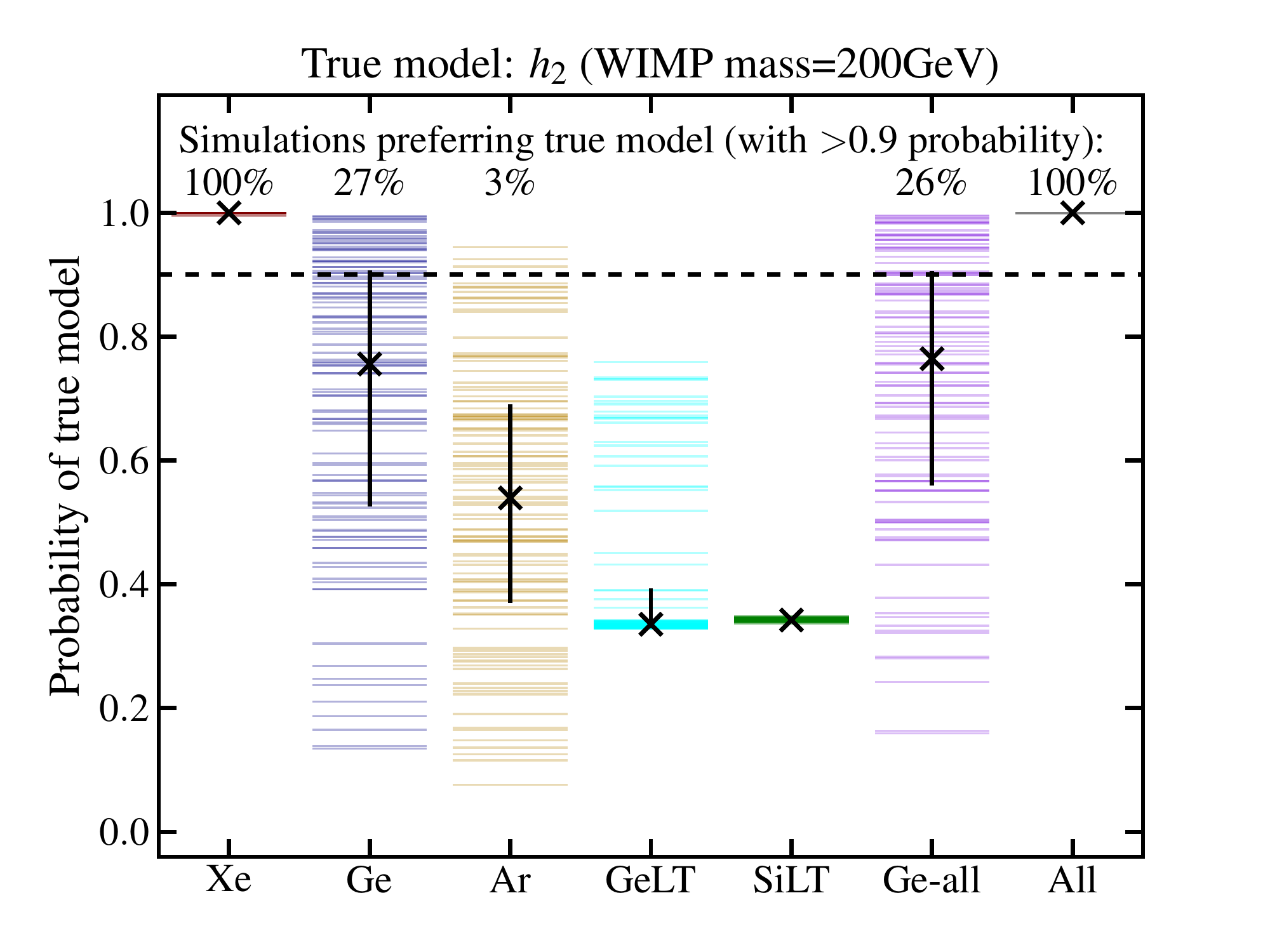}
\includegraphics[width=.48\textwidth,keepaspectratio=true]{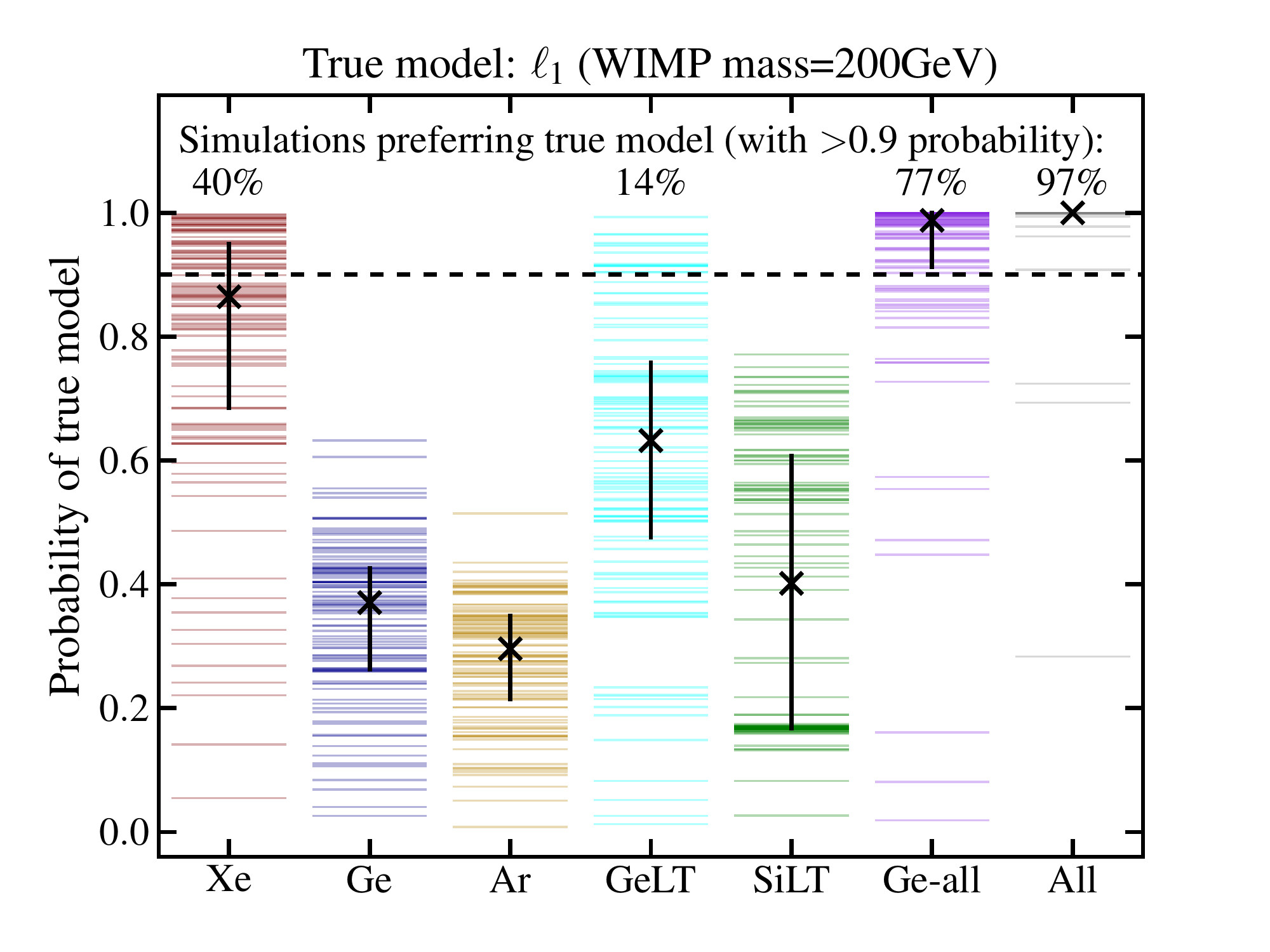}
\includegraphics[width=.48\textwidth,keepaspectratio=true]{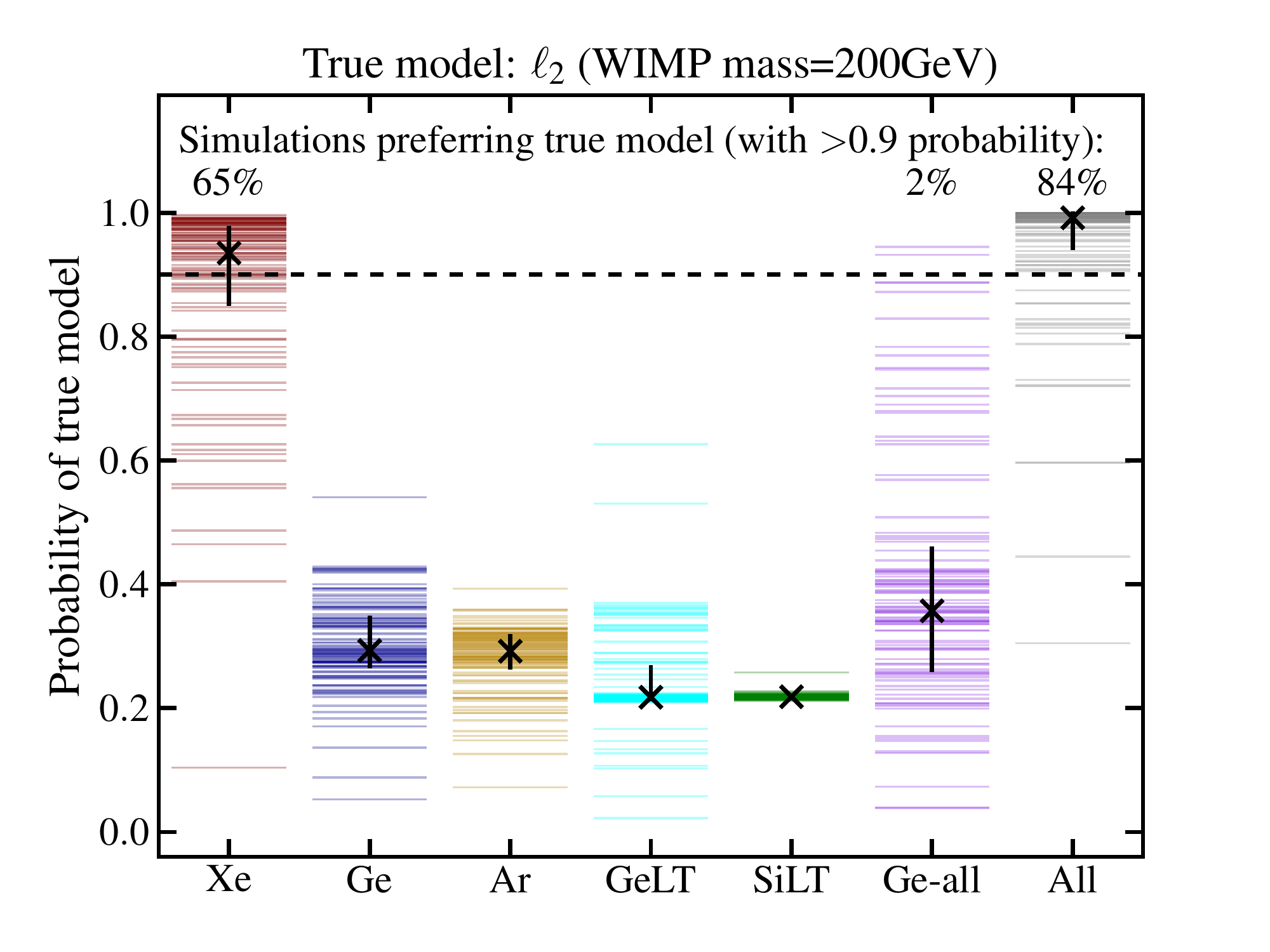}
\caption{Probability of successful model selection with G2 experiments for WIMP mass of 200 GeV (the same description as Figure \ref{fig:lines_m20}; see text for more details.).\label{fig:lines_m200}}
\end{figure*}
We first address the question of how well will data (including Poisson noise) from the upcoming G2 direct-detection experiments be able to individually distinguish momentum-dependent scattering operators from each other. For this purpose, we generate a large number of noisy data realizations where only one operator was ``turned on'' in any given simulation (with the associated effective coupling set to its current upper limit from Table \ref{tab:limits}). Table \ref{tab:nexp} lists the number of expected events for each choice of underlying operator model, for each experiment. We use the \textsc{MultiNest} sampler on this entire suite of simulations, four times per data realization, each time fitting a different underlying operator (where the corresponding coupling coefficient and $m_\chi$ are the free parameters). This way, for every realization we obtain the posterior probability for each of the four operator models, where the underlying ``truth'' is only one of the models. We can then evaluate the evidence ratio of Eq.~(\ref{eq:E_ratio}) for each simulation, and assign probability $p$ to every operator model.  Taking $p(M) >0.9$ as the nominal ``success threshold''\footnote{It is common practice in Bayesian analysis to use, for example, Jeffreys' scale to label the level of success in model selection (see, for instance, Ref.~\cite{2013arXiv1310.5718A}). However, any such scale is too fine for the questions we are addressing. We find a binary choice more appropriate, so we classify probabilities above 0.9 as ``success'', and those below as ``not success''. This threshold is chosen somewhat arbitrarily, and can be changed. A small change in the threshold choice will only affect the summarized interpretation of the results (where we try to give a single-number answer to the question of ``how likely is a given experiment to pick out the right model?'', for example), but will not alter qualitative conclusions of this study, nor the computations behind them.}, we compute the fraction of all data realizations that successfully select the right operator, and interpret it as a probability that future data will do the same. This procedure is applied to simulated data from each experiment individually, and to a combination of data from different experiments. 

The results are shown in Figures \ref{fig:lines_m20}, \ref{fig:lines_m50}, and \ref{fig:lines_m200}, corresponding to three WIMP-mass benchmarks: 20 GeV, 50 GeV, and 200 GeV, respectively. In each figure, columns correspond to either one of the five G2 experiments, or to the combined analysis of some of them. On the vertical axis is the probability of the right underlying model, where each horizontal line is computed from a single data realization. A pile-up of lines around 0.25 signifies that many data realizations do not prefer any of the four models (there is even distribution of probabilities among the four models), while a pile-up toward the top of the vertical axis signifies high success probability. 

Analyzing these results, we come to a number of conclusions. First, we find a large variance in the success rate of individual simulations, brought on by Poisson noise in event counts.  This variance is represented by the span of horizontal colored lines in the three Figures. The $\times$'s mark median values of the probability assigned to the right underlying operator, with the vertical line spanning 50$\%$ of the simulations around it; this is a visual guide towards a typical experimental outcome. A particular future data set could, naturally, be more or less successful than the median (which is why we quantify ``the probability of success'' as explained above).

Furthermore, for the fixed values of the effective couplings, the success of model selection is a function of several factors of roughly equal importance: on one side are the WIMP mass, the mediator mass, and the type of scattering (represented by different scattering operators),  and on the other is the experimental recoil-energy window. All these factors affect the shape of the recoil-energy spectrum (in particular its slope and the prominence and position of spectral features), and their interplay can give rise to strikingly different outcomes for the same number of observed events; indeed, it can even produce situations where a smaller number of events, in some scenarios, bears more information than a larger one in others. For instance, a successful reconstruction is almost guaranteed for ``Ge''+``GeLT''+``SiLT'' combination of experiments, in case of a Coulomb-like interaction ($\ell_1$ operator) for a WIMP mass of 20 GeV, where a total of $\sim$40 events is observed on all these targets. This is because the low WIMP mass and the $\ell_1$ operator both force an extremely steep energy spectrum, producing a very recognizable feature that is difficult to mimic with other operators and WIMP-mass choices. In contrast, the reconstruction is likely unsuccessful (the data most often does not distinguish different underlying operators) in the case of a 50 GeV WIMP with a standard heavy-mediator interaction ($h_1$ operator), when ``Xe'' observes 90 events, and all G2 experiments gather around 100 events total. The same happens in the case of $h_2$ with 20 GeV WIMP, where all G2 experiments combined see around 100 events. In the first case, it happens because the shape of the $h_1$ spectrum is easily mimicked by either $\ell_2$ with a heavier WIMP, or with $h_2$ with a lighter WIMP. In the second case, the telltale feature of the $h_2$ model---a turnover in the recoil spectrum---is below the energy threshold of the G2 experiments, which makes it indistinguishable from other operators with different WIMP-mass choices.

Finally, one of the most striking features of these Figures is the increase in the success probability when data of different experiments are analyzed jointly; even a null result (no events observed) can be valuable additional information that helps raise the success rate in combination with other data sets. This is because null results penalize different models differently. For example, for a Coulomb-like interaction ($\ell_1$ operator) of a 50 GeV WIMP, while it is unlikely that any single experiment will be able to select the right governing operator with certainty, a combination of various experiments ensures success. The expected number of events for ``Ge'', ``GeLT'', and ``SiLT'' is 3, 8, and 2, respectively, meaning that many data realizations for ``Ge'' and ``SiLT'' are null results. However, combining ``Ge'' and ``SiLT'' with ``GeLT'' (which by itself has only 10$\%$ success rate) is enough to boost the sucess rate to $84\%$. The addition of ``Xe'' (successful on its own only 38$\%$ of the times, with $\sim$90 events) guarantees successful selection of the underlying operator. Such complimentarity of different experiments is mainly due to the fact that wide energy windows and very low thresholds (preferably below 1 keV), in combination with large exposures and a variety of target materials, are necessary to distinguish underlying operator models. This is a consequence of the following: the slope of the recoil spectrum in the lowest energy bins is controlled both by the WIMP mass and by how fast the spectrum rises due to momentum dependence of the operator. To break this degeneracy (where, for example, a smaller WIMP mass can mimic a steeper operator model, such as those with light mediators) and ensure successful model selection, a handle on a wide range of energies is necessary. If the goal of direct DM searches is going beyond signal detection into probing details of DM physics, this last point is a very strong argument for future efforts to be geared towards multiple experimental techniques with a variety of targets and energy thresholds rather than to be focused solely on ever-increasing exposure and target mass. 

Let us now examine the outcome of the analysis for the three WIMP-mass regimes separately; these are represented in Figures \ref{fig:lines_m20} (``light''), \ref{fig:lines_m50} (``intermediate''), and \ref{fig:lines_m200} (``heavy'').
For a light WIMP, we see the following. On the one hand, it is not likely that any single experiment will be able to identify the right underlying operator. On the other hand, when several experiments are combined, model selection is successful only if the underlying model is a Coulomb-like interaction ($\ell_1$), while it is mostly unsuccessful in both heavy-mediator cases ($h_1$ and $h_2$), and shows about $50\%$ chance of success if the true model is $\ell_2$. Put another way, even if the signal is just below the current upper limit, it is only in the $\ell_1$ case that G2 will almost certainly distinguish the underlying operator, if the WIMP mass is 20 GeV or smaller.  For an intermediate WIMP mass, the outcome is slightly more optimistic: with a combination of experiments, only in the $\ell_1$ case is success guaranteed, while $h_2$ and $\ell_2$ produce around $50\%$ chance of success, and $h_1$ is unlikely to allow for a successful model selection. For a heavy WIMP (with a mass above 200 GeV) the situation is most optimistic, overall: a single ``Xe''-like experiment is guaranteed a successful model selection only in $h_2$ scenario, and a combination of experiments has a high chance of success, regardless of the governing operator.
\subsection{Distinguishing underlying interactions: heavy vs. light mediator}
\label{sec:distops_hl}
\begin{figure*}
\centering
\includegraphics[width=.48\textwidth,keepaspectratio=true]{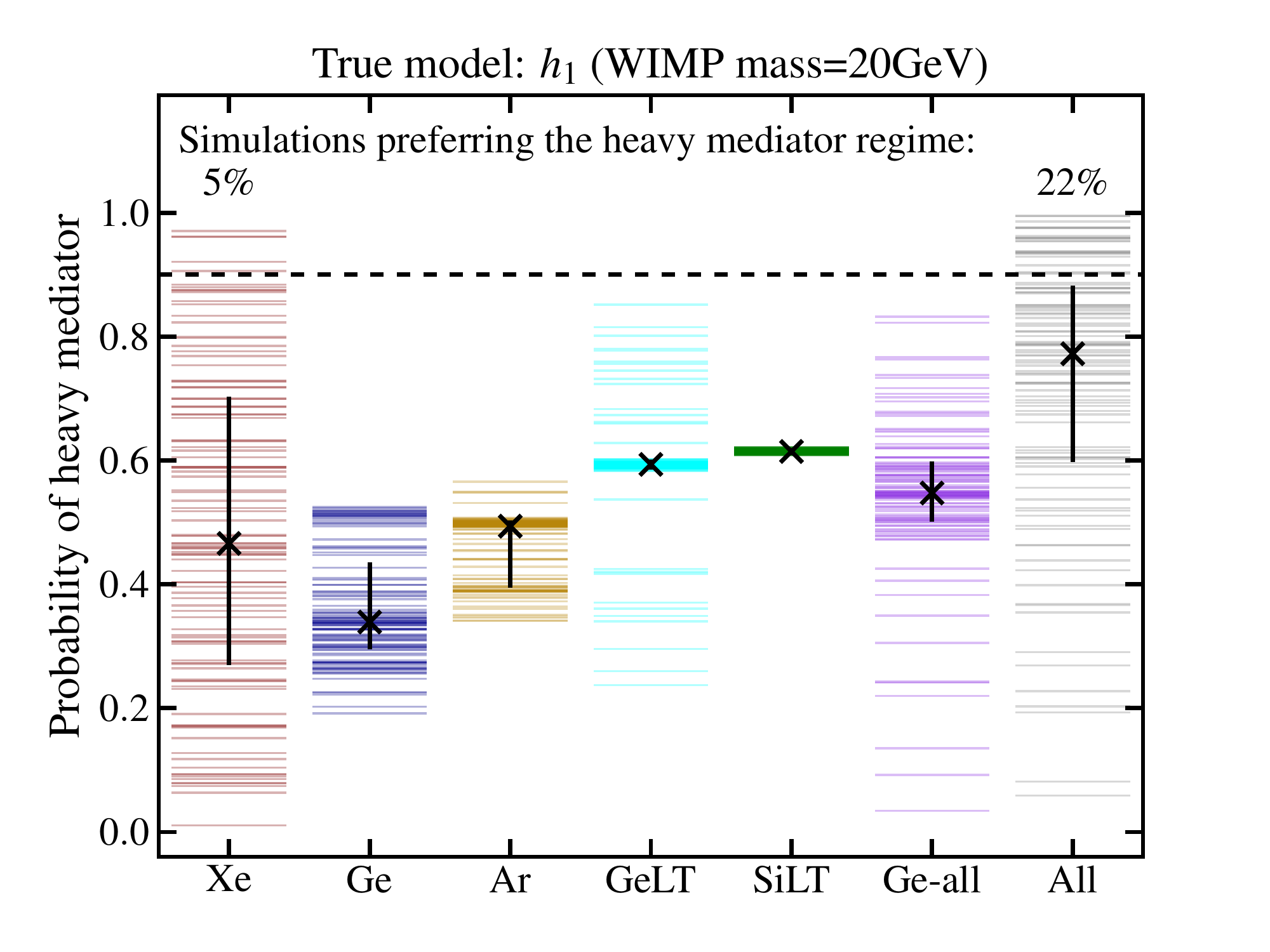}
\includegraphics[width=.48\textwidth,keepaspectratio=true]{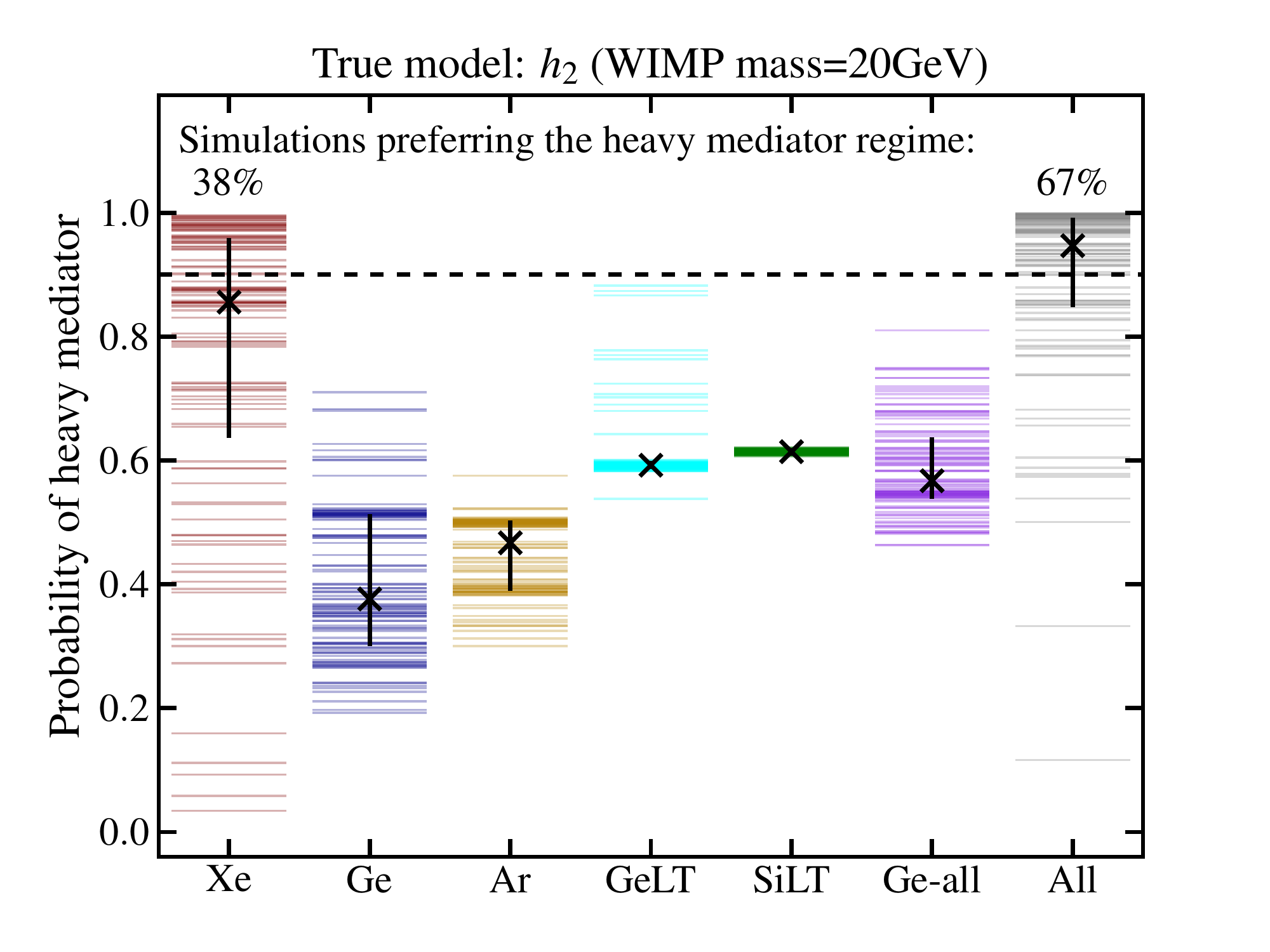}
\includegraphics[width=.48\textwidth,keepaspectratio=true]{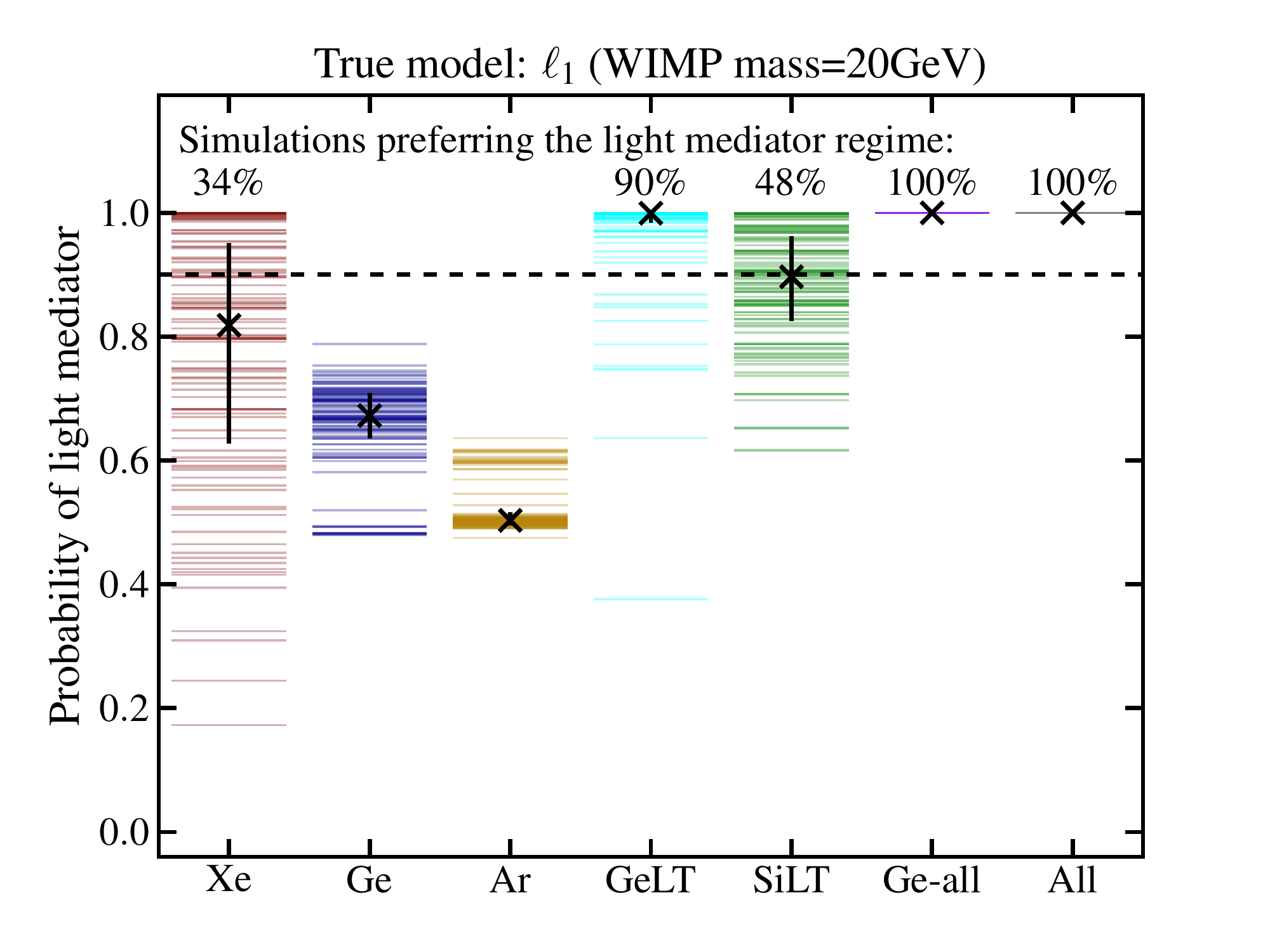}
\includegraphics[width=.48\textwidth,keepaspectratio=true]{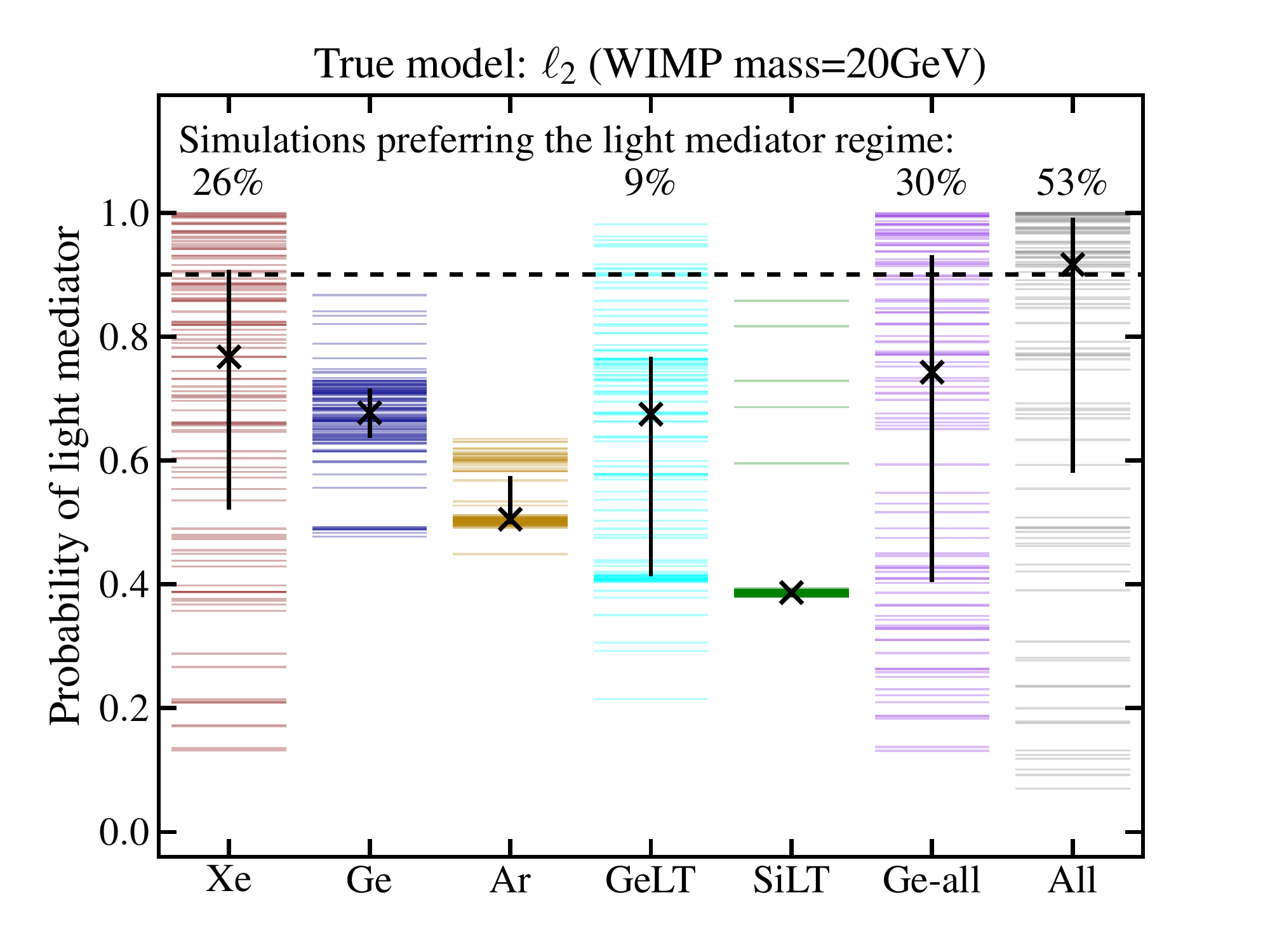}
\caption{Probability of successful selection of the right mediator-mass regime with G2 experiments for WIMP mass of 20 GeV. The plot is the same as Figure \ref{fig:lines_m20}, except that the vertical axis measures the probability of selecting the right mediator-mass scenario, and the percentage on top of each column denotes the fraction of all simulations where evidence-ratio calculation gives more than $90\%$ probability for the right subset of operator models ($\{h_1,h_2\}$ for the heavy-mediator case represented in the two upper panels, and $\{\ell_1,\ell_2\}$ for the light-mediator case of the two lower panels).\label{fig:lines_m20_hl}}
\end{figure*}
\begin{figure*}
\centering
\includegraphics[width=.48\textwidth,keepaspectratio=true]{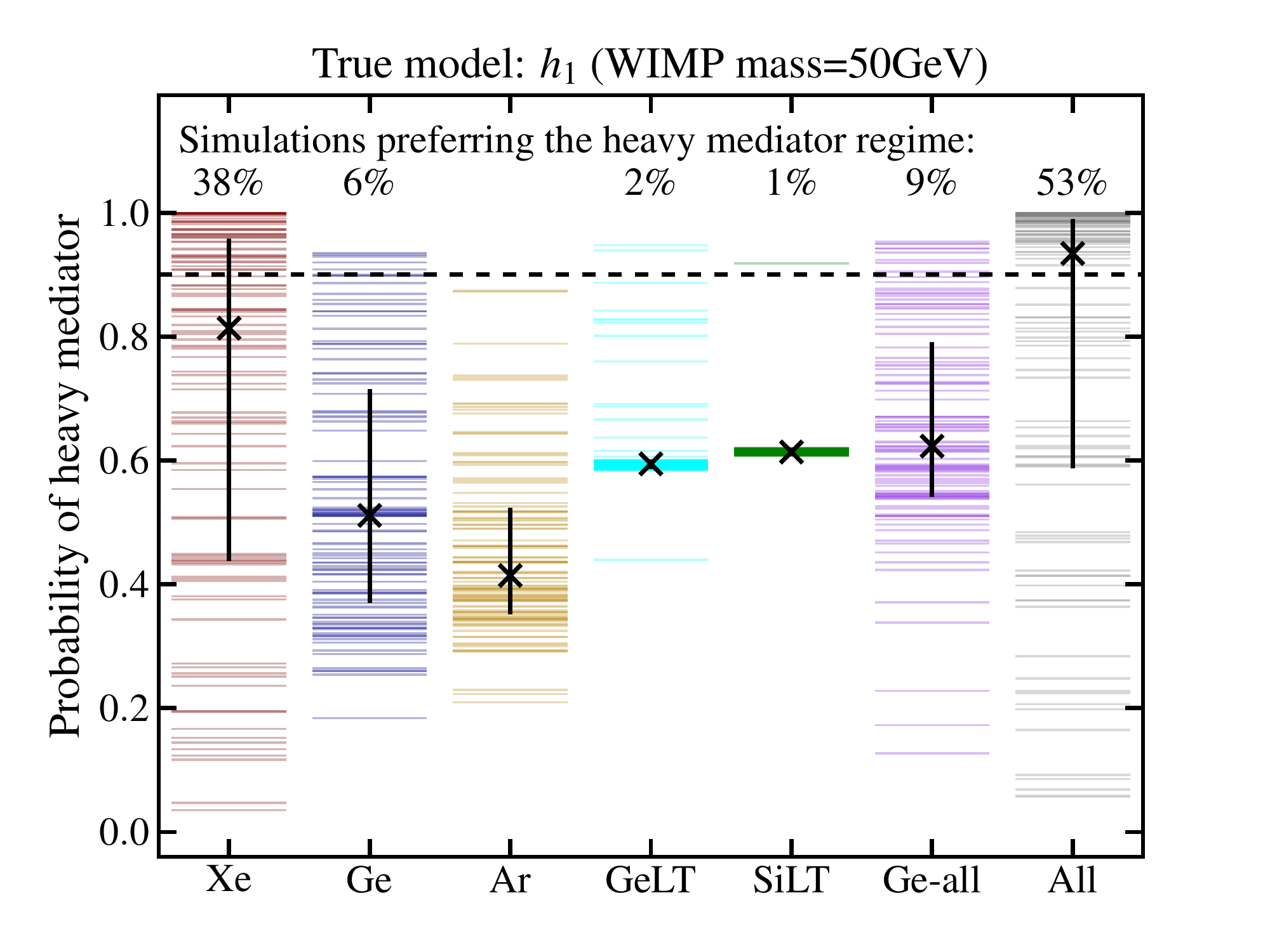}
\includegraphics[width=.48\textwidth,keepaspectratio=true]{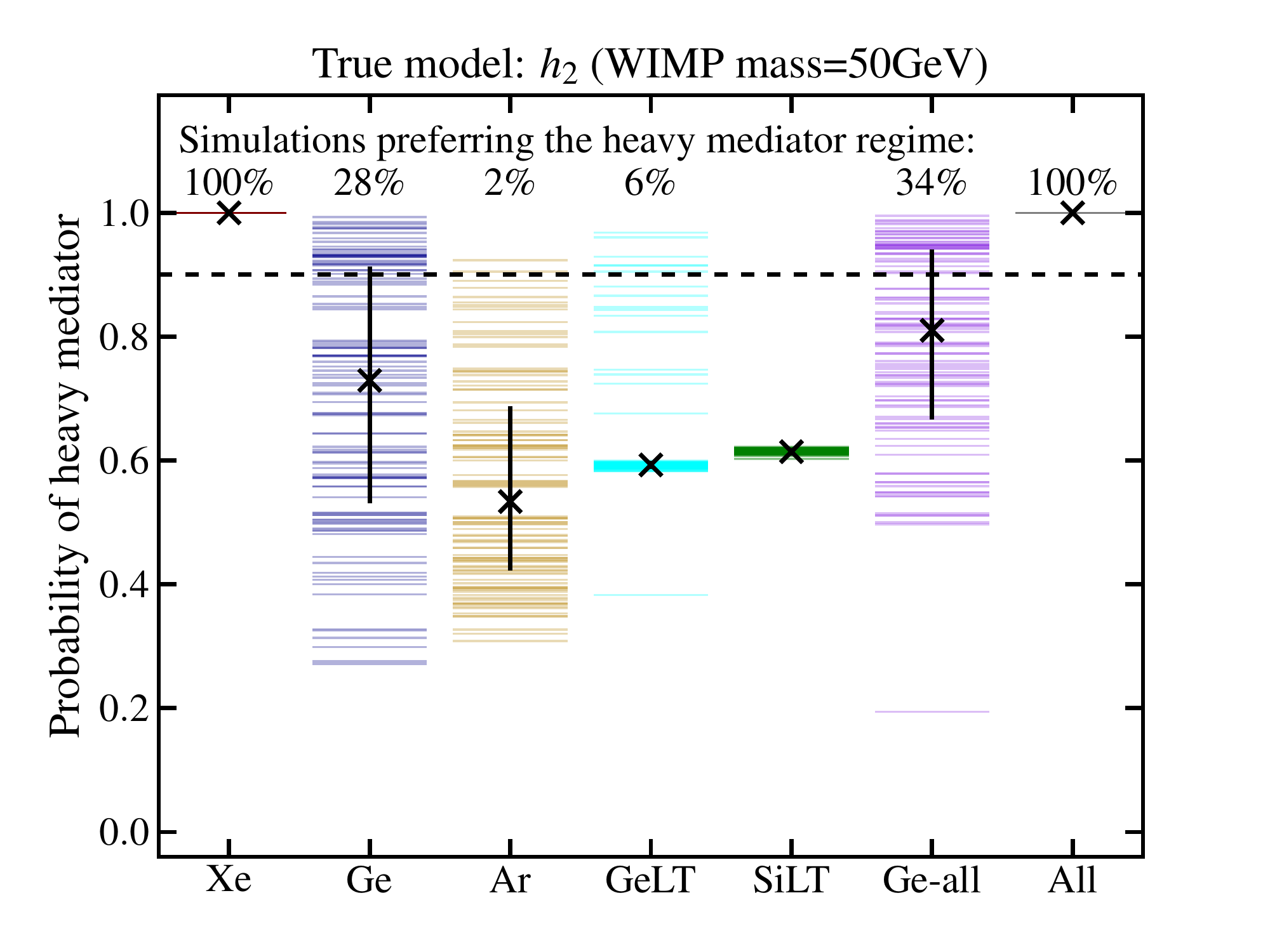}
\includegraphics[width=.48\textwidth,keepaspectratio=true]{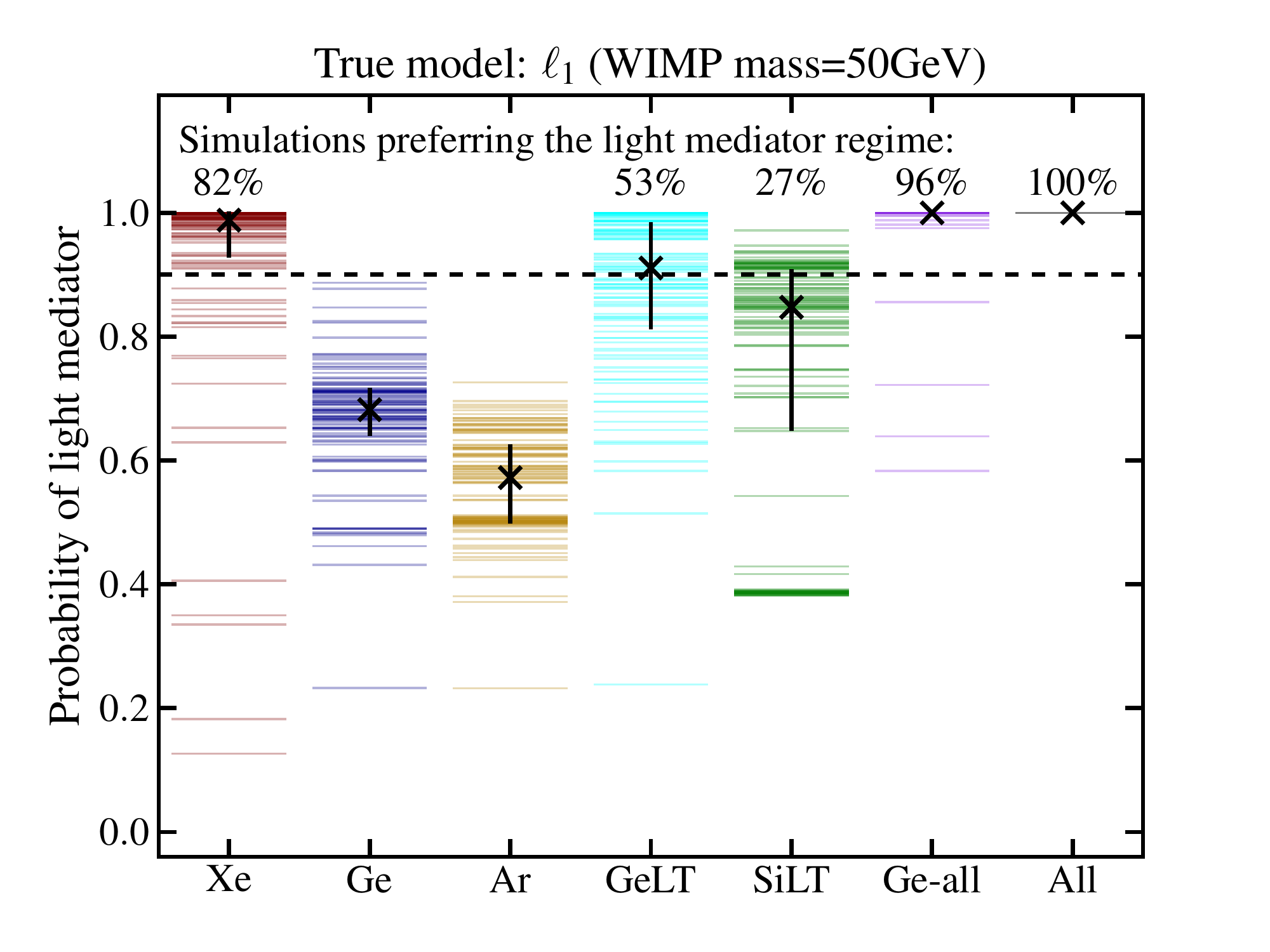}
\includegraphics[width=.48\textwidth,keepaspectratio=true]{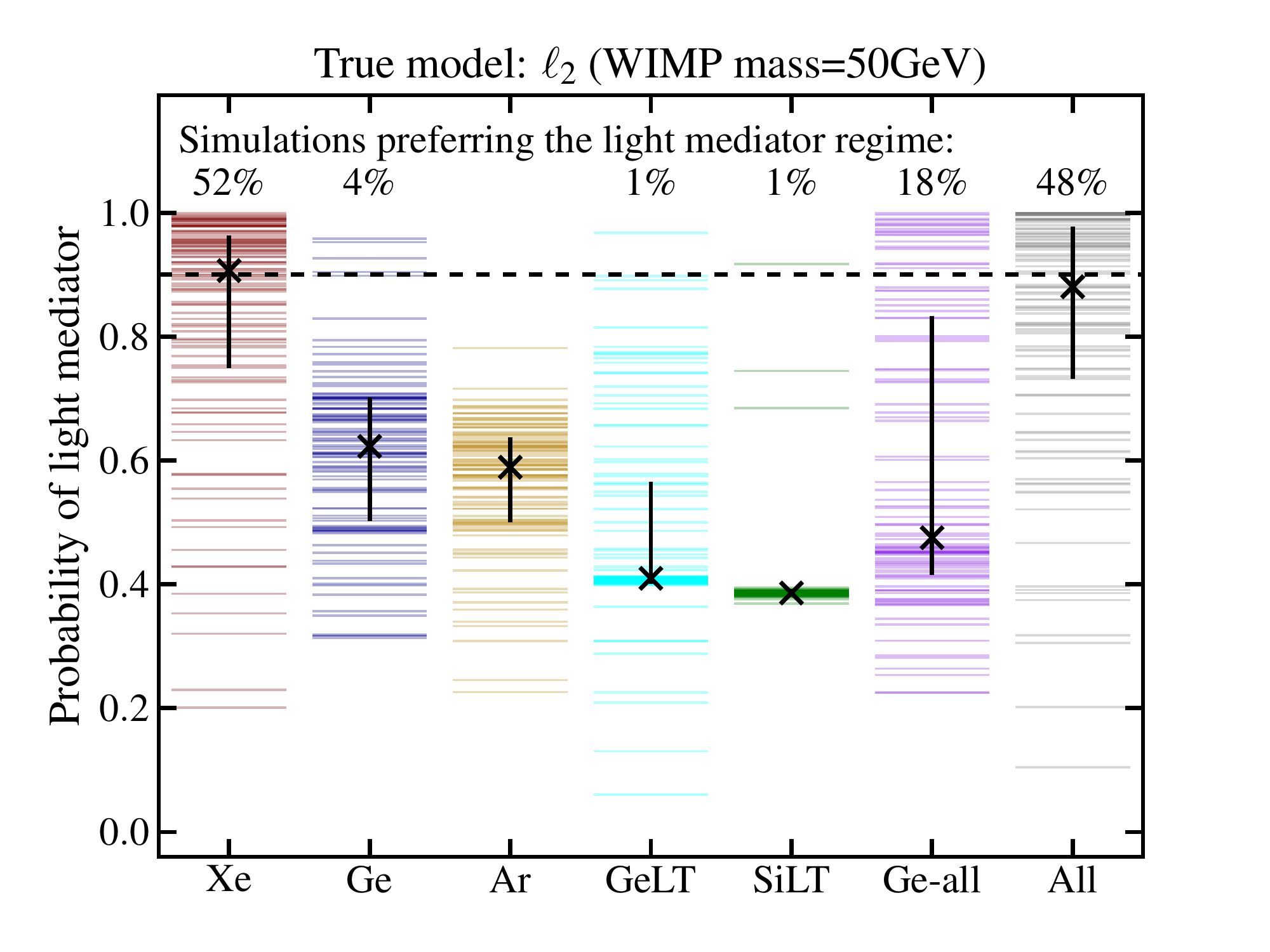}
\caption{Probability of successful selection of the right mediator-mass regime with G2 experiments for WIMP mass of 50 GeV (same description as Figure \ref{fig:lines_m50_hl}).\label{fig:lines_m50_hl}}
\end{figure*}
\begin{figure*}
\centering
\includegraphics[width=.48\textwidth,keepaspectratio=true]{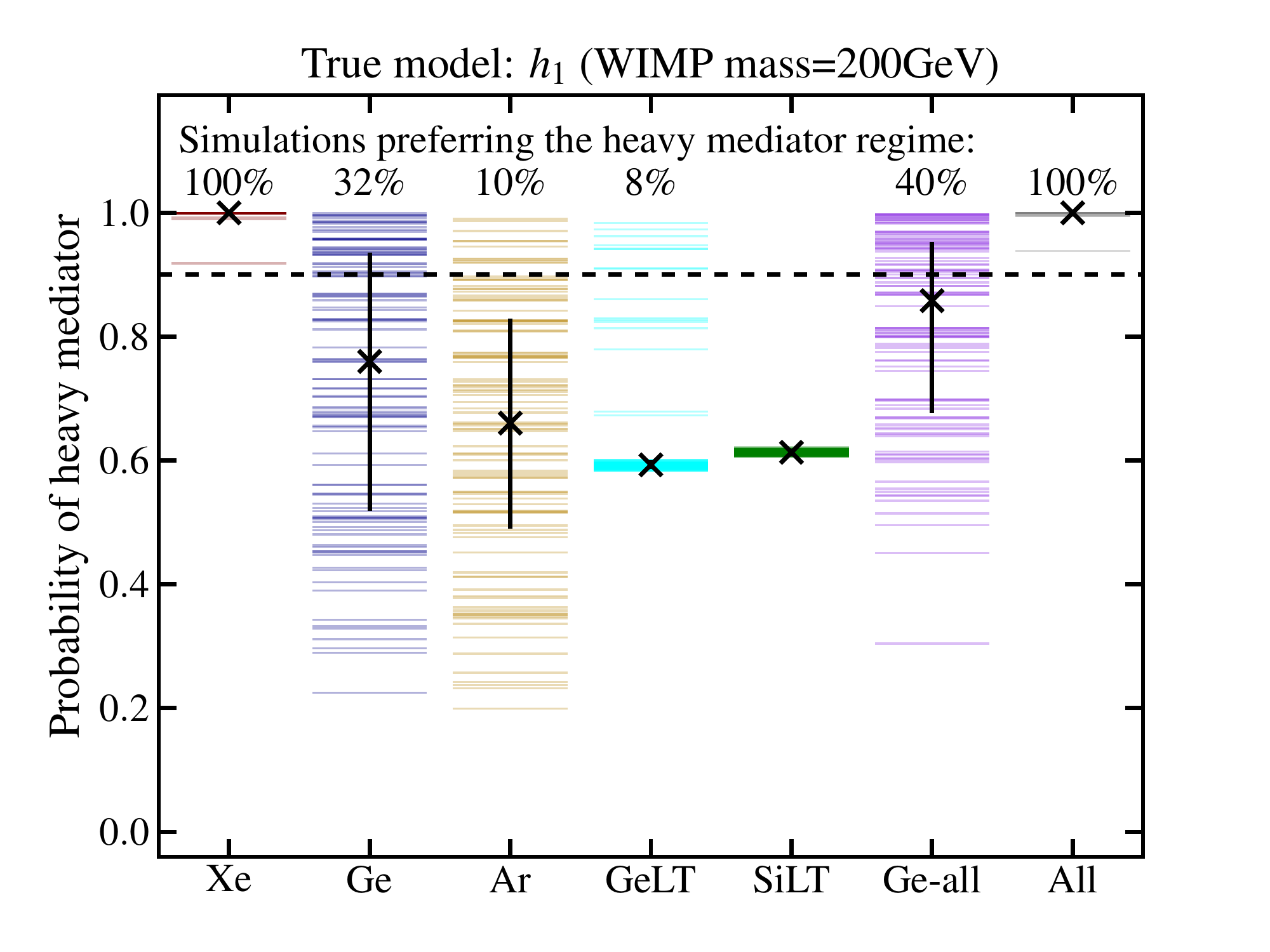}
\includegraphics[width=.48\textwidth,keepaspectratio=true]{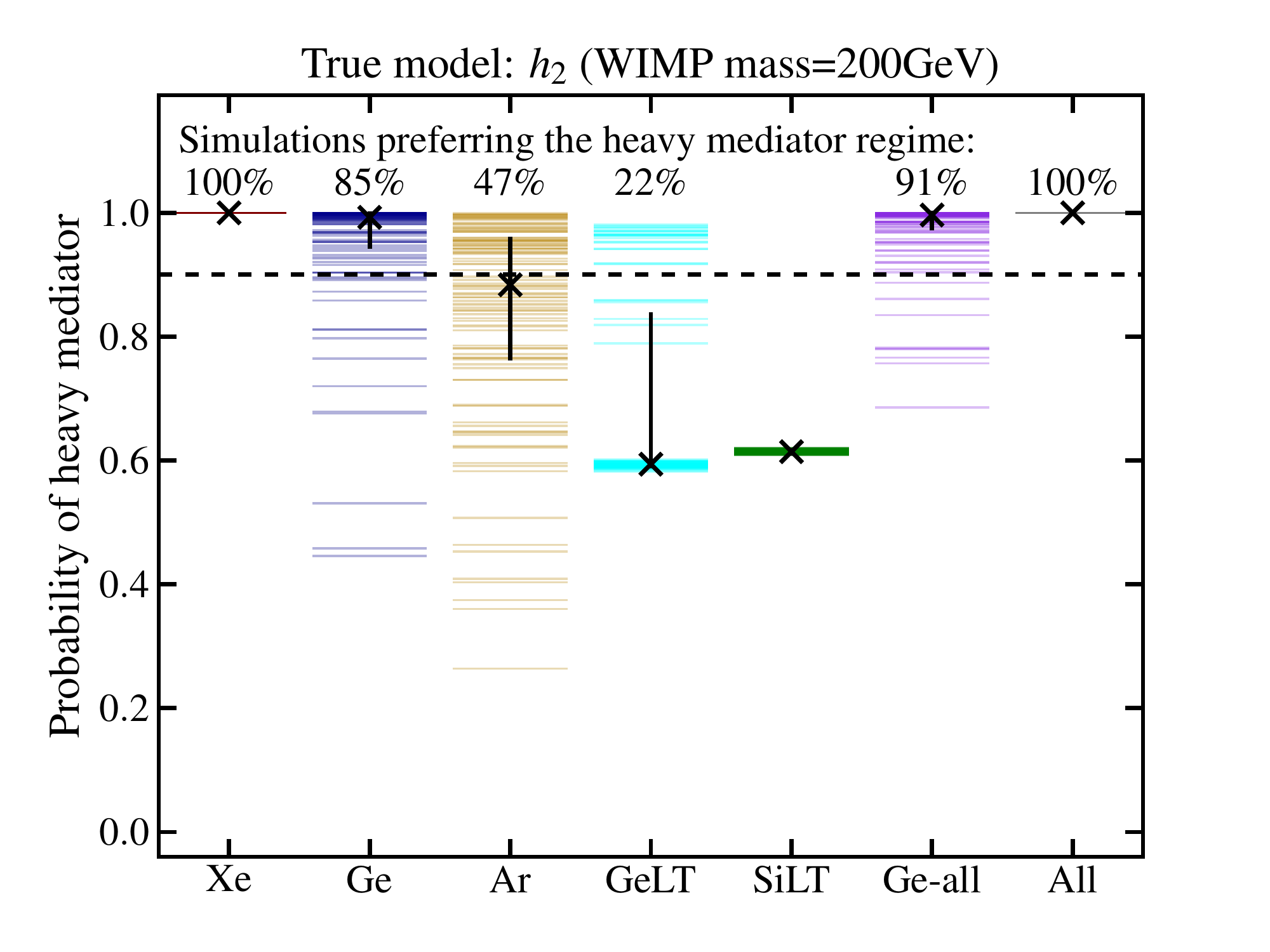}
\includegraphics[width=.48\textwidth,keepaspectratio=true]{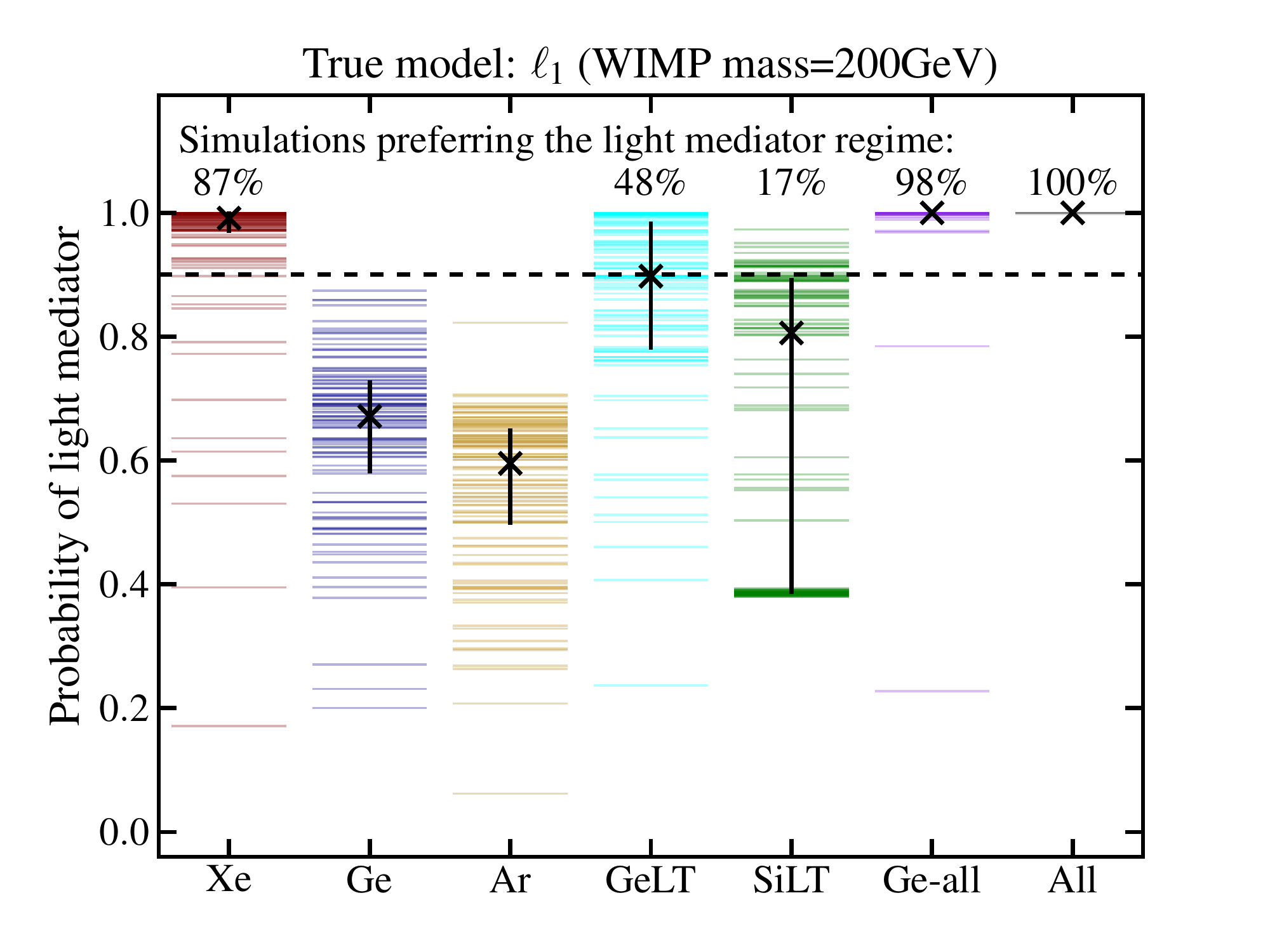}
\includegraphics[width=.48\textwidth,keepaspectratio=true]{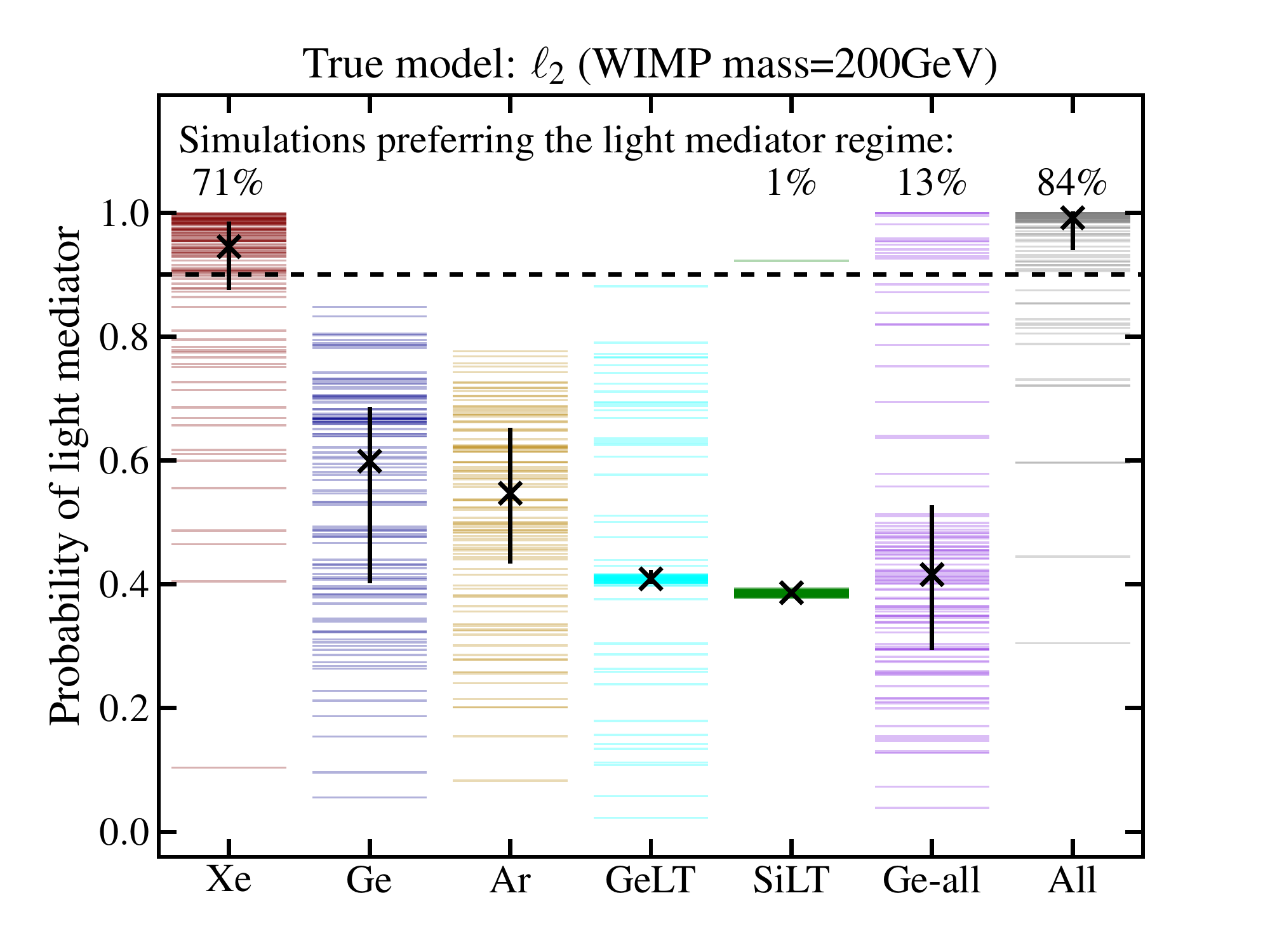}
\caption{Probability of successful selection of the right mediator-mass regime with G2 experiments for WIMP mass of 200 GeV (same description as Figure \ref{fig:lines_m200_hl}).\label{fig:lines_m200_hl}}
\end{figure*}
Given the varying degree of success in model selection we forecast for G2 in the previous section, here we look into a more general question and evaluate the probability that the mediator-mass regime is correctly reconstructed from G2 data. For this purpose, we use the simulations and associated evidence ratios discussed in the previous section (where only one of the underlying operators is turned on in any given simulation, and its coupling coefficient set to its current upper limit), and analyze them in a slightly different way. Namely, we define the probability of selecting the heavy-mediator case as the sum of the probability of $h_1$ and the probability of $h_2$, and similarly for the light-mediator case. The results are shown in Figures \ref{fig:lines_m20_hl}, \ref{fig:lines_m50_hl}, and \ref{fig:lines_m200_hl}.
The overall conclusion here is that the prospects for successful model selection are more optimistic than when individual operators are considered, as expected. The chance of success is on the order of $50\%$, or more, in all scenarios, except in the canonical $h_1$ case for a light WIMP. 
\subsection{Quality of parameter estimation and degeneracies}
\label{sec:parest}
\begin{figure*}
\centering
\includegraphics[width=.32\textwidth,keepaspectratio=true]{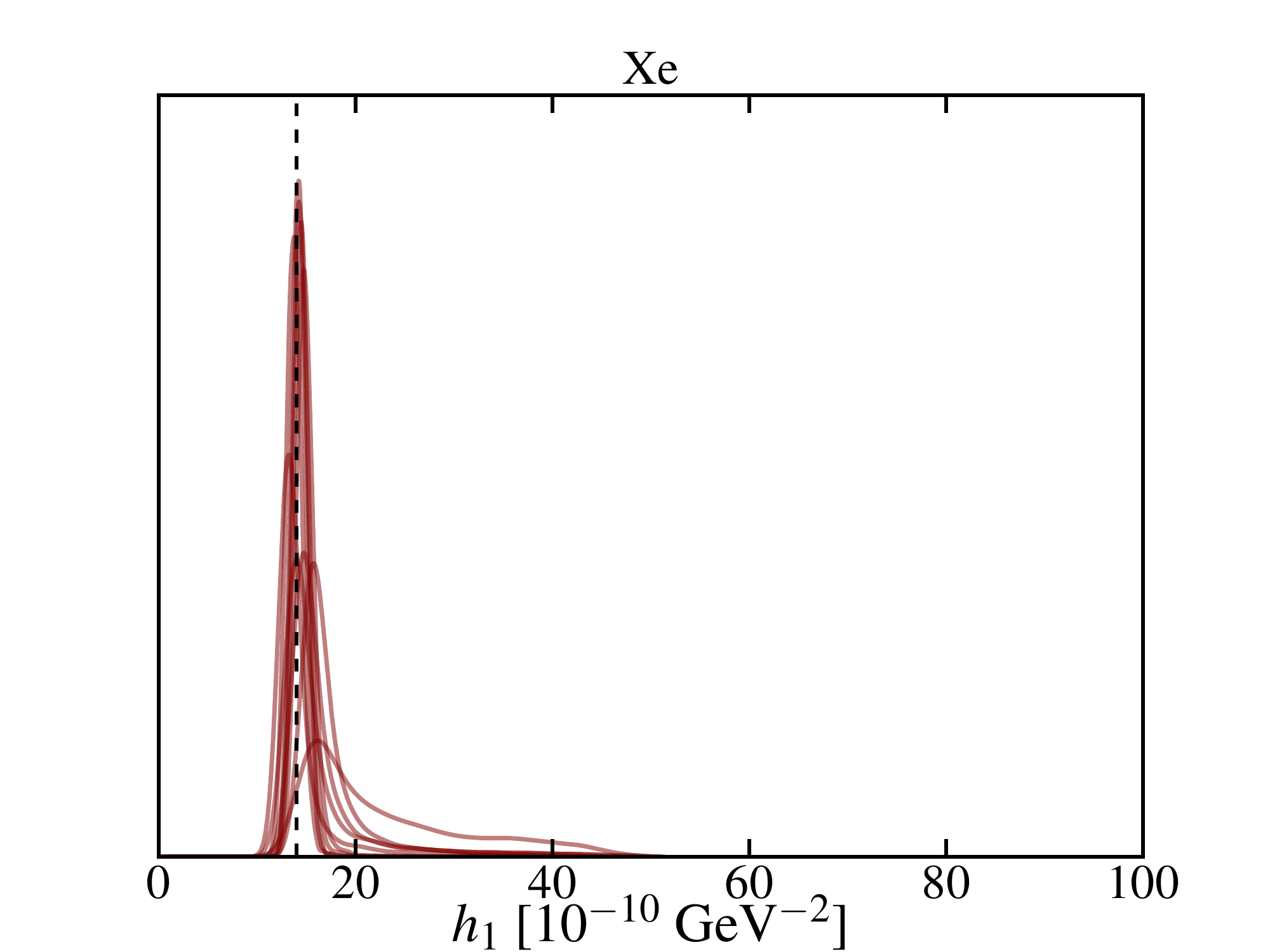}
\includegraphics[width=.32\textwidth,keepaspectratio=true]{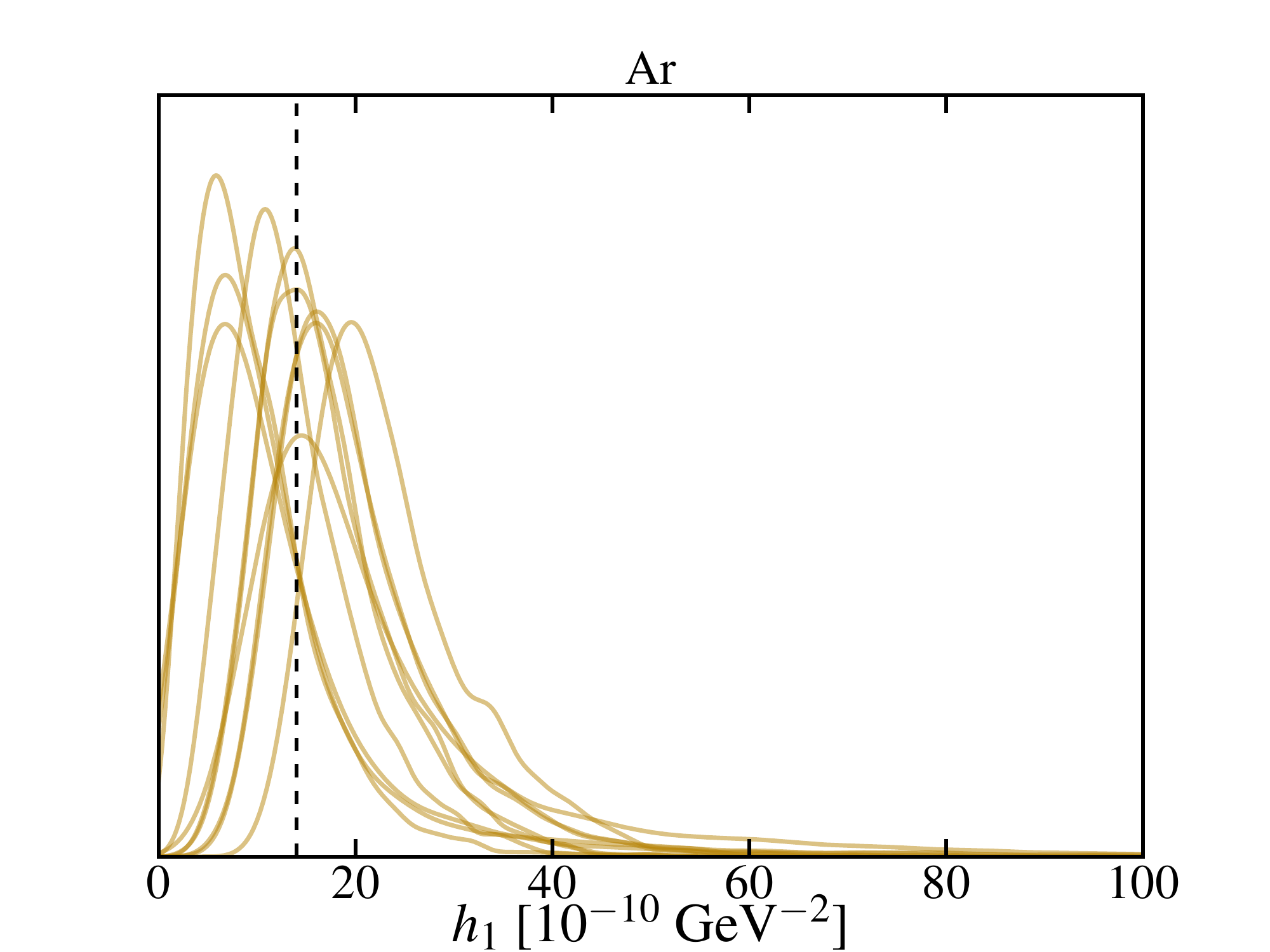}
\includegraphics[width=.32\textwidth,keepaspectratio=true]{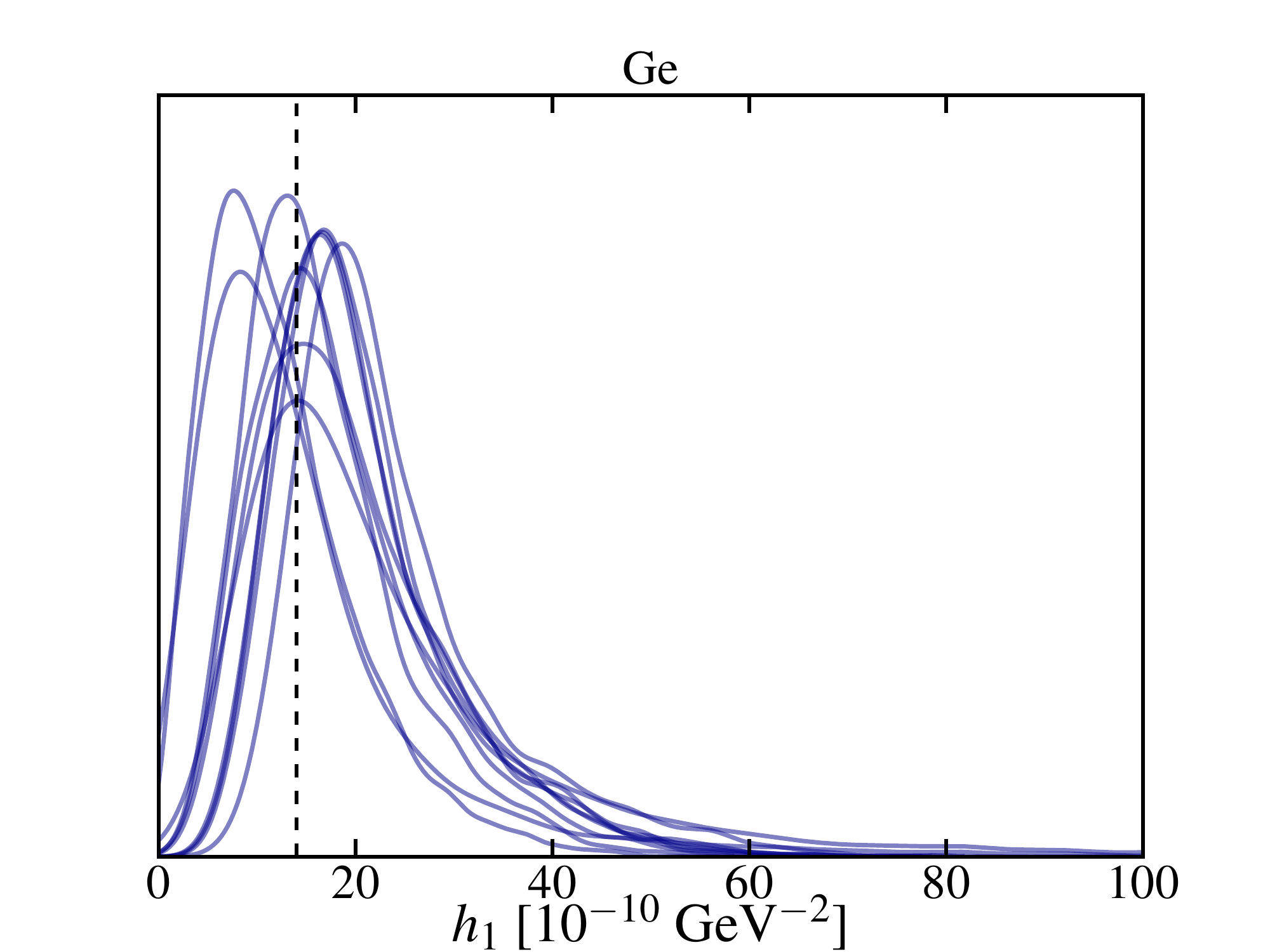}
\includegraphics[width=.32\textwidth,keepaspectratio=true]{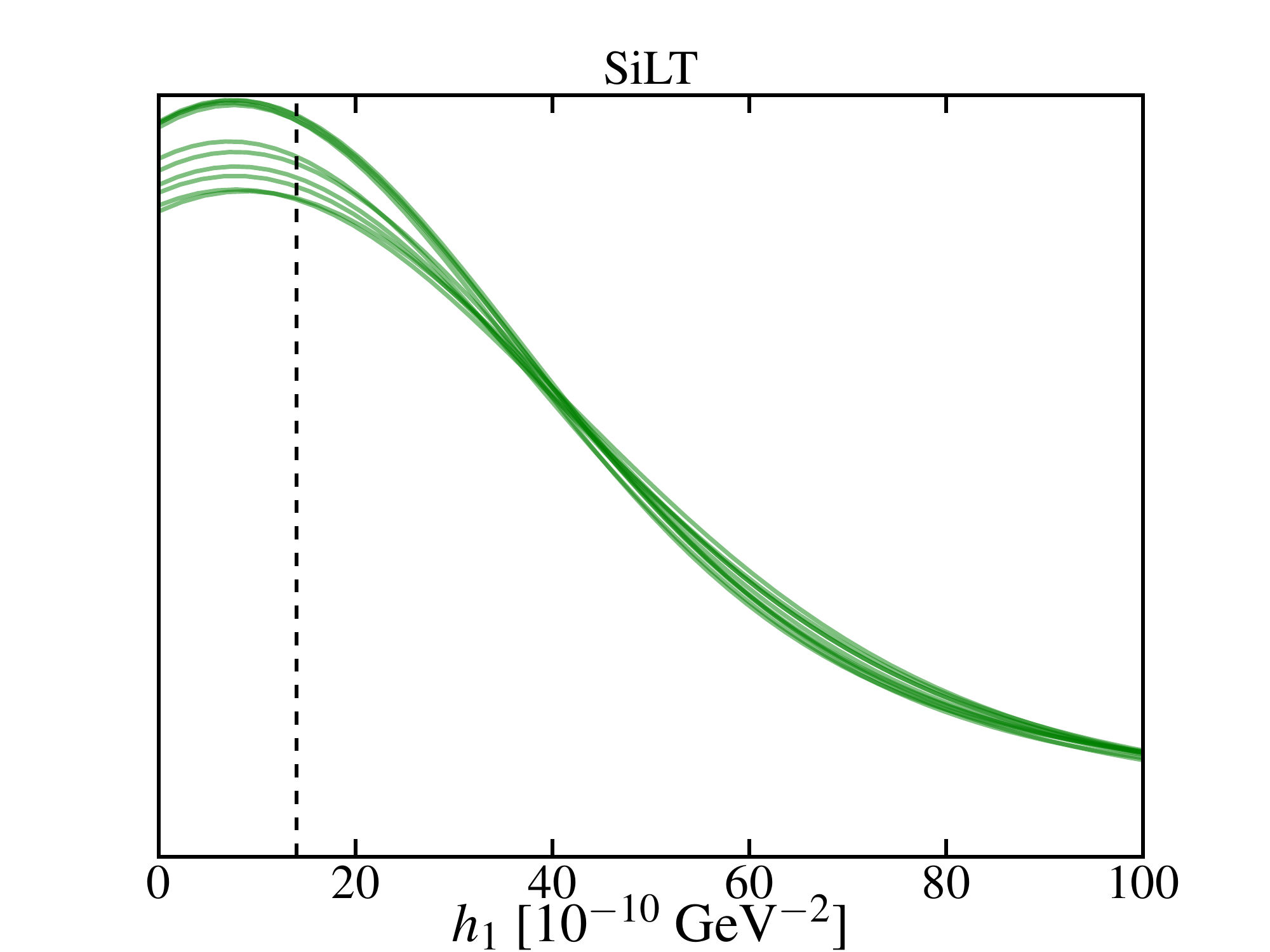}
\includegraphics[width=.32\textwidth,keepaspectratio=true]{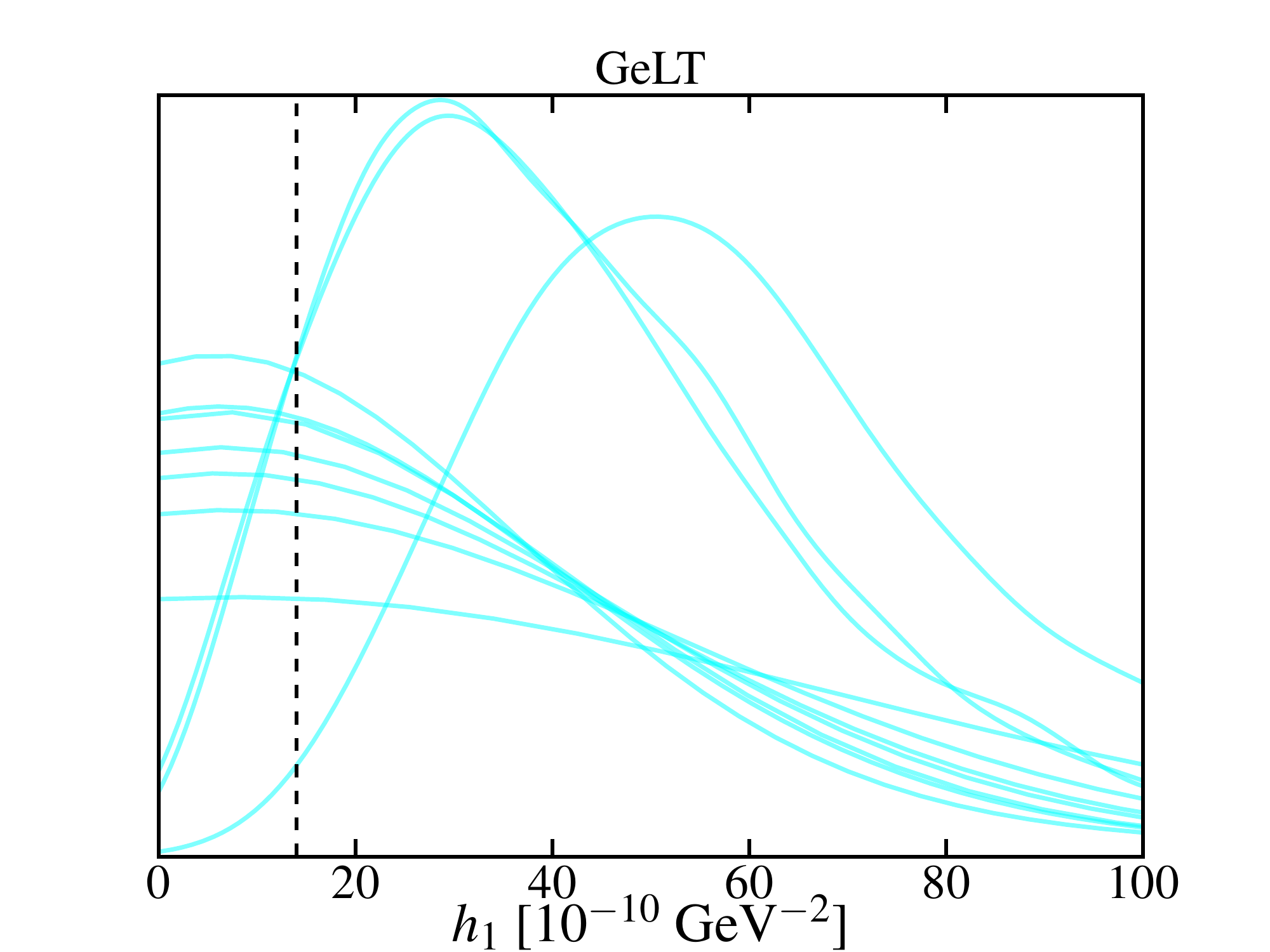}
\includegraphics[width=.32\textwidth,keepaspectratio=true]{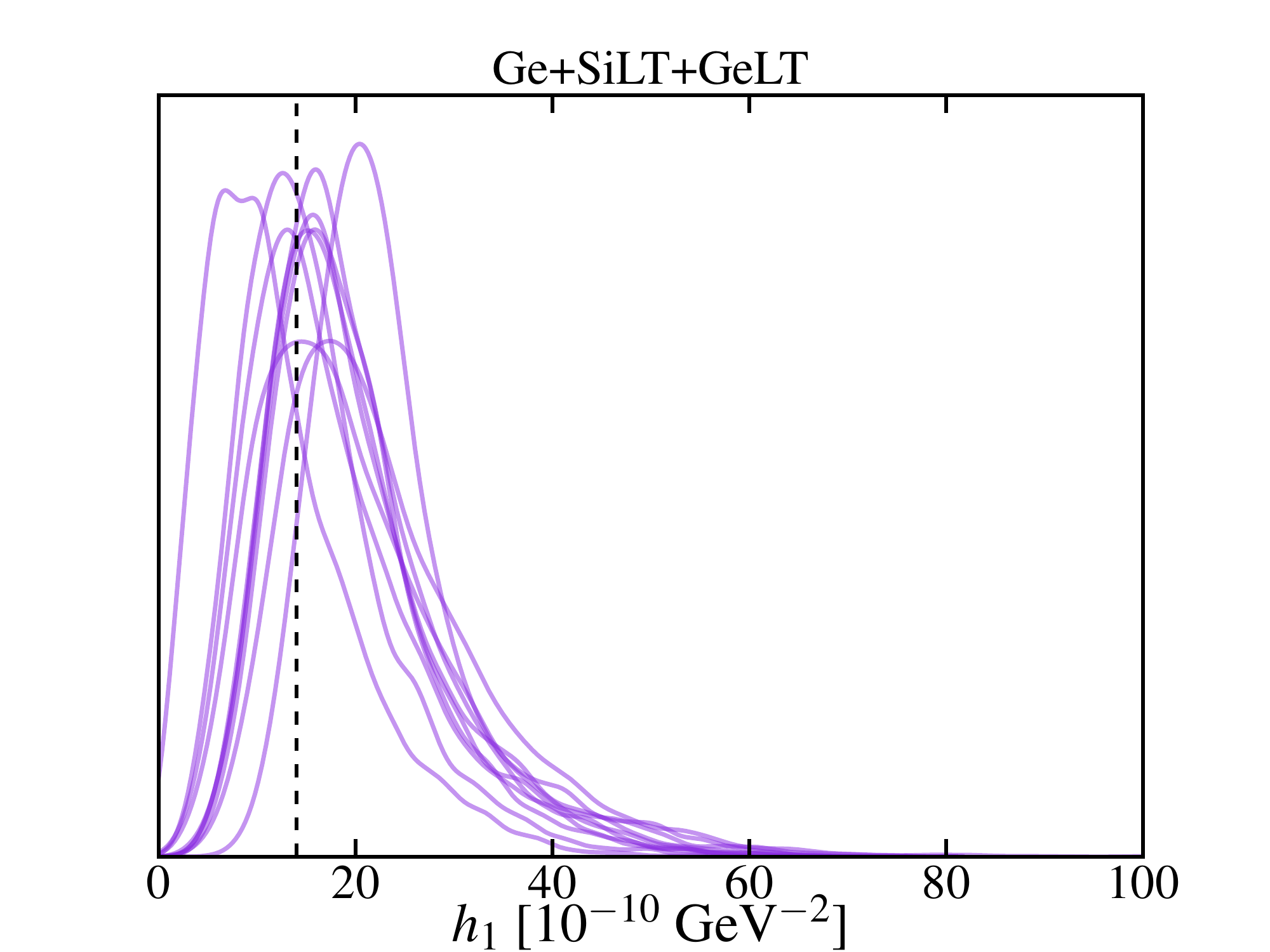}
\includegraphics[width=.32\textwidth,keepaspectratio=true]{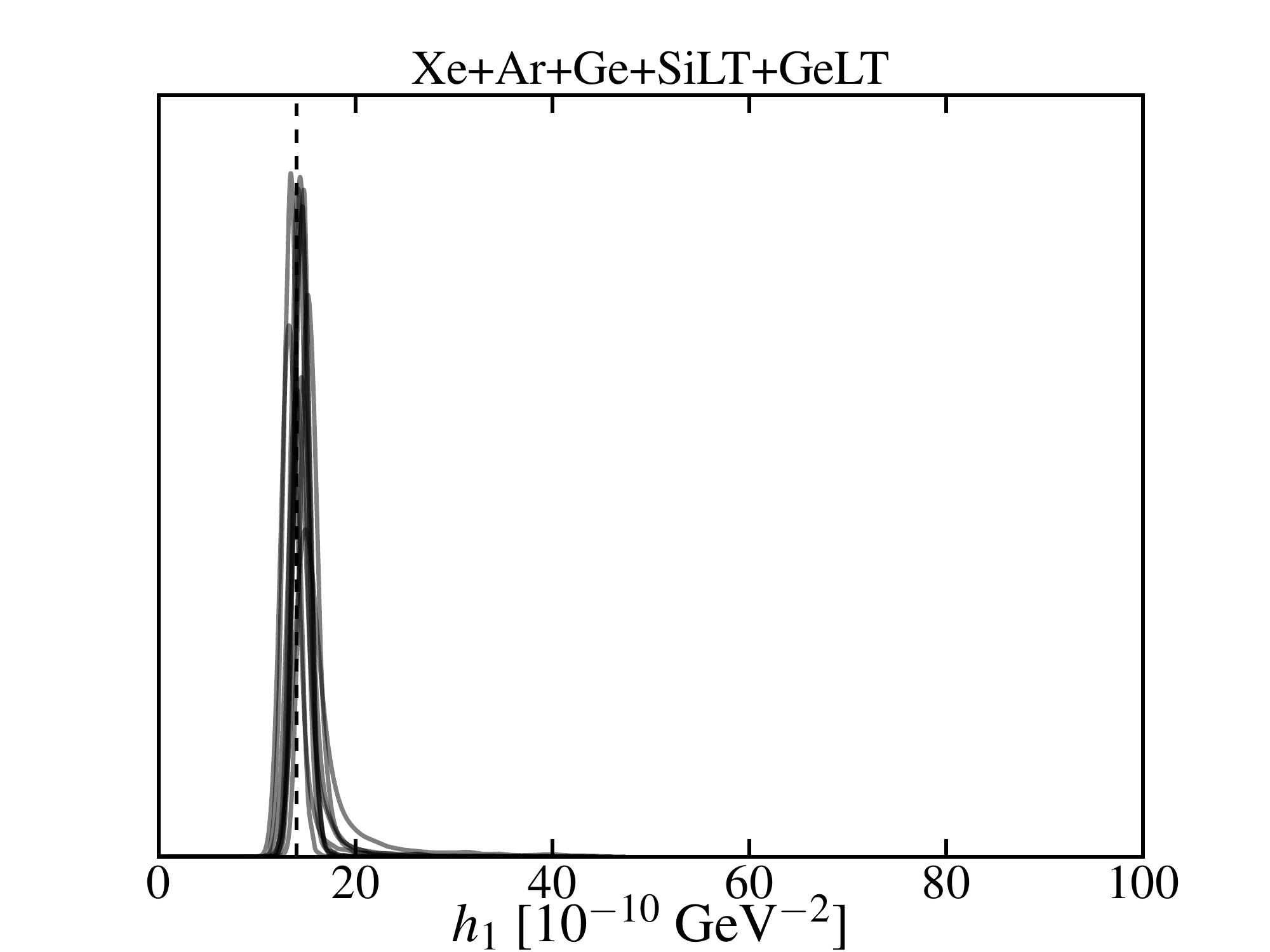}
\caption{Marginalized 1d posteriors for the $h_1$ scenario, for G2 experiments. This Figure represents the results of the posterior reconstruction for a subset of simulations described in \S\ref{sec:distops_single} and analyzed assuming the right underlying operator. Results for all G2 experiments are shown, as well as the results from a joint analysis of some and all of them. The parameter estimation is generally best with the experiment of the largest exposure (here, ``Xe''), and the addition of other experiments leads to further improvement.\label{fig:marginals_G2}}
\end{figure*}
\begin{figure*}
\centering
\includegraphics[width=.48\textwidth,keepaspectratio=true]{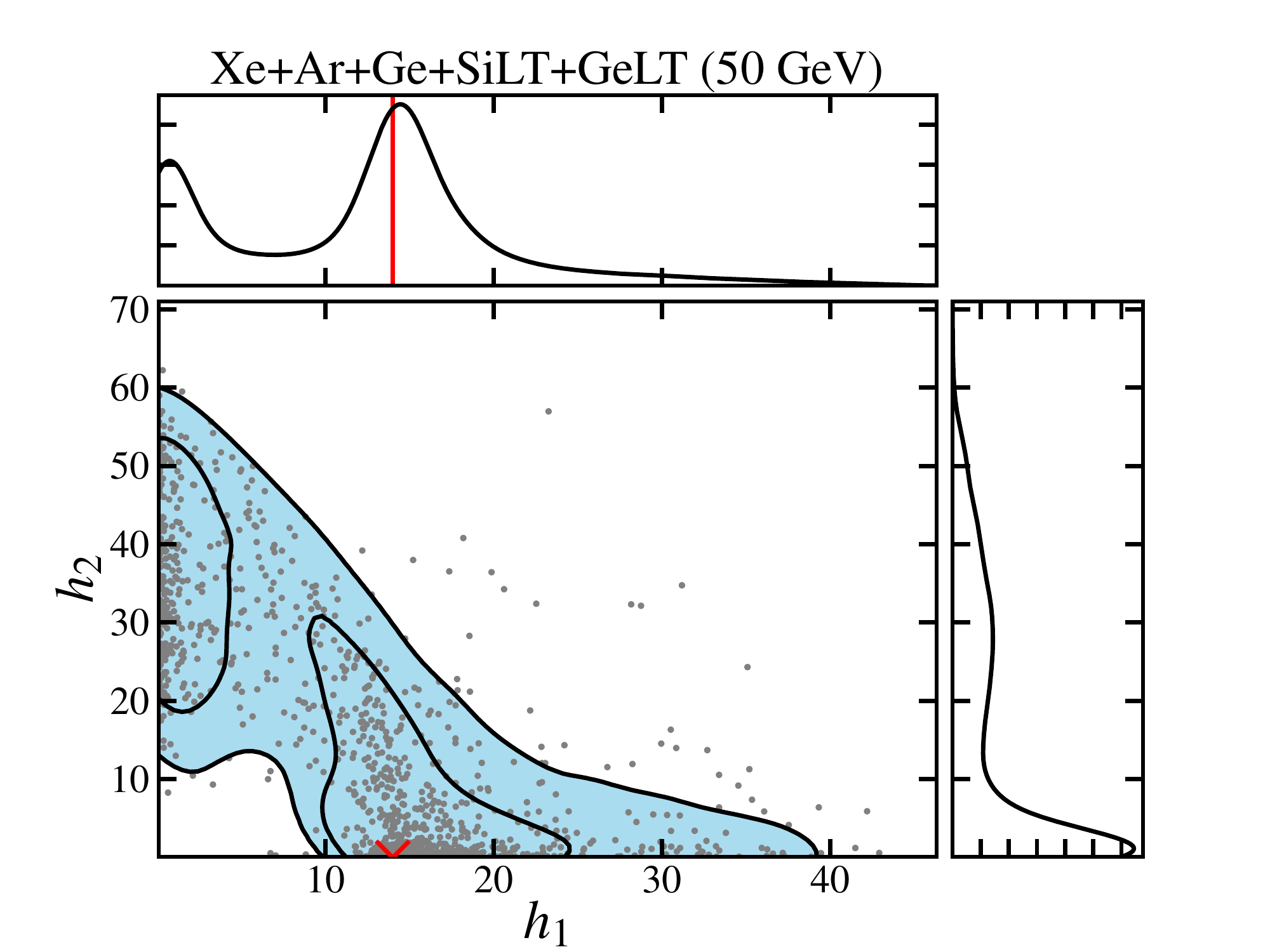}
\includegraphics[width=.48\textwidth,keepaspectratio=true]{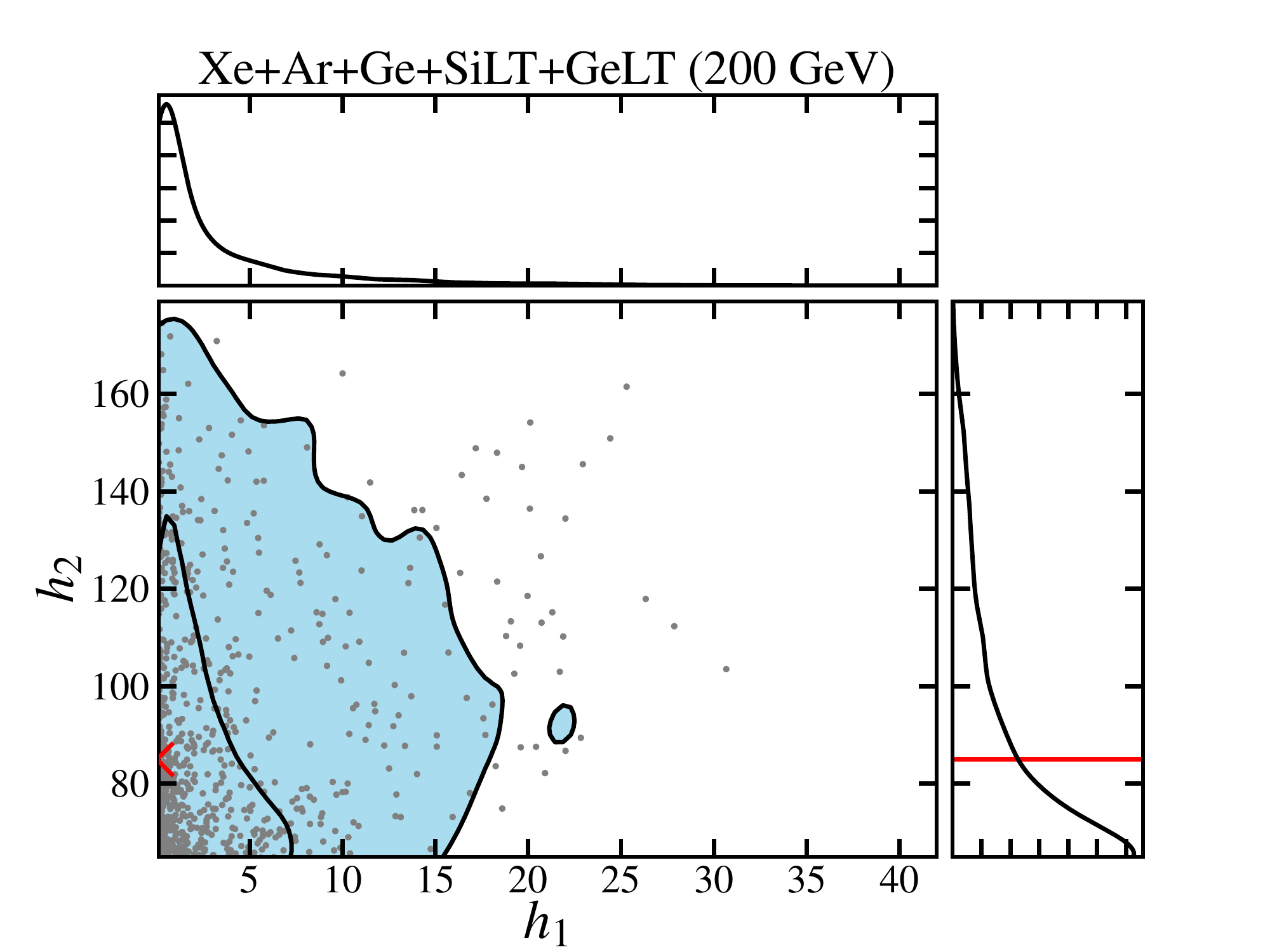}
\includegraphics[width=.48\textwidth,keepaspectratio=true]{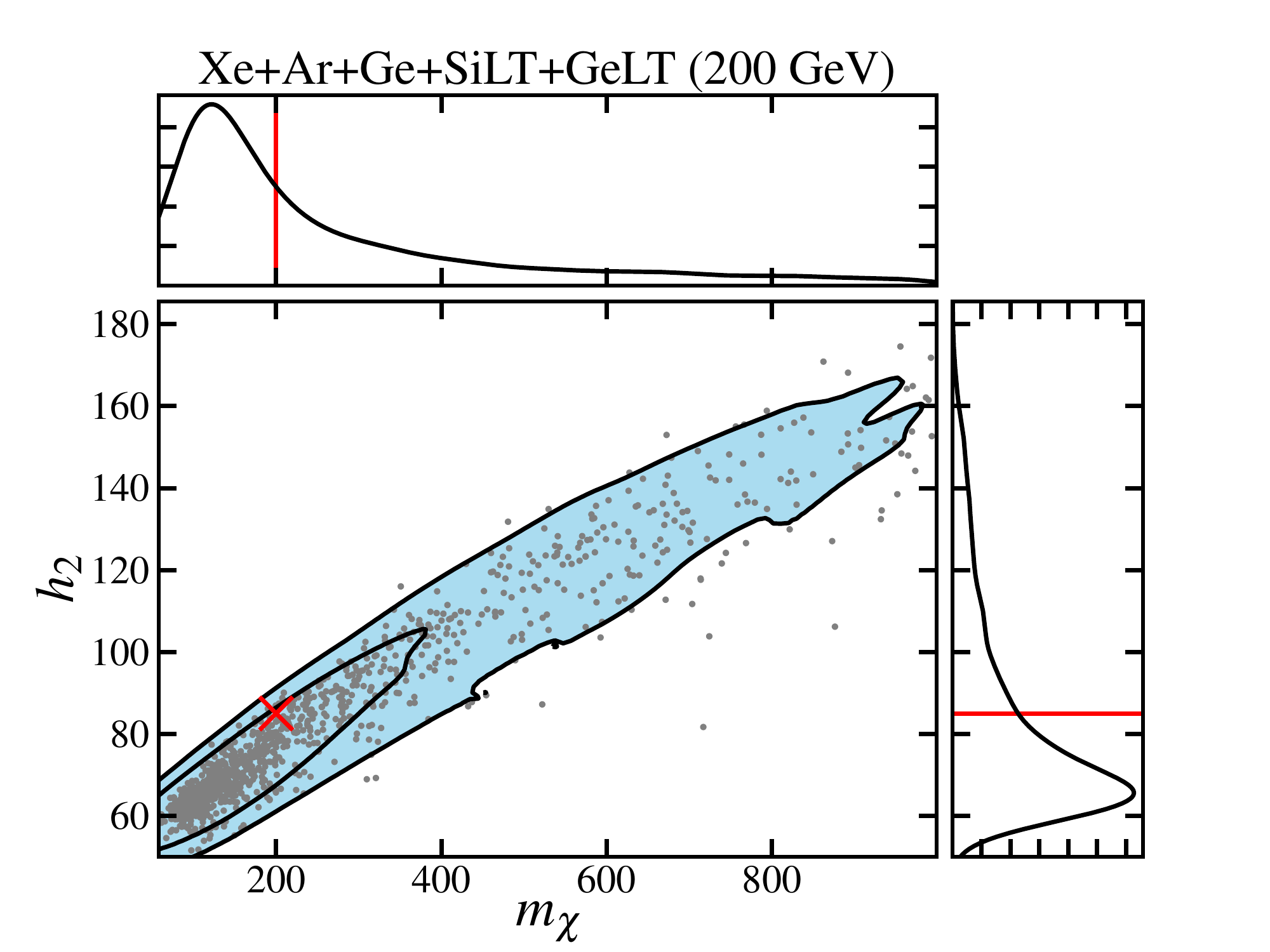}
\includegraphics[width=.48\textwidth,keepaspectratio=true]{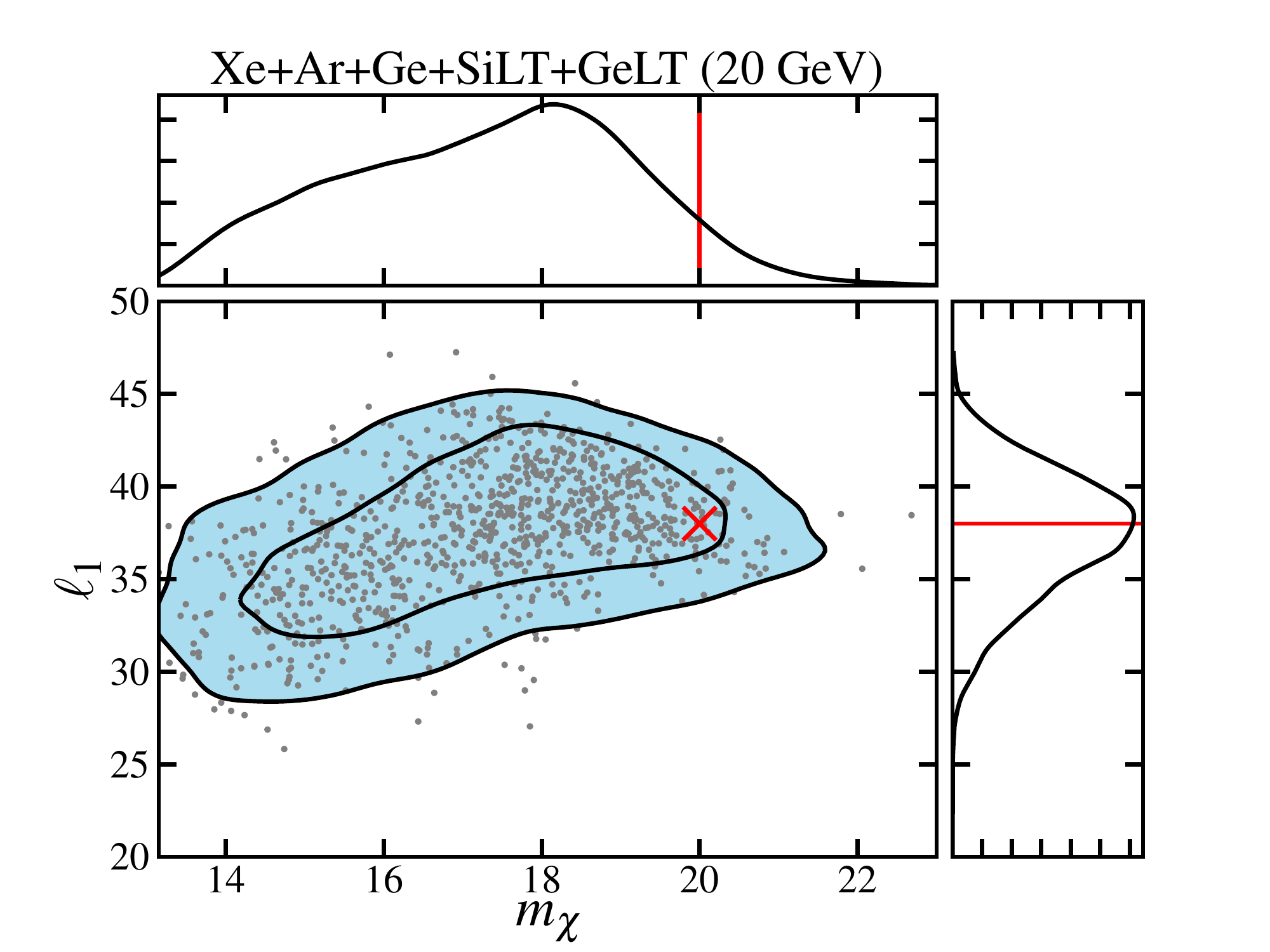}
\caption{Marginalized 2d posterior probability distributions for G2 experiments (with $68\%$ and $95\%$ credible intervals that contain $68\%$ and $95\%$ of the probability around the global maximum, respectively), for typical data realizations generated using a single scattering operator, analyzed agnostically (by fitting the full set of four Wilson coefficients + WIMP mass). Only one-tenth of the sample points used to reconstruct the posterior is shown, for clarity. \textit{Top left}: $h_1$ simulation for a WIMP mass of 50 GeV; this is a case in which the data cannot pick out the dominant operator.  \textit{Top right and bottom left}: $h_2$ simulation for a 200 GeV WIMP; this is an example where model selection is generally successful. \textit{Bottom right}: $\ell_1$ simulation for a 20 GeV WIMP; this is another example of successful model selection.\label{fig:multimodal_broad_contours}}
\end{figure*}
\begin{figure*}
\centering
\includegraphics[width=.7\textwidth,keepaspectratio=true]{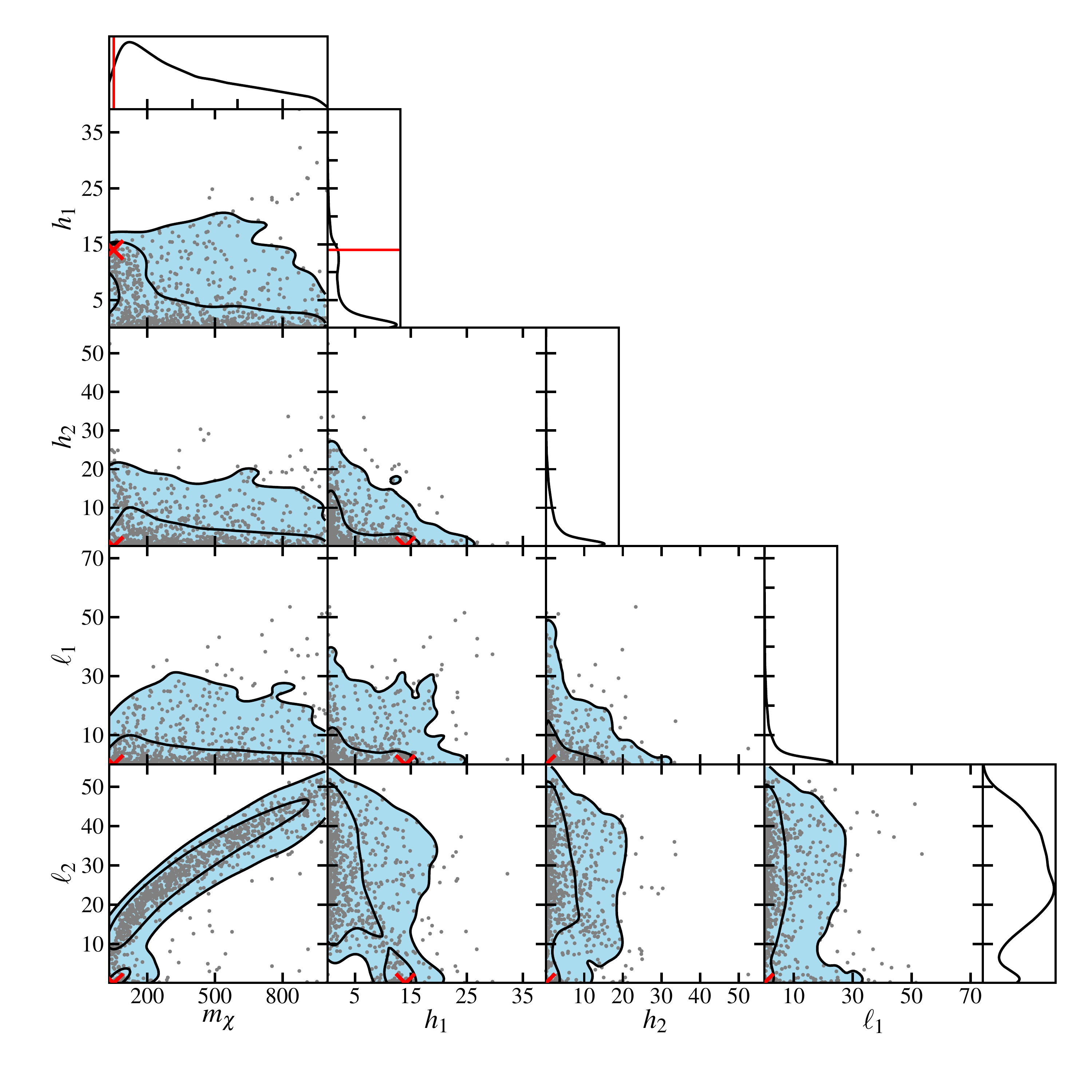}
\caption{Marginalized 2d posterior probability distribution for a typical G2 simulation. A single operator $h_1$ was turned on, and the associated Wilson coefficient set to its current upper limit, and all G2 data sets are analyzed jointly and agnostically (by fitting data with all four Wilson coefficients + WIMP mass). Only one-tenth of the sample points used to reconstruct the posterior is shown, for clarity. \label{fig:triangle_G2}}
\end{figure*}
\begin{figure*}
\centering
\includegraphics[width=.7\textwidth,keepaspectratio=true]{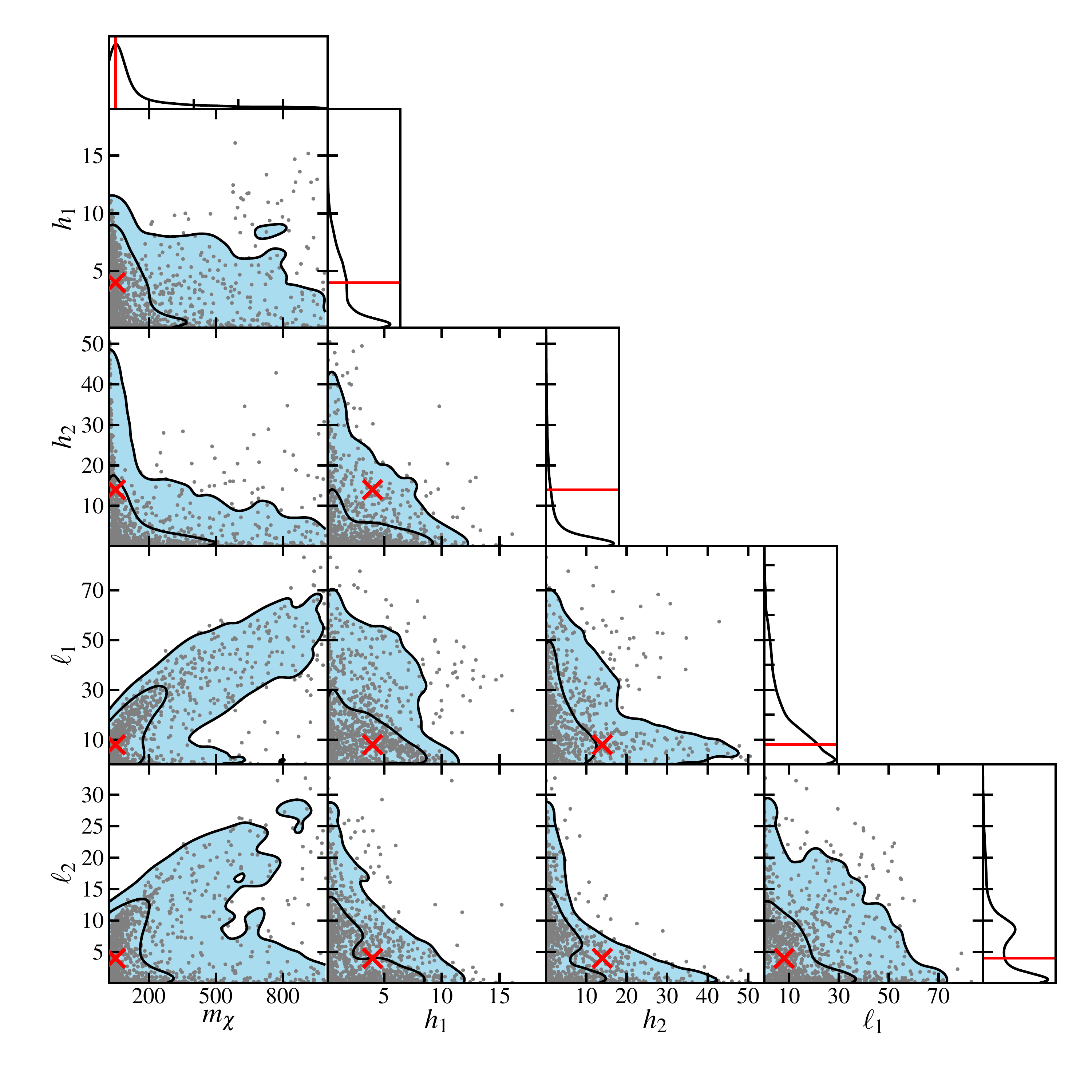}
\caption{Marginalized 2d posterior probability distribution for a typical G2 simulation. Similar to Figure \ref{fig:triangle_G2}, with the difference being that all four operators are turned on in the simulation, and their Wilson coefficients set to a quarter of the appropriate upper limit each.\label{fig:triangle_all4_G2}}
\end{figure*}
To illustrate the quality of parameter estimation with G2 experiments for a WIMP signal just below the current limits, we again turn to the same set of simulations described in \S\ref{sec:distops_single} (where one operator is used for a single simulation) and examine the associated posterior probability distributions. To visualize the results, we produce marginalized posteriors by integrating over some of the free parameters, producing the 1-dimensional (1d) and 2-dimensional (2d) marginalized posteriors, as needed. 

A typical set of 1d marginalized posteriors produced by fitting the right underlying operator model to the simulated data is shown in Figure \ref{fig:marginals_G2}. When the right model is assumed, the experiment with the largest exposure (in this case ``Xe'') will generally deliver the best parameter estimates. Addition of other experiments visibly improves the estimation by suppressing the long tail of the probability distribution.

We now take a closer look at the full five-dimensional posterior space in order to gain a sense of parameter degeneracies arising under different operators. For this purpose, we first use the same single-operator simulations, but this time analyze them agnostically, without assuming the underlying operator, i.e.~we fit the full set of five parameters (WIMP mass + four coupling coefficients). A typical sample of 2d posteriors produced with this procedure is shown in Figure \ref{fig:multimodal_broad_contours}, where the red ``$\times$'' denotes the input values to the simulation at hand. This Figure highlights three notable features of the posteriors. 

First, we see that in cases where the data are generally not able to discern the underlying model, like in the case of a 50 GeV WIMP and $h_1$ interaction (shown in the upper left panel), in spite of a local maximum appearing around the input values, the posterior tends to have a complicated multimodal shape. Interestingly, in this specific case, the $68\%$ credible intervals are peaked at the $h_1 = 0$ and $h_2=0$ values, suggesting that the data prefer that only one operator dominates, but is indecisive on which one. 
 
Second, even in the cases where model selection is $100\%$ successful, like in the case of a 200 GeV WIMP with $h_2$ interaction (shown in the upper right and lower left panels), there are large degeneracies; for example, between $h_2$ and the WIMP mass (see the lower left panel). This is a well-known phenomenon from the classical literature on direct and indirect detection: the WIMP-nuclear reduced mass asymptotes to a constant value for large WIMP masses, and the only dependence of the energy spectrum on the WIMP mass remains in the prefactor (see Eq.~(\ref{eq:drdq_general})).  Thus, in this regime, the actions of increasing the coupling coefficient (also a multiplier factor to $dR/dE_R$) and decreasing the WIMP mass have the same effect on the recoil spectra, which produces the observed degeneracy. Moreover, when such data is analyzed agnostically, even though the reconstruction of the parameters is not biased, this degeneracy shows up in all subsets of the full parameter space (see the upper right panel).
  
And third, in the light-mediator scenario, such as the case of a 20 GeV WIMP with $\ell_1$ interaction (shown in the lower right panel), the reconstruction of the WIMP mass is quite accurate, even when all five parameters are allowed to vary in the fit.  This is important---as we show in the next section, when we assume that only one operator dominates the scattering, if a wrong choice of the operator is made, this last feature can give rise to significant biases in mass estimates.  

More detailed features of all 2d marginalized posteriors in a typical data realization of the G2 set can be seen in Figures \ref{fig:triangle_G2} and \ref{fig:triangle_all4_G2}. The former shows one of the simulations of the set described in the previous section, created using the $h_1$ operator at its maximal strength. For the latter one, a different simulation is used than in all previous discussion, and in it all four operators are turned on, and the four Wilson coefficients are set to a quarter of their respective current upper limit each. For both Figures, the simulations are analyzed agnostically, by fitting the WIMP mass and all four effective couplings, simultaneously. In the latter case, as expected, parameter estimation degrades compared to the case where one operator is at its maximal strength, but the degeneracy regions in both cases are very large.
\subsection{Reliability of mass estimates}  
\label{sec:massbias}
\begin{figure*}[tbp]
\centering
\includegraphics[width=.4\textwidth,keepaspectratio=true]{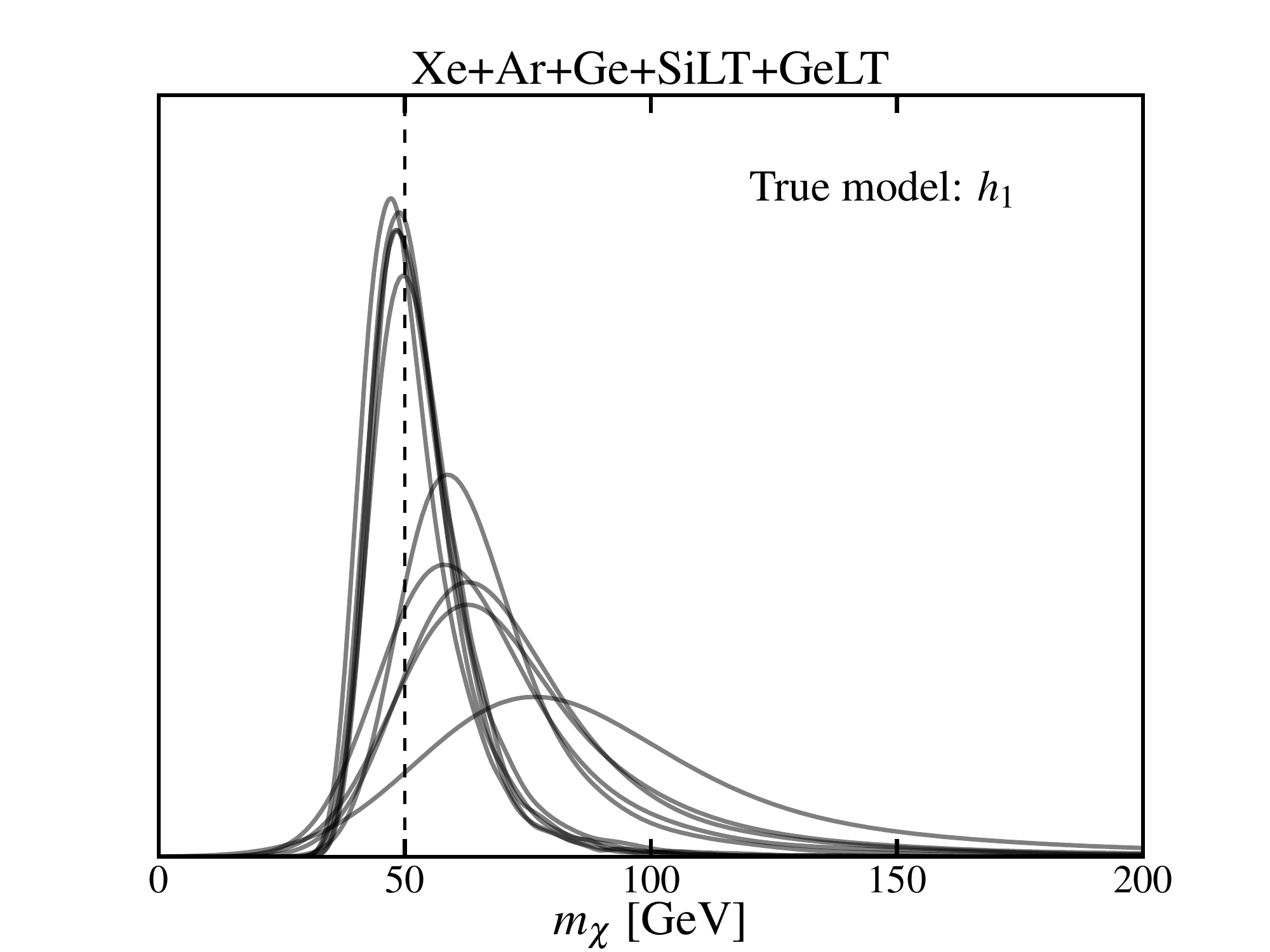}
\includegraphics[width=.4\textwidth,keepaspectratio=true]{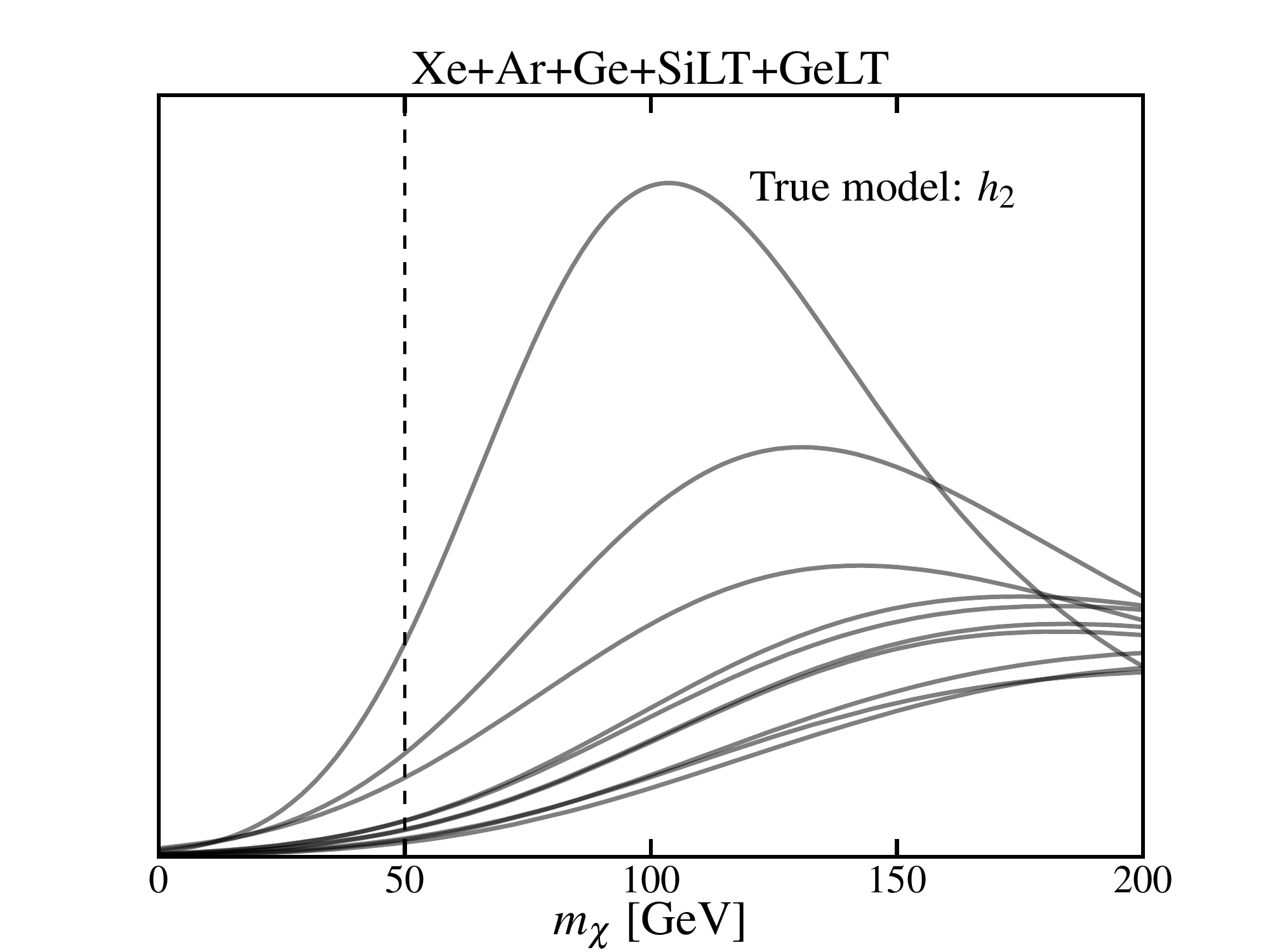}
\includegraphics[width=.4\textwidth,keepaspectratio=true]{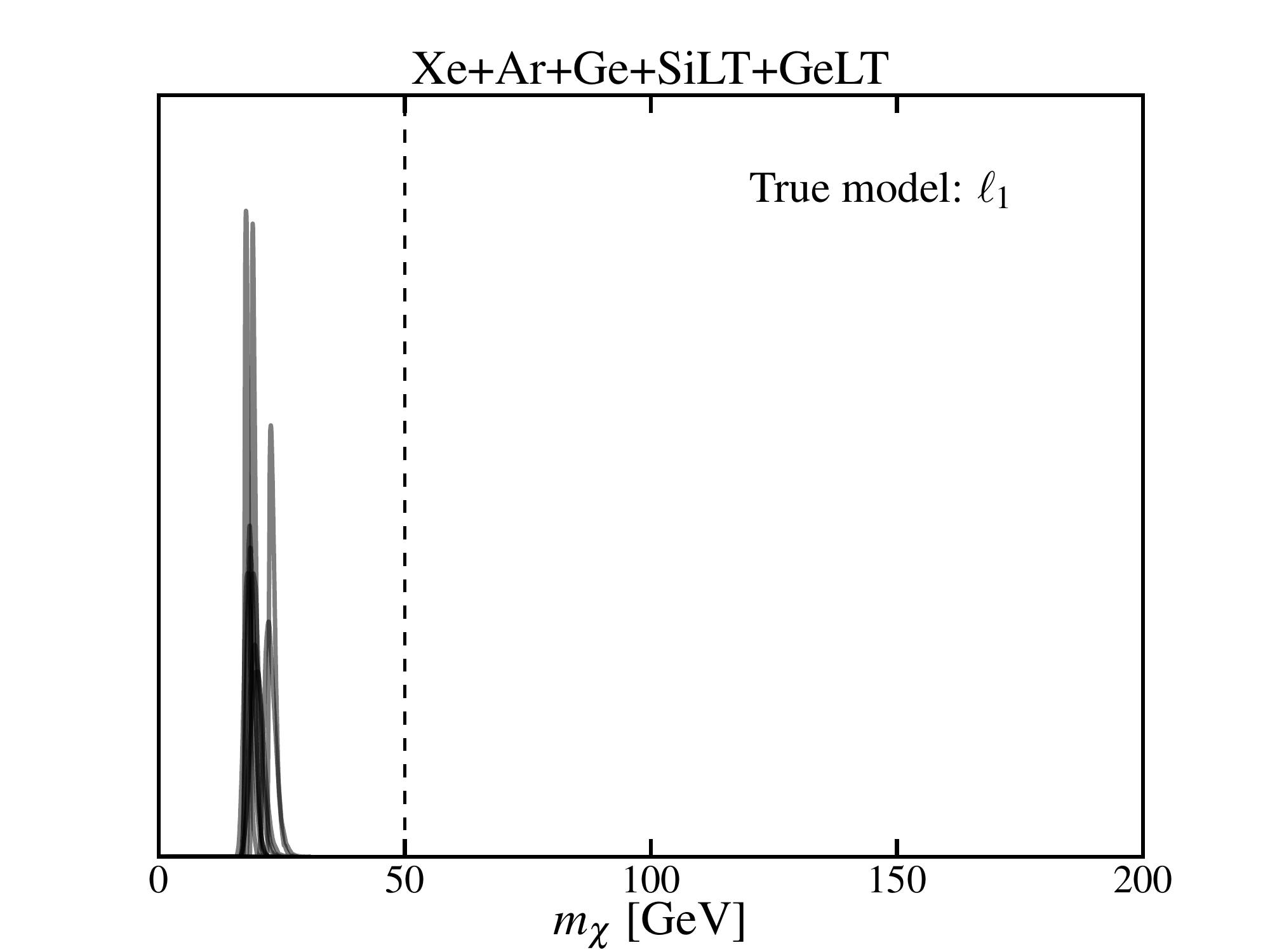}
\includegraphics[width=.4\textwidth,keepaspectratio=true]{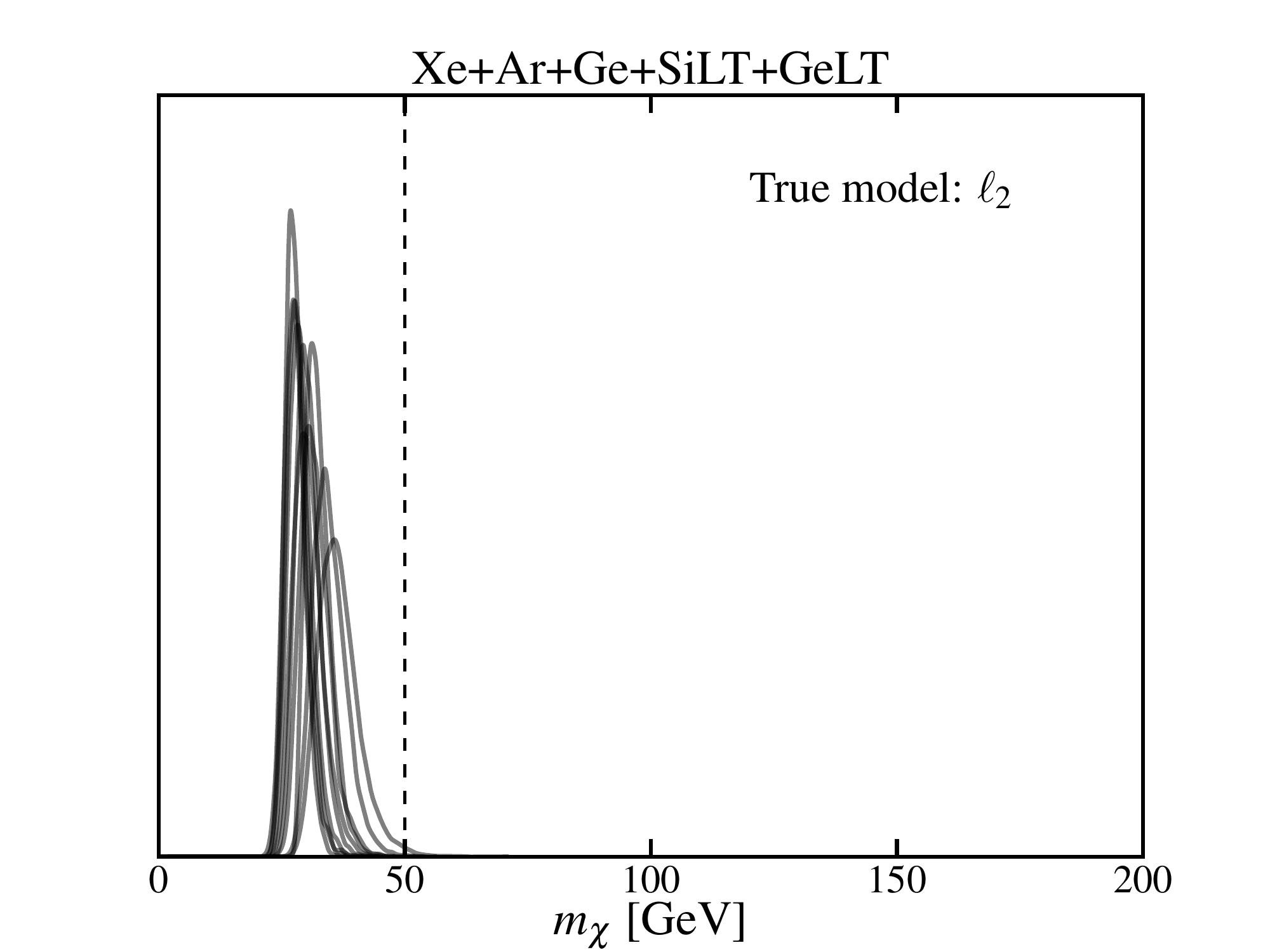}
\caption{Marginal 1d posterior probability distribution for the WIMP mass. The posterior distributions shown here represent a subset of 10 typical single-operator simulations discussed in \S\ref{sec:distops_single} (each panel represents simulations generated with a different underlying operator). All simulations are fit with the canonical $h_1$ model. If the assumption of $h_1$ is wrong (the case in all panels, except the top left panel), the mass estimate is biased. \label{fig:biased_margs_mass}}
\end{figure*}
Given that the WIMP mass is the main parameter that direct, indirect, and collider searches all aim to measure, it is important to understand how robust are the mass measurements in direct searches. In particular, in this Section we investigate the dependence of the mass estimate on the common assumption that the scattering is governed by a momentum-independent interaction through a heavy mediator (the $h_1$ scenario). For this purpose, we choose a subset of single-operator simulations described in \S\ref{sec:distops_single}, and on each of them perform four \textsc{MultiNest} runs, in each run fitting a different operator model. The resulting marginalized posteriors are shown in Figure \ref{fig:biased_margs_mass}. The conclusion we draw from this Figure is that if the standard $h_1$ scenario is assumed, while the underlying truth is something else (in either light- or heavy-mediator case), mass estimates are severely biased (even though the uncertainties are very small, in some cases). This effect can be understood in terms of the degeneracy between the model choice and the WIMP mass, which both control the slope of the recoil-energy spectrum. For example, if the spectrum is assumed to be shallower (by choosing $h_1$) for a given WIMP mass, while the true model (like $\ell_1$) forces it to be steeper for the same mass, then the mass estimate is pushed toward lower masses to compensate for the wrong assumption about the underlying operator. Therefore, the ability to robustly measure the WIMP mass crucially relies on the correct assumption about the momentum dependence of WIMP-baryon scattering. Therefore, it is advisable that future data analysis be done either in an agnostic way (by considering a larger set of underlying operators), or, preferably, that it includes model selection as an integral first step.
\subsection{Ultimate reach of (background-free) direct detection}
\label{sec:ultimate}
\begin{figure} [tbp]
\centering
\includegraphics[width=.5\textwidth,keepaspectratio=true]{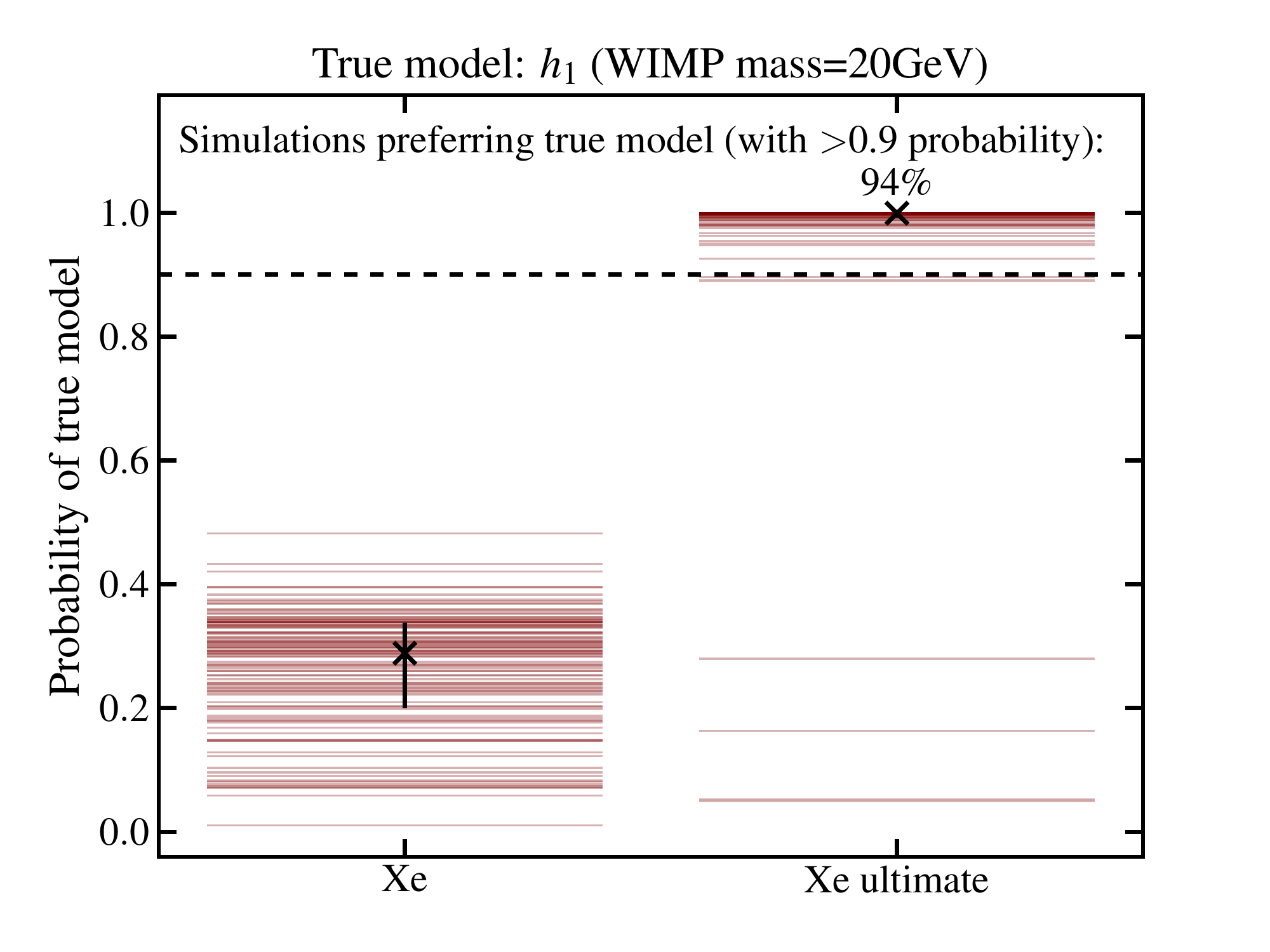}
\caption{Model-selection prospects for xenon-based detectors. Improvement in the prospects for correctly identifying the true effective operator from G2 ``Xe'' to ``Xe ultimate'' is striking. The case we chose to illustrate the improvement corresponds to the most pessimistic scenario in which model selection is unlikely for G2 (the case of a standard operator $h_1$, for a low WIMP mass).}
\label{fig:ultimateselection}
\end{figure}
\begin{figure*}[tbp]
\centering
\includegraphics[width=.48\textwidth,keepaspectratio=true]{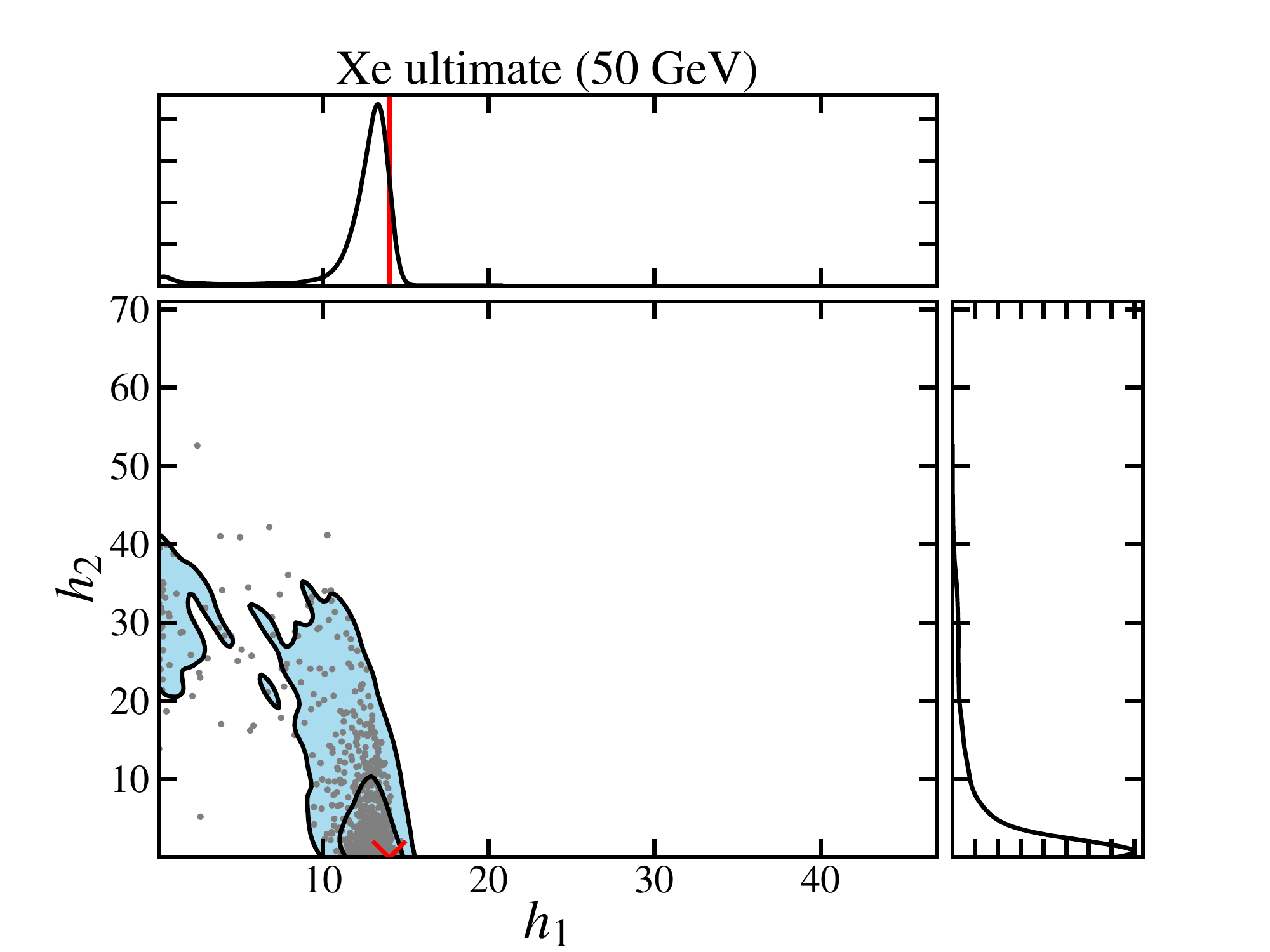}
\includegraphics[width=.48\textwidth,keepaspectratio=true]{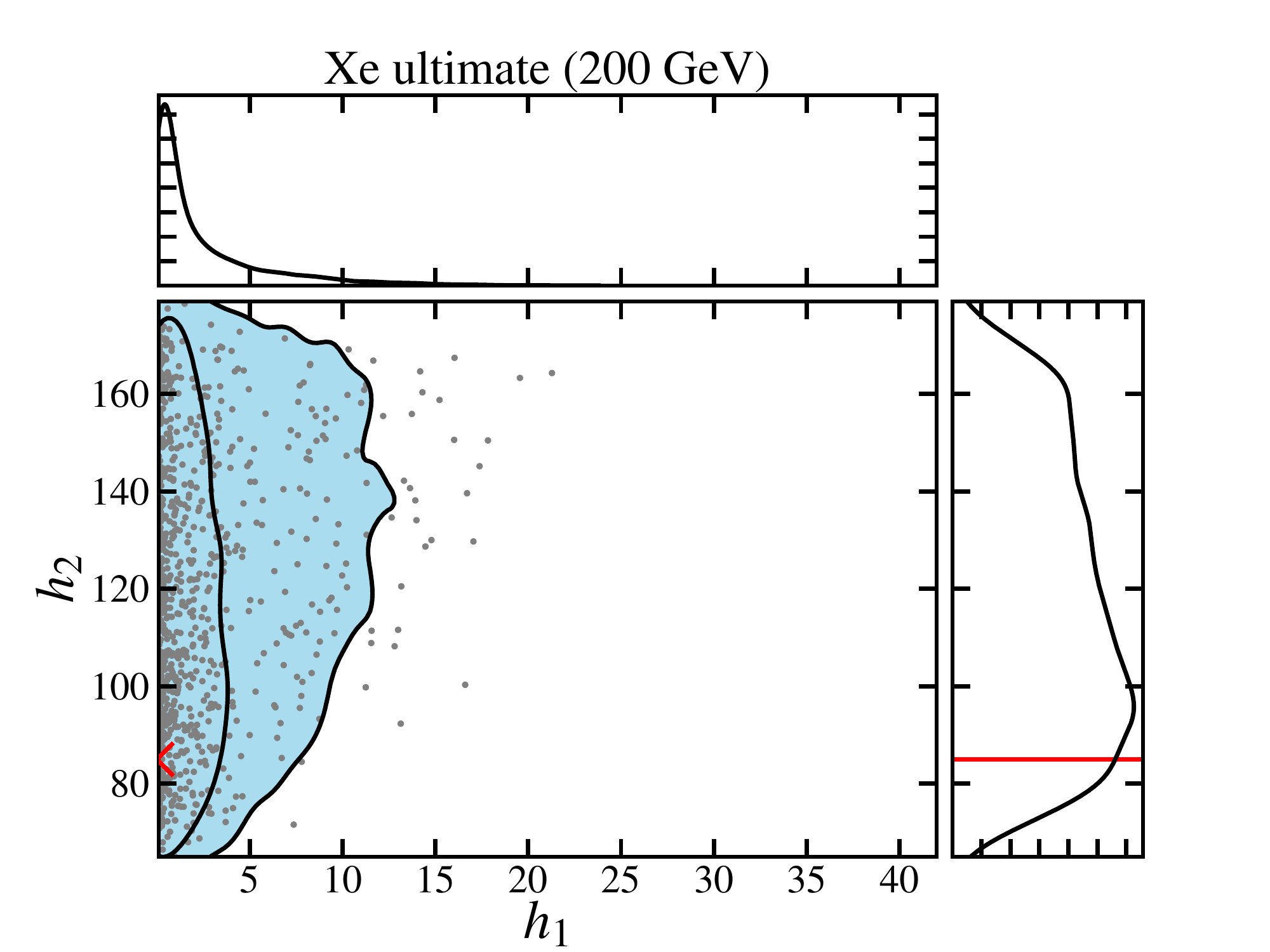}
\includegraphics[width=.48\textwidth,keepaspectratio=true]{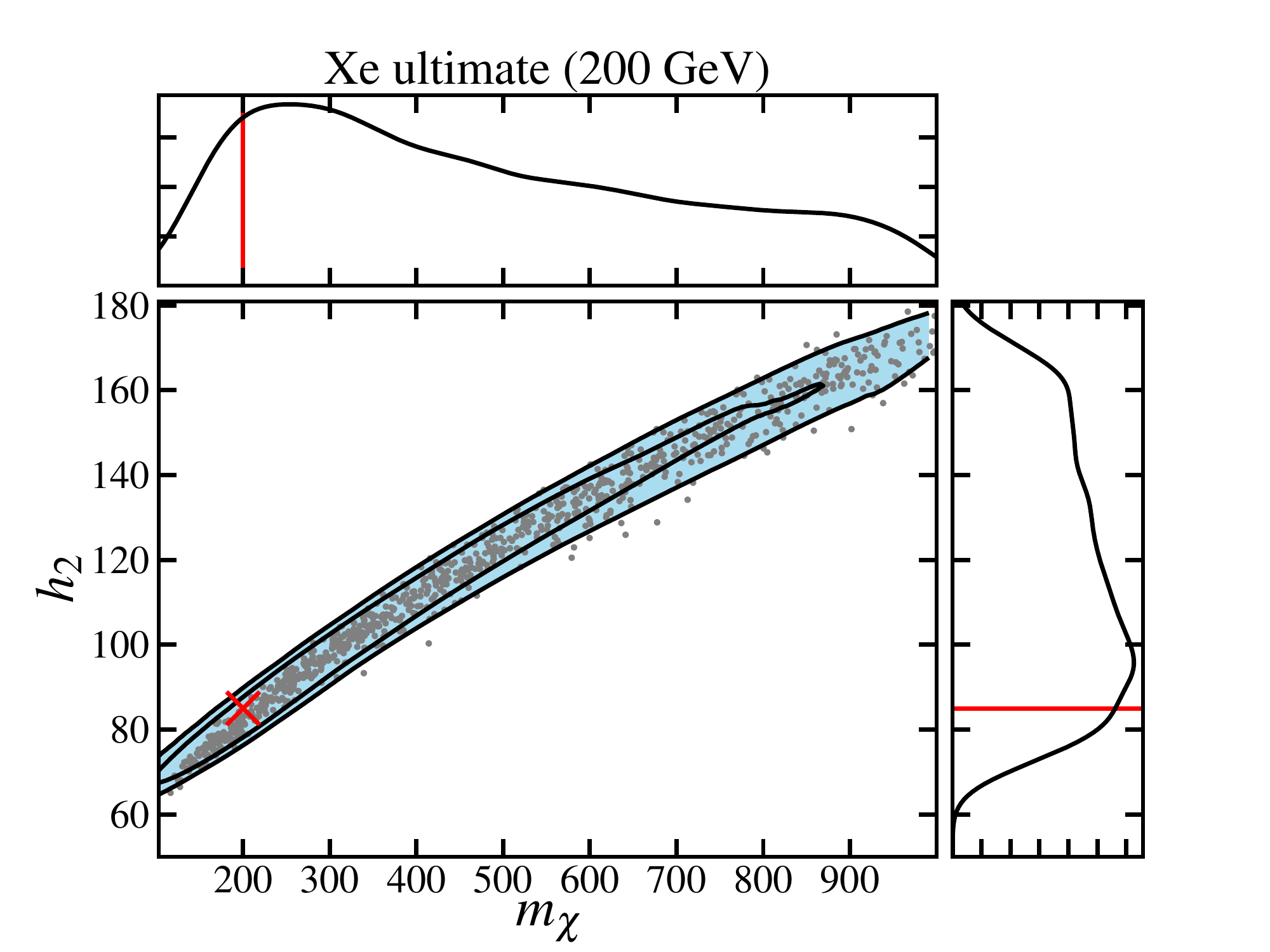}
\includegraphics[width=.48\textwidth,keepaspectratio=true]{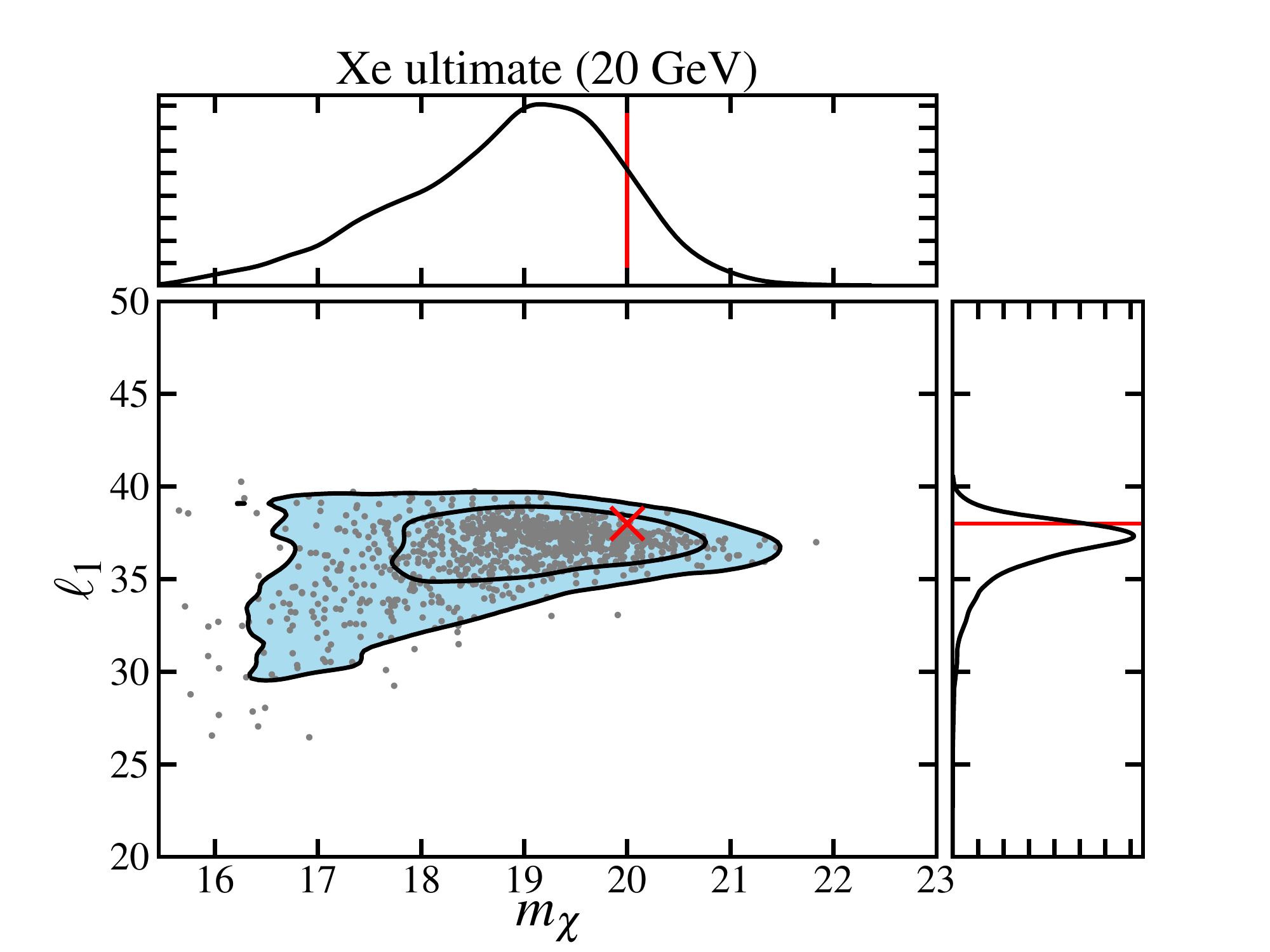}
\caption{Same as Figure \ref{fig:multimodal_broad_contours}, but for the ``Xe ultimate'' experiment that reaches the neutrino-background floor. \label{fig:multimodal_broad_contours_ultimate}}
\end{figure*}
\begin{figure*}[tbp]
\centering
\includegraphics[width=.7\textwidth,keepaspectratio=true]{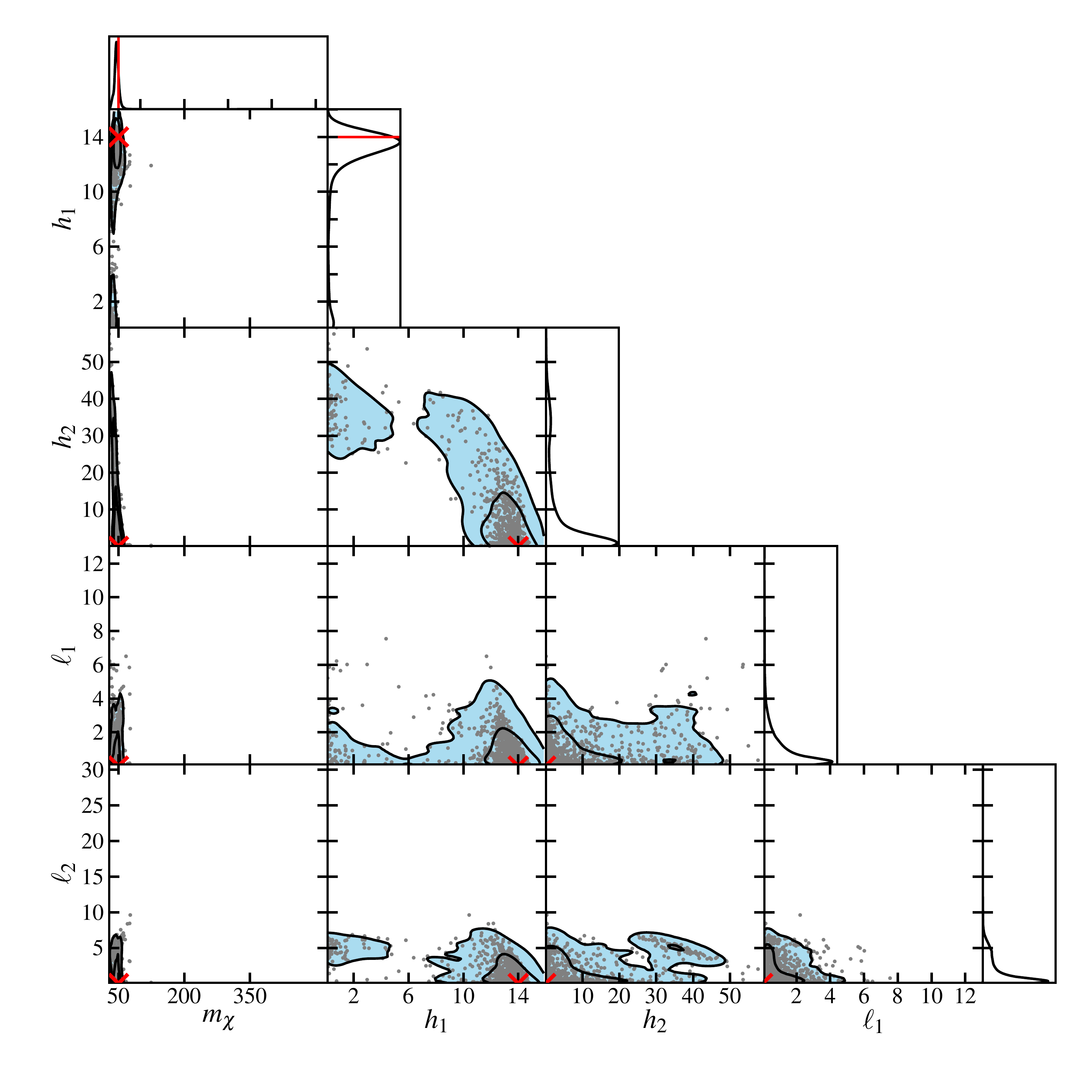}
\caption{Same as Figure \ref{fig:triangle_G2}, but for the ``Xe ultimate'' experiment that reaches the neutrino-background floor.\label{fig:triangle_ultimate}}
\end{figure*}
\begin{figure*}[tbp] 
\centering
\includegraphics[width=.7\textwidth,keepaspectratio=true]{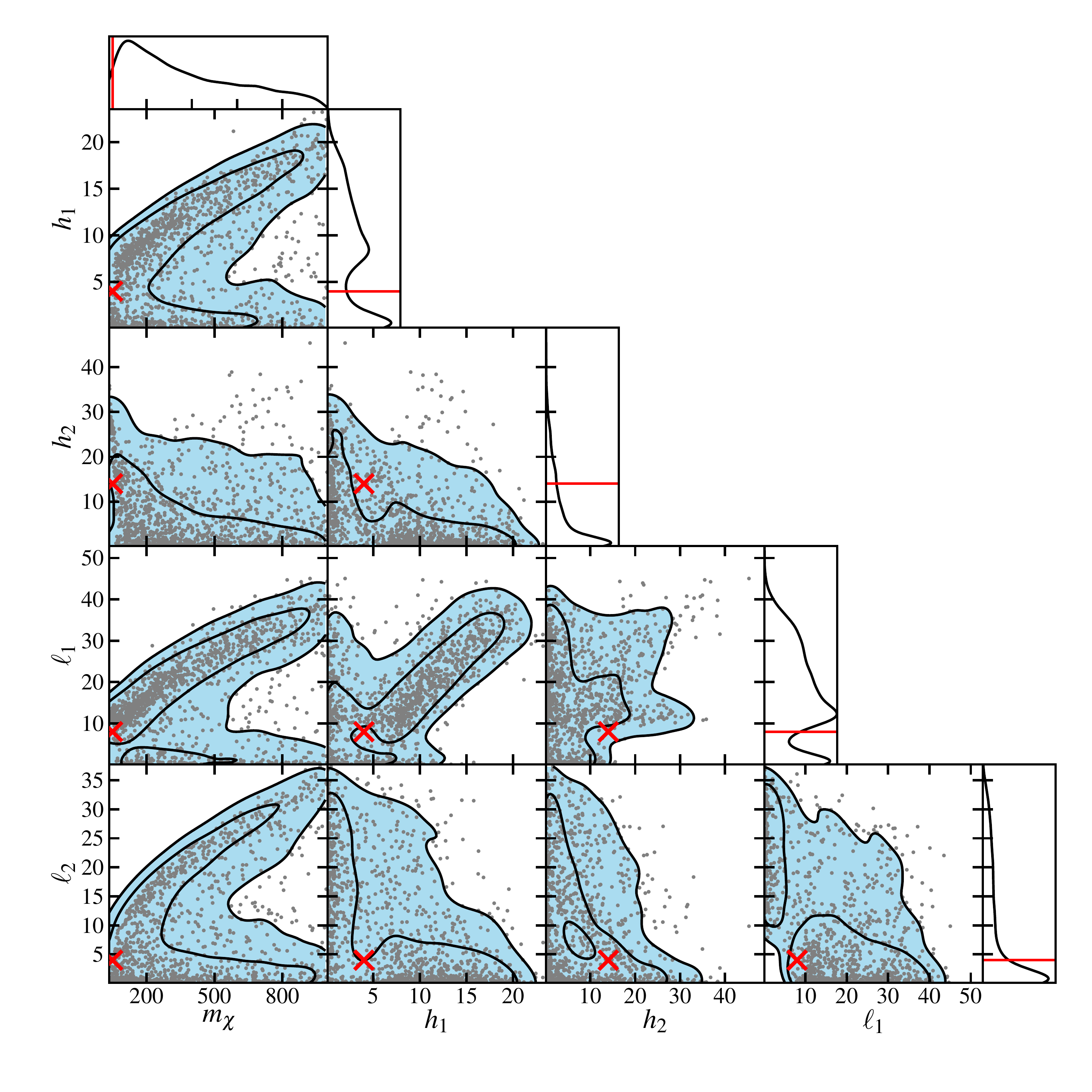}
\caption{Same as Figure \ref{fig:triangle_all4_G2}, but for the ``Xe ultimate'' experiment that reaches the neutrino-background floor. \label{fig:triangle_all4_ultimate}}
\end{figure*}
\begin{figure*}[h]
\centering
\includegraphics[width=.48\textwidth,keepaspectratio=true]{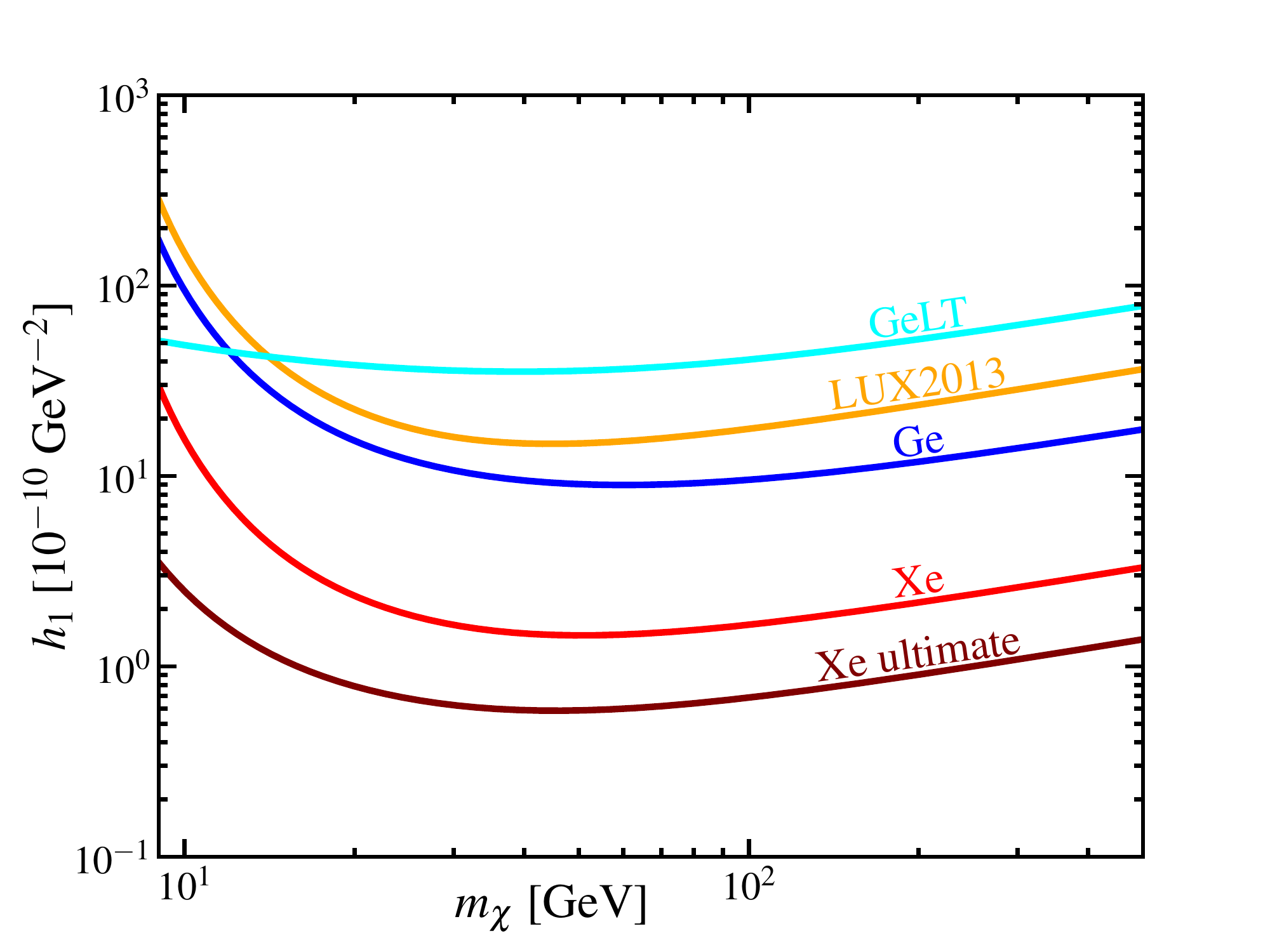}
\includegraphics[width=.48\textwidth,keepaspectratio=true]{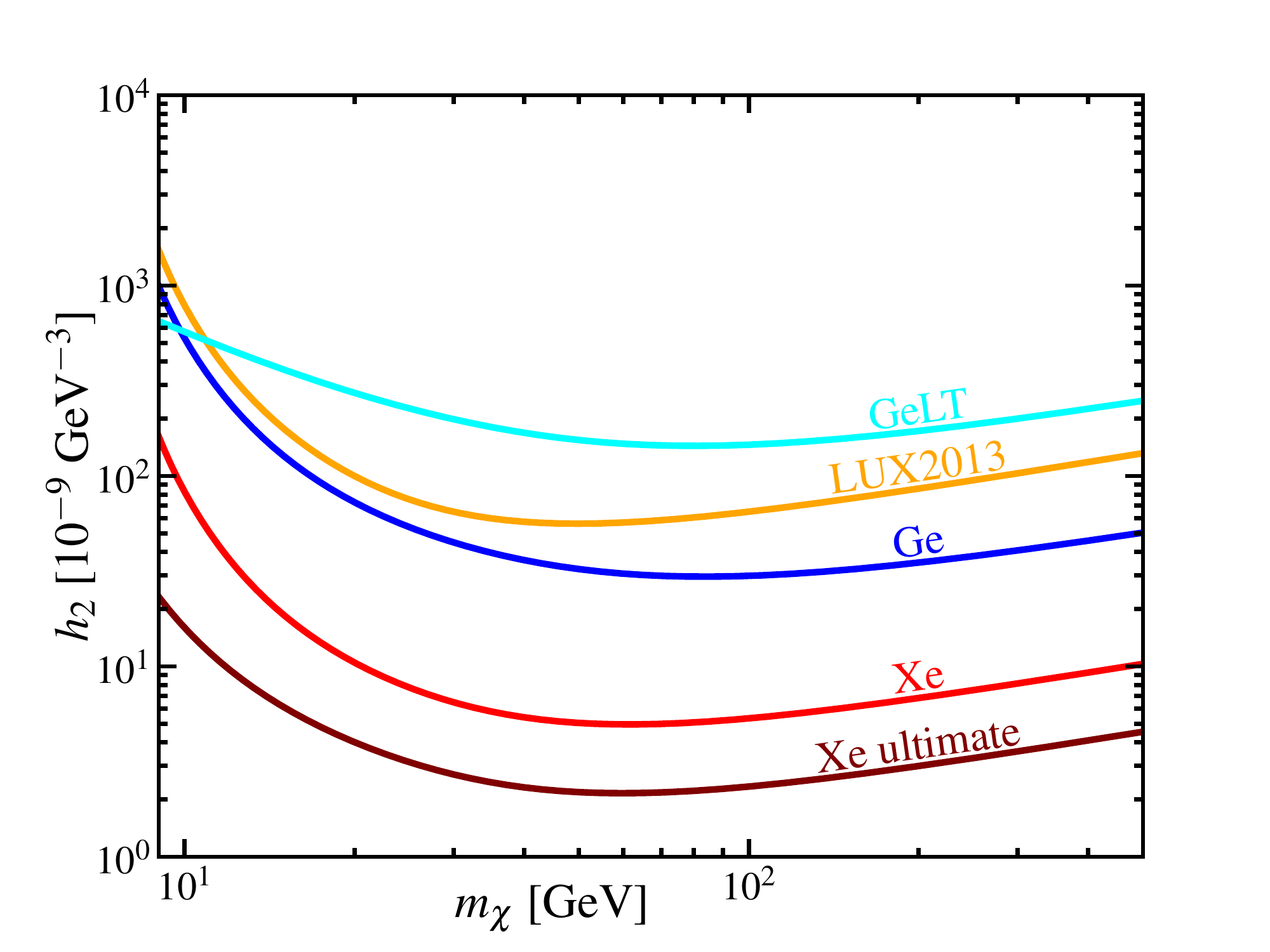}
\includegraphics[width=.48\textwidth,keepaspectratio=true]{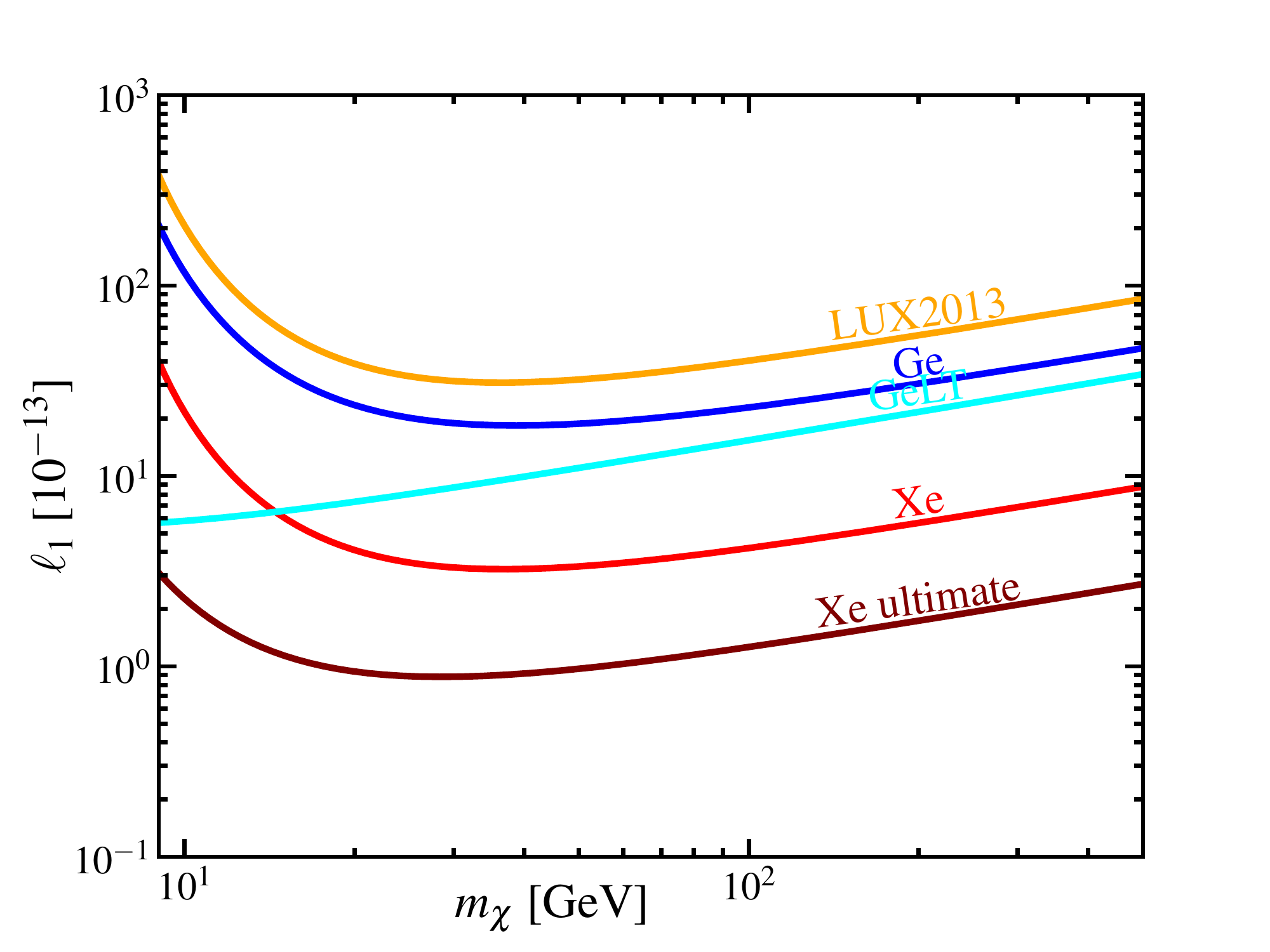}
\includegraphics[width=.48\textwidth,keepaspectratio=true]{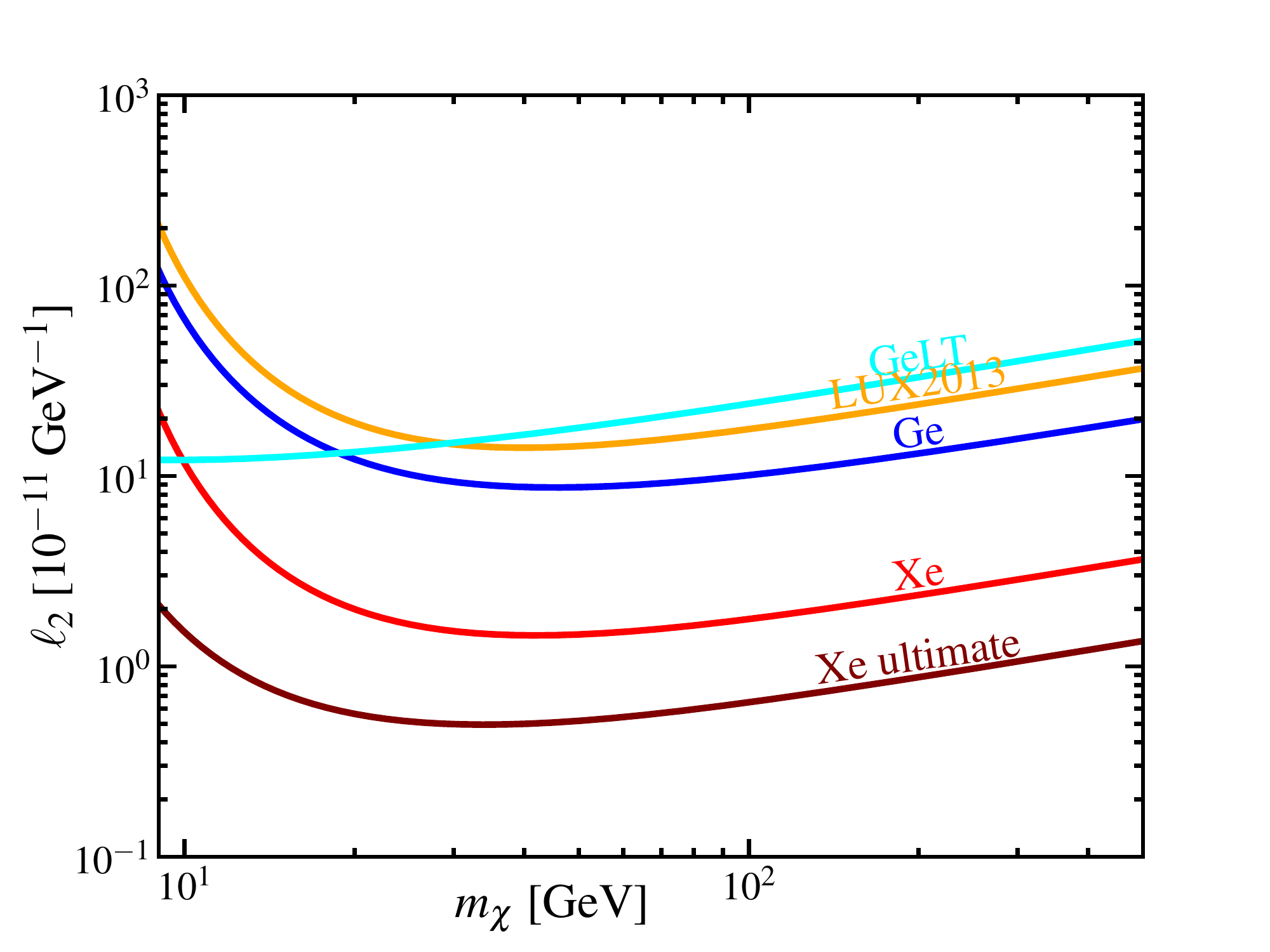}
\caption{Projected exclusion plots. Projected exclusion plots for each of the four effective operators considered in \S\ref{sec:eft}, for an ultimate 10-ton scale liquid-xenon experiment that would reach the irreducible neutrino floor are compared to the corresponding projected exclusions for some of the G2 experiments; see text and Table \ref{tab:experiments} for more details.\label{fig:ultimate_exclusion}}
\end{figure*}
\begin{table*}[h]
  \setlength{\extrarowheight}{3pt}
  \setlength{\tabcolsep}{5pt}
  \begin{center}
	\begin{tabular}{c|m{3cm}m{3cm}m{2cm}m{3cm}}
	$m_\chi$ & $h_1$ [$10^{-10}$ GeV$^{-2}$] & $h_2$ [$10^{-9}$ GeV$^{-3}$] & $\ell_1$ [$10^{-13}$] & $\ell_2$ [$10^{-11}$ GeV$^{-1}$]\\
	\hline\hline
	20 GeV & 0.8 & 4& 0.9& 0.6\\
	50 GeV & 0.6 & 2& 1& 0.5\\
	200 GeV & 0.9 & 3& 2& 0.9\\
	\end{tabular}
  \end{center}
\caption{Projected upper limits on the effective coupling coefficients for scattering operators considered in \S\ref{sec:eft}, computed for the ``Xe ultimate'' experiment that reaches the irreducible neutrino backgrounds. Entries in this Table are extracted from Figure \ref{fig:ultimate_exclusion}.
\label{tab:ultimate_exclusion}}
\end{table*}
In this Section, we explore the prospects of operator selection and parameter estimation for an ``ultimate'' direct-detection experiment that will reach the irreducible neutrino background for WIMP masses above a few GeV. For this purpose, we assume a single experiment with exposure of 10 000 kg-years on a xenon target, labeled ``Xe ultimate'' in Table \ref{tab:experiments}. We first create and analyze single-operator simulations for this experiment---the same simulations as those described in \S\ref{sec:distops_single} where each coupling coefficient is set to its current upper limit, and only one operator used to generate a single simulation. From the analysis analogous to what we used to produce Figures  \ref{fig:lines_m20}, \ref{fig:lines_m50}, and \ref{fig:lines_m200} for G2 experiments, we find that, for ``Xe ultimate'', there is a high chance of future data correctly identifying the momentum dependence of the scattering interaction, if the WIMP signal is just beyond the current experimental reach. This probability is greater than $80\%$ for all WIMP masses above 20 GeV, under all operator models. In Figure \ref{fig:ultimateselection}, we illustrate the improvement in model selection prospects from ``Xe'' to ``Xe ultimate'', for a light WIMP with the $h_1$ operator---the case where model selection is most challenging with G2 experiments (see Figure \ref{fig:lines_m20}).  From this Figure, we see that in this case the chance of correct model selection increases from zero to 94\% when experimental exposure is increased by a factor 5 (and the energy threshold lowered slightly, from 5 keV to 3 keV). On the other hand, in the case that G2 experiments do not detect a WIMP signal (and thus set even more stringent upper limits on the relevant couplings), we find that the chance of successful model selection with ``Xe ultimate'' is vanishingly small. This is because the difference in exposure between LUX (which sets the current constraints) and the ``Xe'' experiment is greater than the increase in exposure ``Xe'' would still need to reach the neutrino floor. 

As for parameter estimation with ``Xe ultimate'', Figures \ref{fig:multimodal_broad_contours_ultimate}, \ref{fig:triangle_ultimate}, and \ref{fig:triangle_all4_ultimate} illustrate similar calculations as Figures \ref{fig:multimodal_broad_contours}, \ref{fig:triangle_G2}, and \ref{fig:triangle_all4_G2} for G2 experiments. The improvement in parameter estimation with respect to G2 is visible, but large parameter degeneracies still remain. Some of these degeneracies could be broken with the addition of other multi-ton experiments with different target materials.  However, xenon-based technologies are reaching better sensitivities faster than other technologies.  If we are restricted to xenon-based experiments in the long-term, most of what is achievable in terms of parameter estimation (prior to reaching neutrino backgrounds) will already be delivered by the G2 experiments, with only small advances possible with the next stage of multi-ton targets.  

Finally, as a guide for future experimental efforts, we calculate approximate ``ultimate'' exclusion curves for each of the four effective couplings considered in this work, in case no WIMP signal is seen when the neutrino limit is reached. For reference, we compare these exclusion curves to those that some of the Generation 2 experiments would be able to produce in case of no detection. The results are shown in Figure \ref{fig:ultimate_exclusion} and Table \ref{tab:ultimate_exclusion}. These exclusions are obtained by setting the total recoil rate for each individual operator to be less than 1, in a given experiment.\footnote{This is a simplified calculation that does not consider any subtle experimental effects and is, furthermore, not done using the Bayesian framework, which was the method of choice for the rest of this study. However, we find this simplified calculation to produce a very effective illustration of the reach of different generations of direct detection experiments, not far from the results a more detailed procedure would yield; for an example, see Ref.~\cite{2014arXiv1406.0524C}.} 
\section{Summary and Discussion}
\label{sec:conclusions}

In this work, we explore the reach of direct-detection experiments in identifying and constraining different non-relativistic effective operators for WIMP-baryon scattering beyond the traditional spin- and momentum-independent contact-interaction operator.  We first quantify how likely Generation-2 experiments are to select out the correct operator.  The main results are shown in Figures \ref{fig:lines_m20}, \ref{fig:lines_m50}, and \ref{fig:lines_m200}), in which we consider the best-case scenario where the WIMP signal is just below the current upper limit set by the recent LUX results. We focus on the impact of the fundamental Poisson-noise limitation only, and ignore backgrounds (discussed in more detail below). We find that discerning the right operator is only likely with a combination of data from different experiments (covering a wide recoil-energy range) in two cases: either when the true model is $\ell_1$ (Coulomb-like interaction through a light mediator), or in the case of a heavy WIMP (with a mass of 200GeV or larger). Other scenarios yield varying degrees of success probability, which is the result of an interplay of several factors (such as momentum-dependence of the operator governing the scattering and the energy threshold of the experiment), rather than a simple monotonic function of the number of observed events. The prospects for a simpler task of distinguishing heavy- from light-mediator scenarios are more optimistic: the chance of success is on the order of $50\%$ (when experimental data is combined) or more, for all underlying operators, except in the canonical $h_1$ case when the WIMP is light. 

We also evaluate the ``ultimate'' prospects of probing WIMP interactions before reaching irreducible neutrino backgrounds. For this purpose, we consider a future 10 ton-year exposure on a xenon target with 3 keV threshold (inspired by the proposed final stage of LZ and the DARWIN experiment). We find that if the signal is just below the current limit, such an experiment will almost certainly be able to reliably distinguish between possible underlying scattering operators, for a large span of WIMP masses (anything above 20 GeV or so), under the assumption that the astrophysical distribution of WIMPs is known perfectly. If Generation 2 does not see a WIMP signal, however, the prospects for distinguishing underlying operators before hitting the neutrino floor are slim. We also evaluate the ultimate upper limits this future experiment would place on each of the four effective couplings discussed in Ref.~\cite{2010JCAP...11..042F}, should it see no WIMP-induced recoil events.

Finally, we probe the shape of the 5-dimensional posterior space where the WIMP mass and all four Wilson coefficients are considered. We find that the posterior tends to be multimodal, with large degeneracies between some parameters. We also find that analyzing data under the assumption of the canonical contact interaction through a heavy mediator leads to a large bias in the WIMP mass estimate in the scenario where the underlying truth is something else. This point raises a note of caution for the discovery stage of WIMP searches, indicating the need for a more agnostic analysis (where several well-motivated effective operators are allowed, and marginalized over), or for performing model selection to establish the right underlying operator prior to estimating the WIMP mass. We also evaluate the quality of parameter estimation both with the upcoming G2 experiments and with a future 10-ton-year xenon-target experiment, in case the signal is just below the current detection limit. 
 
In order to gain an understanding of how experimental parameters (especially the choice of the target material, exposure, and energy window) and the fundamental limitation of Poisson noise affect the potential to probe physics of dark matter, we made several simplifying assumptions in this work, which we now discuss in more detail. First, we assumed a low-background regime for all experiments considered in this study. This assumption is yet to be tested as the new experiments come online. Should its validity be compromised, the optimism of our best-case conclusions will decrease. As shown in previous work \cite{peter2011,2012PhRvD..86f5027K,2012JCAP...01..024F,
2013JCAP...02..041P,2013PhRvL.111c1302K,2013JCAP...10..026D,
2013arXiv1310.5718A}, the addition of realistic backgrounds will weaken WIMP parameter constraints, but not fatally so.  The same will happen if the WIMP signal is far below the upper limits we consider here---the probability of successful model selection, as well as the quality of parameter estimation, will degrade accordingly. 

Second, we assume perfect energy resolution for all experiments. Ref.~\cite{2013arXiv1310.7039P} confirms that parameter estimation from recoil spectra is not significantly affected by finite energy resolution at the level of experiments considered in this study, so we expect that qualitative results from the best-case-scenario analysis presented in this work hold within a more rigorous treatment of experimental and theoretical systematics. 

Third, we ignored all interactions which depend on nuclear spin.  In reality, even minimal extensions to the Standard Model generically predict spin-dependent interactions.  Other work has shown that the ability to distinguish spin-independent from spin-dependent interactions is highly sensitive to the set of target materials available \cite{2011JCAP...10..035P,2013arXiv1310.7039P}.  Finally, following the standard data-analysis approach in the direct-detection community, we ignored astrophysical uncertainties. If the velocity distribution of the WIMP population near Earth is very different from the Standard Halo Model, it can significantly complicate the interpretation of future data sets and change the ability of direct-detection experiments to probe details of dark-matter particle physics \cite{2013arXiv1310.7039P}. The uncertainty in the WIMP velocity distribution will weaken the model-selection probabilities from what we show in this work, but a detailed treatment of this is left for future work.

As a final point in the concluding comments, let us briefly discuss related previous work of Ref.~\cite{2012PhRvD..85l3507M}. This work differs from that presented here in several important ways. First, it considers a set of arbitrarily chosen futuristic experiments, not taking into account current and planned experimental capabilities, and parametrizes possible outcome of future data analysis in terms of the number of events observed. In contrast, this work deals with evaluating the prospects for Generation 2 and futuristic experiments that would reach the neutrino-background floor, and focuses on the currently allowed parameter space for each of the effective operators. Secondly, Ref.~\cite{2012PhRvD..85l3507M} focuses on a single noisy realization of data, claiming that different noise realizations do not affect the ability to extract information from data. However, we do find significant variation in possible outcomes depending on the noise realization, even in the case where the signal is as strong as currently allowed. We therefore particularly focus on quantifying this variation; i.e., ~we evaluate the probability and implications of different analysis outcomes, in an attempt to construct a guideline for future experimental efforts. Thirdly, we observe a strong dependence of the results on the experimental energy window, as well as on the true underlying model, which we quantify in detail. To contrast our results with those of Ref.~\cite{2012PhRvD..85l3507M}, we highlight one example: Ref.~\cite{2012PhRvD..85l3507M} claims the need for 300 events per target for WIMP masses above 50 GeV, and 1000 or more events per target for lower masses, in order to successfully identify the underlying operator. In contrast, while we generally see some positive trend in distinguishing power with the number of observed events, we also find very strong dependence on the mediator and WIMP masses, and the underlying momentum dependence of the scattering. For instance, a successful reconstruction is guaranteed in case of a Coulomb-like interaction with as few as $90$ events per target, even for a WIMP mass of 20 GeV, while the reconstruction is unsuccessful (the data does not distinguish different underlying operators) in the case of a 50 GeV WIMP with a standard heavy-mediator interaction, with the same number of events per target.  Finally, in the context of the currently allowed parameter space, we specifically examine the ``ultimate'' reach of direct-detection searches in terms of probing momentum-dependence of WIMP-baryon scattering before entering the irreducible-background regime.

In conclusion, this study is constructed to serve as a roadmap for direct-detection experiments within the next decade or so (before further progress becomes limited by irreducible neutrino backgrounds), quantifying the prospects and requirements for probing the effective theory of dark matter-baryon interactions. We show that if Generation 2  experiments report null results, the chances of correctly identifying the dominant scattering operator before irreducible backgrounds are reached is vanishingly small. On the other hand, if the WIMP signal is just below the current limit, the prospects are optimistic. In either case, we find that a combination of data from complementary experimental technologies drastically increases the information that can ultimately be gathered from direct searches.

\acknowledgments
We thank M. Reece, J. Fan, Enectali Figueroa, and Daniel McKinsey for useful conversations.  VG gratefully acknowleges support from the Friends of the Institute for Advanced Study. AHGP was supported by NASA grant NNX09AD09G and a McCue Fellowship awarded through the Center for Cosmology at UC Irvine during the early stages of this work.  All the computations in this work were performed at the Institute for Advanced Study cluster facility.
\paragraph{Note added.} During the final stages of preparation of this manuscript, Ref.~\cite{2014arXiv1406.0524C} was submitted to the arXiv.  It too simulates future experiments with a broad set of non-relativistic effective operators. There is, however, a significant difference in the emphasis of the two studies: our work focuses heavily on model selection, a topic which is absent in Ref.~\cite{2014arXiv1406.0524C}.
\bibliographystyle{JHEP}
\bibliography{operators_roadmap_Gluscevic}
\end{document}